\documentclass[aps,showpacs]{revtex4}

\usepackage{graphicx}
\usepackage{epsfig}
\usepackage{bm}
\usepackage{color}
\usepackage{amsmath}

\newcommand{\be}{\begin{eqnarray}}
\newcommand{\ee}{\end{eqnarray}}
\newcommand{\no}{\nonumber \\}
 \newcommand{\gsim}{\mathrel{\hbox{\rlap{\lower.55ex \hbox {$\sim$}}
                   \kern-.3em \raise.4ex \hbox{$>$}}}}
\newcommand{\lsim}{\mathrel{\hbox{\rlap{\lower.55ex \hbox {$\sim$}}
                   \kern-.3em \raise.4ex \hbox{$<$}}}}

\usepackage{amssymb}

\begin{document}


\title{The ``ripples'' on relativistically expanding fluid }
\author{Shuzhe Shi$^{1}$, Jinfeng Liao$^{2,3}$, Pengfei Zhuang$^{1}$}
\address{$^1$ Physics Department, Tsinghua University and Collaborative Innovation Center of Quantum Matter, Beijing 100084, China\\
$^2$ Physics Department and Center for Exploration of Energy and Matter,
Indiana University, 2401 N Milo B. Sampson Lane, Bloomington, IN 47408, USA\\
$^3$ RIKEN BNL Research Center, Bldg. 510A, Brookhaven National Laboratory, Upton, NY 11973, USA}
\date{\today}

\begin{abstract}
Recent studies have shown that fluctuations of various types play important roles in the evolution of the fireball created in relativistic heavy ion collisions  and bear many phenomenological consequences for experimental observables. The bulk dynamics of the fireball is well described by relativistic hydrodynamic expansion and the fluctuations on top of such expanding background can be studied within the linearized hydrodynamic framework. In this paper we present complete and analytic sound wave solutions on top of both Bjorken flow and Hubble flow backgrounds.
\end{abstract}


\maketitle

\section{Introduction}

In relativistic heavy ion collisions, a hot deconfined form of strongly interacting matter, predicted by Quantum Chromodynamics (QCD) and known as the quark-gluon plasma (QGP), has been discovered \cite{sQGP_review,sQGP_ES,BraunMunzinger:2009zz,Fodor:2009ax}. Such fireball of hot QCD matter with high initial energy density, once created in a heavy ion collision, will violently explode outwards and become cooler and more dilute with time. The system will eventually experience the ``freeze-out'' after which thousands of produced hadrons individually fly away from the collision zone. Dedicated detectors have been built to detect the identity and momentum information of these hadrons. Such heavy ion collision experiments are now done at both the Relativistic Heavy Ion Collider (RHIC)~\cite{Adams:2005dq,Adcox:2004mh,Arsene:2004fa,Back:2004je} and the Large Hadron Collider (LHC)~\cite{Muller:2012zq}.

As it turns out, a significant part of the rather complicated dynamical evolution of the created fireball can be very successfully described by relativistic hydrodynamics with the inclusion of viscous corrections~\cite{Heinz:2009xj,Heinz:2013th,Ollitrault:2008zz,Schaefer:2014awa,Song:2013gia}. In view of the smallish size of the fireball itself at $10-$fm scale, this phenomenological success is highly nontrivial, implying an extremely short dissipative length scale in the fluid. Significant efforts have been made to quantify the dissipative transport properties e.g. the shear viscosity normalized by entropy density $\eta/s$ which is found to be much smaller than other known substances~\cite{Csernai:2006zz,Lacey:2006bc,Liao:2009gb}.  Dating back to the famous boost-invariant solution commonly known as Bjorken flow~\cite{Hwa:1974gn,Bjorken:1982qr}, there have also been persistent efforts in finding analytic solutions to relativistic hydrodynamic equations that may bear relevance to certain features of the expanding fluid in heavy ion collisions \cite{Bialas:2007iu,Csorgo:2006ax,Nagy:2007xn,Csorgo:2013ksa,Liao:2009zg,Lin:2009kv,Gubser:2010ze,Gubser:2010ui,Gubser:2012gy,Hatta:2014gqa,Marrochio:2013wla,Pedraza:2014moa,Csanad:2014dpa}.

More recently there have been a lot of interests in studying the influence of fluctuations in such a relativistically expanding fluid~\cite{Kapusta:2011gt,Ling:2013ksb,Springer:2012iz,Kapusta:2014dja,Staig:2010pn,Staig:2011wj,Shuryak:2013uaa,Shuryak:2013cja,Alver:2010gr,Alver:2010dn,Sorensen:2008bf, Teaney:2010vd,Qiu:2011iv,Takahashi:2009na,Schenke:2010rr,Xu:2010du,Qin:2010pf,Schenke:2012wb,Zhang:2012ie,Bloczynski:2012en,Fogaca:2014gwa}. As clearly categorized in \cite{Kapusta:2011gt}, there are different sources of fluctuations: initial-state fluctuations occurring at the very beginning of hydrodynamic expansion; hydrodynamic fluctuations arising from intrinsic thermal fluctuations of finite local fluid cells and happening all along the evolution; fluctuations induced by depositions from processes ``external'' to the fluid such as a penetrating jet; and finally freeze-out fluctuations occurring at the very end of hydrodynamic expansion when the fluid ``falls apart''. Except the freeze-out fluctuations, all the other three types of fluctuations create ``ripples''  propagating as sound waves on top of the expanding fluids, like the ripples created by throwing a stone into a stream.  These ripples co-evolve with the bulk flow toward the end and lead to measurable effects such as specific rapidly and azimuthal angle correlation patterns in heavy ion experiments~\cite{Wang:2013qca,Wang:2014sfa}. The most extensively studied fluctuation is probably the initial-state fluctuation which has been found to manifest through the so-called ``ridge'' correlation~\cite{Staig:2010pn,Staig:2011wj,Shuryak:2013uaa,Shuryak:2013cja,Alver:2010gr,Alver:2010dn,Sorensen:2008bf,Teaney:2010vd,Qiu:2011iv,Takahashi:2009na,Schenke:2010rr,Xu:2010du, Qin:2010pf,Schenke:2012wb,Wang:2013qca}. There has also been a lot of investigations on the hard-soft particle correlations as a result of fluctuations induced by jet energy loss (see e.g. \cite{Wang:2014sfa} and references therein). The influence of fluctuations from stochastic  hydrodynamics on observables has only been explored  very recently~\cite{Kapusta:2011gt,Ling:2013ksb,Springer:2012iz}.

To be specific, there are two interesting problems in studying such fluctuations. The first is how such fluctuations come about, i.e. what physics generates/dominates those fluctuations and how to quantify them. To this end, different types of fluctuations pertain to quite different physical processes and need to be studied separately. For example, the initial-state fluctuations depend on the correct description of initial nuclear wave functions~\cite{Iancu:2002xk,Gelis:2010nm} as well as the pre-equilibrium evolution process~\cite{Gelis:2012ri,Berges:2012ks,Huang:2014iwa}, and there are different initial-state models predicting different amounts of fluctuations. The hydrodynamic fluctuations on the other hand should be treated by a proper account of stochastic fluctuations in local fluid cell~\cite{Kapusta:2011gt}.

The second problem is {\it how such fluctuations, once created, evolves with the relativistically expanding fluid}. This latter problem is universal for different types of fluctuations and can be well studied by treating the fluctuations as perturbations within the linearized hydrodynamics framework. The propagation of such ripples (i.e. the sound waves) critically depends on the background flow: e.g. the ripples from a thrown stone would look very different in a static pond than that in a flowing stream. It is of great interest to know how these fluctuations evolve on top of various relativistically expanding backgrounds. It is also highly desirable to have analytic solutions which allow convenient applications to studying the phenomenological consequences from various types of fluctuations. In the present work, we focus on this problem and present complete and analytic sound wave solutions on top of both Bjorken flow and Hubble flow backgrounds.

The paper is organized as follows. In Section \ref{s2} we present the general formalism to treat the evolution of fluctuations on top of ideal hydrodynamics. In sections \ref{s3} and \ref{s4} we analytically derive the space-time evolution of any fluctuation on top of Bjorken flow and Hubble flow and especially consider the Gaussian perturbation as an example. The numerical results and discussions in these two cases are shown in Section \ref{s5}. We summarize in Section \ref{s6}.

\section{Generalities}
\label{s2}
Let us first set up the linearized hydrodynamics framework to be used later. In a general space-time coordinate frame, the   hydrodynamic  system with vanishing conserved charge density can be described by the energy-momenta conservation equations:
\begin{eqnarray} \label{eq_hydro}
 {T^{\mu \nu}}_{;\mu}=0
\end{eqnarray}
The energy-momentum tensor in the ideal hydrodynamic limit, which we shall adopt throughout the present paper, is given by the following form
\begin{eqnarray} \label{eq_hydro_2}
 T^{\mu \nu}=(\epsilon + p) u^\mu u^\nu- g^{\mu \nu}p,
\end{eqnarray}
where $\epsilon$ is the energy density and $p$ is the pressure, related to each other by the  equation of state for the underlying system, $p=c_s^2 \epsilon$ with $c_s$ the speed of sound. For generality and to clearly reveal the role of $c_s$, we will keep $c_s$ symbolically in most of the derivation. For applications and numerical results related to high temperature quark-gluon plasma, we use $c_s=1/\sqrt3$ which applies to an ideal relativistic plasma as well as any conformal fluid.
 The four-velocity of the fluid is $u^\mu = \gamma (1,\vec v)$ with $\vec v$ the three-velocity and $\gamma=1/\sqrt{1-\vec v^2}$ the Lorentz factor. The four velocity is subject to the constraint $u^\mu u_\mu =1$. In hydrodynamics the degrees of freedom are these five fields (energy density, pressure, and three independent velocity components) which satisfy five equations: four equations from energy and momentum conservation (\ref{eq_hydro}) and one from the equation of state, forming a closed equation system. Finally $g^{\mu \nu}$ is the metric tensor in the used coordinate.
 Note that in general coordinates, the derivative  $_{;\mu}$ in the hydro equations should be a covariant one, defined as
\begin{eqnarray}
T^{\mu\nu}_{~~~;\mu} &=& T^{\mu\nu}_{~~~,\mu}
+ \Gamma^\mu_{~\rho\mu}T^{\rho\nu} + \Gamma^\nu_{~\rho\mu}T^{\mu\rho} \nonumber\\
&=& \partial_\mu T^{\mu\nu}
+ \Gamma^\mu_{~\rho\mu}T^{\rho\nu} + \Gamma^\nu_{~\rho\mu}T^{\mu\rho}
\end{eqnarray}•
where the affine connections are $\Gamma^{\rho}_{~\mu\nu}
=\frac{1}{2}g^{\rho\sigma} ( g_{\sigma\mu,\nu}+g_{\sigma\nu,\mu}-g_{\mu\nu,\sigma} ) $.

The sound wave is a collective excitation arising from small density and pressure fluctuations on top of certain background. One may treat such fluctuations as a perturbation which shall satisfy the linearized hydrodynamic equations. To do that, consider   certain background flow described by $p_0$ and $u_0^\mu$, already satisfying the hydrodynamic equations in (\ref{eq_hydro}).
 Let us then add a sound wave as a perturbation, with the hydro fields now being $p=p_0+p_1,~u^\mu=u_0^\mu+u_1^\mu$.
Hence, to the linear order in the perturbation, the energy-momentum tensor can be written as
\begin{eqnarray}
T^{\mu\nu} &=& {(1+c_s^{-2})} p u^\mu u^\nu - g^{\mu\nu}p \nonumber\\
&=& {(1+c_s^{-2})} (p_0+p_1) (u_0^\mu+u_1^\mu)(u_0^\nu+u_1^\nu) - g^{\mu\nu}(p_0+p_1) \nonumber\\
&\approx&  \left\{ T_0^{\mu\nu} \right \}  +   \left\{ T_1^{\mu\nu} \right \} \nonumber \\
&=& \left \{ {(1+c_s^{-2})} p_0 u_0^\mu u_0^\nu - g^{\mu\nu}p_0 \right \} + \left \{ {(1+c_s^{-2})} u_0^\mu u_0^\nu p_1 + {(1+c_s^{-2})} p_0 u_0^\nu u_1^\mu + {(1+c_s^{-2})} p_0 u_0^\mu u_1^\nu - g^{\mu\nu} p_1 \right\}.
\end{eqnarray}
Note that the velocity perturbation is subject to the constrain  $u_0^\mu {u_1}_\mu=0$. Thus the linearized hydrodynamic equations are simply given by
 \begin{eqnarray} \label{eq_hydro_linear}
 {T_1^{\mu \nu}}_{;\mu}=0.
\end{eqnarray}
The solutions to such linearized hydrodynamic equations are the sound waves on a general background. While the sound wave solutions on static and homogeneous background are familiar, the extension to expanding background is highly nontrivial. Clearly the sound wave solution depends on the background flow which itself shall be a solution to hydrodynamic equation in the first place. In what follows, we will find sound wave solutions to the linearized hydro equations above, based on known exact solutions as the background flow. As will become evident in the explicit examples later, for a given background flow there will be multiple sound wave solutions in general, and a given (arbitrary) initial perturbation will trigger a certain superposition of these solutions which subsequently propagate independently. These sound waves generated by a common source perturbation propagate away and create correlations over the spatial range of such sound propagation.

\section{Sound waves on top of 1D Bjorken flow}
\label{s3}
As the first example let us consider the linearized hydrodynamic equation on the background solution known as the boost-invariant Bjorken flow, which provides a good description of hot QCD fluid undergoing longitudinal expansion at relatively early time in a  heavy ion collision.

\subsection{The linearized hydrodynamic equations }
As is well known, the Bjorken flow is a 1-dimensional flow along the longitudinal direction, with only $z$-direction flow velocity $v_z=z/t$. To describe this background flow, it is most  convenient to use the following coordinates:
\begin{eqnarray}
\tau &=& \sqrt{t^2-z^2} ,~~~~~~
\eta = \frac{1}{2} \ln \frac{t+z}{t-z}, \nonumber\\
\rho &=& \sqrt{x^2+y^2} ,~~~~~
\phi = \frac{1}{2i} \ln \frac{x+iy}{x-iy},
\end{eqnarray}•
with inverse transformation
\begin{eqnarray}
t &=& \tau~\cosh\eta, ~~~~~
z ~=~ \tau~\sinh\eta,\nonumber\\
x &=& \rho~\cos\phi, ~~~~~~~
y ~=~ \rho~\sin\phi.
\end{eqnarray}•

In the $(\tau, \eta, \rho, \phi)$ coordinates, the metric tensor $g_{\mu\nu}$ is given by
\begin{eqnarray}
g_{\mu\nu}=\mathrm{Diag}(1,-\tau^2,-1,-\rho^2),\nonumber\\
g^{\mu\nu}=\mathrm{Diag}(1,-\frac{1}{\tau^2},-1,-\frac{1}{\rho^2})
\end{eqnarray}
and the non-vanishing connections are
\begin{eqnarray}
\Gamma^{\tau}_{\eta\eta}=\tau,~~~\Gamma^{\eta}_{\eta\tau}=\Gamma^{\eta}_{\tau\eta}=\frac{1}{\tau},\nonumber\\
\Gamma^{\rho}_{\phi\phi}=-\rho,~~~\Gamma^{\phi}_{\phi\rho}=\Gamma^{\phi}_{\rho\phi}=\frac{1}{\rho}.
\end{eqnarray}•

The hydrodynamic equations (\ref{eq_hydro}) in this coordinate system take the following form:
\begin{eqnarray}
0&=&T^{\tau\tau}_{~~,\tau}+T^{\tau\eta}_{~~,\eta}+T^{\tau\rho}_{~~,\rho}+T^{\tau\phi}_{~~,\phi}+\tau T^{\eta\eta}+\frac{1}{\tau} T^{\tau\tau}+\frac{1}{\rho}T^{\rho\tau},\nonumber\\
0&=&T^{\eta\tau}_{~~,\tau}+T^{\eta\eta}_{~~,\eta}+T^{\eta\rho}_{~~,\rho}+T^{\eta\phi}_{~~,\phi}+\frac{3}{\tau}T^{\eta\tau}+\frac{1}{\rho}T^{\rho\eta},\nonumber\\
0&=&T^{\rho\tau}_{~~,\tau}+T^{\rho\eta}_{~~,\eta}+T^{\rho\rho}_{~~,\rho}+T^{\rho\phi}_{~~,\phi}-\rho T^{\phi\phi}+\frac{1}{\rho}T^{\rho\rho}+\frac{1}{\tau}T^{\tau\rho},\nonumber\\
0&=&T^{\phi\tau}_{~~,\tau}+T^{\phi\eta}_{~~,\eta}+T^{\phi\rho}_{~~,\rho}+T^{\phi\phi}_{~~,\phi}+\frac{3}{\rho}T^{\rho\phi}+\frac{1}{\tau}T^{\tau\phi}.
\end{eqnarray}
The background Bjorken flow is a solution to the above equations, specified by pressure field ${p_0(\tau) = p(\tau_0) \tau_0^{1+c_s^2}/\tau^{1+c_s^2}}$ and velocity field $u_0^\mu(\tau)=(1,0,0,0)$ in this coordinate.

Let us then consider a small perturbation on top of the Bjorken flow, $p=p_0 + p_1$ and $u^\mu=u_0^\mu+u_1^\mu$. The velocity field constraint $g^{\mu\nu}u_\mu u_\nu=1$ requires $u_1^\tau=0$.   The linearized hydrodynamic equations (\ref{eq_hydro_linear}) are then given by
\begin{eqnarray}\label{eq_bj_linear}
0&=& \frac{p_0 u_1^\rho}{\rho} +\frac{p_1}{\tau} +{\frac{1}{1+c_s^2}}p_{1,\tau} +p_0(u^\eta_{1,\eta}+u^\rho_{1,\rho}+u^\phi_{1,\phi}),\nonumber\\
0&=& p_0 u^\eta_{1,\tau}+{\frac{2-c_s^2}{\tau}}p_0 u_1^\eta + {\frac{c_s^2}{1+c_s^2}} \frac{p_{1,\eta}}{\tau^2},\nonumber\\
0&=& p_0 u^\rho_{1,\tau}-{\frac{c_s^2}{\tau}}p_0 u_1^\rho + {\frac{c_s^2}{1+c_s^2}}p_{1,\rho},\nonumber\\
0&=& p_0 u^\phi_{1,\tau}-{\frac{c_s^2}{\tau}}p_0 u_1^\phi + {\frac{c_s^2}{1+c_s^2}}\frac{p_{1,\phi}}{\rho^2}.
\end{eqnarray}

In what follows we will find solutions to the above equations, describing the sound waves propagating on top of the background Bjorken flow. The background flow clearly differentiates the longitudinal and transverse directions: diluting out in the former while remaining static in the latter. It is physically interesting to first examine sound waves propagating solely in the transverse or longitudinal directions to gain intuitions on these waves, as will be done in the next two subsections. In the last subsection we will then find the general analytic solutions for all sound waves.

\subsection{Transverse sound wave solutions}

Let us first consider solutions for transverse sound wave that travels on the plane perpendicular to the background flow's longitudinal expansion direction. To do that, we find solutions with vanishing longitudinal velocity i.e. $u_1^\eta=0$. This condition simplifies the Eqs.(\ref{eq_bj_linear}) into the following:
\begin{eqnarray}
0&=& \frac{p_0 u_1^\rho}{\rho} +\frac{p_1}{\tau} +{\frac{1}{1+c_s^2}}p_{1,\tau} +p_0(u^\rho_{1,\rho}+u^\phi_{1,\phi}), \label{bj1} \\
0&=& p_{1,\eta}, \label{bj2} \\
0&=& p_0 u^\rho_{1,\tau}-{\frac{c_s^2}{\tau}}p_0 u_1^\rho + {\frac{c_s^2}{1+c_s^2}} p_{1,\rho}, \label{bj3} \\
0&=& p_0 u^\phi_{1,\tau}-{\frac{c_s^2}{\tau}}p_0 u_1^\phi +{\frac{c_s^2}{1+c_s^2}}\frac{p_{1,\phi}}{\rho^2}. \label{bj4}
\end{eqnarray}
The Eq.(\ref{bj2}) can be trivially solved by having all quantities independent of $\eta$ i.e. being boost-invariant.  One strategy to solve the remaining equations is to manipulate the equations into a form allowing variable separation procedures (with the ``price'' of elevating to second order differentiations). To see that, one combines them via (\ref{bj1})$_{,\tau}-$(\ref{bj3})$_{,\rho}-$(\ref{bj4})$_{,\phi}+$(\ref{bj1})$ /\tau-$(\ref{bj3})$/{\rho}$ and obtains
\begin{eqnarray}
c_s^{-2} p_{1,\tau\tau}+ (1+2c_s^{-2})\frac{p_{1,\tau}}{\tau}=p_{1,\rho\rho}+\frac{p_{1,\rho}}{\rho}+\frac{p_{1,\phi,\phi}}{\rho^2}.
\end{eqnarray}•
The above equation can then be further solved by usual variable separation. First by doing Fourier expansion of angle dependence, $p_1=\sum_m p_m \mathrm{e}^{i m \phi}$, we can get
\begin{eqnarray}
 c_s^{-2} \, p_{m,\tau\tau}+ (1+2c_s^{-2}) \frac{p_{m,\tau}}{\tau}=p_{m,\rho\rho}+\frac{p_{m,\rho}}{\rho}-\frac{m^2}{\rho^2} p_m .
\end{eqnarray}
A further separation procedure leads to two decoupled second order differential equations for $\tau$ and $\rho$ dependence, and both are easily solved. At the end, we obtain the following solution:
\begin{eqnarray}\label{eq_bj_t}
p_1 &=& p_0 \Big( \frac{\tau}{\tau'}\Big)^{{\frac{1+c_s^2}{2}}} \sum_{m=0,\pm 1, \pm 2, ...} \int
\Big[a_{m,\omega}J_{{\frac{1+c_s^2}{2}}}(c_s\omega\tau) + b_{m,\omega}J_{{-\frac{1+c_s^2}{2}}}(c_s\omega\tau) \Big]
 J_{m}(\omega \rho) \mathrm{e}^{i m \phi} \mathrm{d}\omega ,\nonumber\\
u_1^\rho &=&  u_1^\rho(\tau')  + {\frac{c_s^2}{1+c_s^2}}u_{\perp,\rho},\nonumber\\
u_1^\phi &=& u_1^\phi(\tau') +  {\frac{c_s^2}{1+c_s^2}}\frac{1}{\rho^2}u_{\perp,\phi},\nonumber\\
u_1^\tau&=&u_1^\eta~=~0,
\end{eqnarray}
where $p_0(\tau)$ is the background solution from Bjorken flow,  and the auxiliary field $u_\perp$ is given by
\begin{eqnarray}
u_\perp &= & \Big( \frac{\tau}{\tau'}\Big)^{{c_s^2}} \sum_m \int
\left[{(\frac{\tau}{\tau'})^{\frac{1-c_s^2}{2}}} \, \frac{a_{m,\omega}}{c_s \omega} \,
J_{{\frac{c_s^2-1}{2}}}(c_s\omega\tau)
-{(\frac{\tau}{\tau'})^{\frac{1-c_s^2}{2}}} \, \frac{b_{m,\omega}}{c_s\omega} \,
J_{{\frac{1-c_s^2}{2}}}(c_s\omega\tau) \right]{\bigg |}^{\tau}_{\tau'}
 J_{m}(\omega \rho) \mathrm{e}^{i m \phi} \mathrm{d}\omega
\end{eqnarray}
which satisfies $u_\perp (\tau \to \tau') \to 0$. Note that to ensure the solutions to be real numbers, one has the constraints $a^*_{-m,\omega} = (-1)^m\, a_{m,\omega}$ and $b^*_{-m,\omega} = (-1)^m\, b_{m,\omega}$. The parameter $\tau'$ has the meaning of initial time when the perturbation is introduced, and $u_1^{\rho,\phi}(\tau')$ shall be matched to the initial velocity field perturbation. The coefficients $a_{m,\omega}$ and $b_{m,\omega}$ shall be determined from initial pressure field perturbation and velocity perturbation. 
It should be noted that the transverse sound wave solutions found here are different from those for a completely static background. Here the longitudinally expanding background flow induces dilution of density which affects the sound propagation even in transverse direction. As a result of this nontrivial interplay between the background and the sound wave, nontrivial time dependence appears in the above solution. In contrast, a transverse solution on a completely static background would have its time dependence as simply $\sim e^{\pm i  c_s \omega \tau}$.

It would be interesting to examine the asymptotic behavior of the solution. The Bessel functions behave as $J_m(x)\sim x^{-1/2}\cos(x-\pi/4-m\pi/2)$ when $x\to\infty$. As such one can infer that in the limit of infinite time and distance $\tau \to \infty$ and $\rho \to \infty$, the solution (\ref{eq_bj_t}) takes the following form
\begin{eqnarray}
\frac{p_1}{p_0} \propto \tau^{c_s^2/2} \rho^{-1/2} \cos(c_s\omega\tau-\pi/4\mp\pi/3)\cos(\omega\rho-\pi/4-m\pi/2)
\end{eqnarray}
which appears as a ``standing wave'' from mixture of inbound and outbound sound waves with phase velocity $\frac{\delta\rho}{\delta\tau}=\pm c_s$. From $p_0(\tau)\sim 1/\tau^{-(1+c_s^2)}$,   $p_1$ itself behaves as $\sim 1/\tau^{-(1+c_s^2/2)}$ at late time limit and decreases in time.

Let us now give an explicit example of the solutions, with the initial Gaussian-shape perturbation for the pressure and vanishing velocity fluctuations,
\begin{eqnarray}
\label{gaussian}
p_1(\tau') &=& p_0(\tau')\frac{\xi}{2\pi\sigma^2} e^{-\frac{\rho^2+\rho'^2-2\rho\rho'\cos(\phi-\phi')}{2\sigma^2\tau'^2}},\nonumber\\
u_1(\tau') &=& 0,
\end{eqnarray}
where there are two dimensionless parameters, $\sigma$ controlling the width of the Gaussian perturbation and $\xi$ governing the magnitude of the perturbation relative to the background pressure. By matching the solution (\ref{eq_bj_t}) with the above initial condition at $\tau'$, we determine all coefficients, and the resulting sound wave from such Gaussian fluctuation is described by
\begin{eqnarray}
\frac{p_1(\tau,\rho,\phi)}{p_0} &=& \frac{\xi \, \tau'^2}{2\pi}\Big( \frac{\tau}{\tau'}\Big)^{{\frac{1+c_s^2}{2}}} \sum_m  \int_0^\infty \frac{J_{{\frac{1-c_s^2}{2}}}(c_s \omega\tau')J_{{\frac{1+c_s^2}{2}}}(c_s \omega\tau)
+J_{-{\frac{1-c_s^2}{2}}}(c_s \omega\tau')J_{-{\frac{1+c_s^2}{2}}}(c_s \omega\tau)}
{J_{{\frac{1-c_s^2}{2}}}(c_s \omega\tau')J_{{\frac{1+c_s^2}{2}}}(c_s \omega\tau')
+J_{-{\frac{1-c_s^2}{2}}}(c_s \omega\tau')J_{-{\frac{1+c_s^2}{2}}}(c_s \omega\tau')}
\no&&~~~\times e^{-\frac{\sigma^2\omega^2\tau'^2}{2}}
 J_{m}(\omega \rho) J_{m}(\omega \rho') \mathrm{e}^{i m (\phi-\phi')}
\omega~ \mathrm{d}\omega .  \label{eq.BT}
\end{eqnarray}
With the very useful sum rule for Bessel functions \cite{handbook},
\begin{eqnarray}
\sum_{m=-\infty}^{\infty} J_m(x)J_m(x')e^{im\phi} = J_0(\sqrt{x^2+x'^2-2xx'\cos\phi})
\equiv J_0(|\vec x-\vec x'|) , .
\end{eqnarray}
we can perform the summation in the solution and obtain
\begin{eqnarray}
\frac{p_1(\tau,\rho,\phi)}{p_0} &=& \frac{\xi \, \tau'^2}{2\pi}\Big( \frac{\tau}{\tau'}\Big)^{\frac{2}{3}}   \int_0^\infty \frac{J_{{\frac{1-c_s^2}{2}}}(c_s \omega\tau')J_{{\frac{1+c_s^2}{2}}}(c_s \omega\tau)
+J_{-{\frac{1-c_s^2}{2}}}(c_s \omega\tau')J_{-{\frac{1+c_s^2}{2}}}(c_s \omega\tau)}
{J_{{\frac{1-c_s^2}{2}}}(c_s \omega\tau')J_{{\frac{1+c_s^2}{2}}}(c_s \omega\tau')+J_{-{\frac{1-c_s^2}{2}}}(c_s \omega\tau')J_{-{\frac{1+c_s^2}{2}}}(c_s \omega\tau')}
  e^{-\frac{\sigma^2\omega^2\tau'^2}{2}}
 J_{0}(\omega \bar\rho)   \omega  \mathrm{d}\omega ,  \quad \label{eq_BT}
\end{eqnarray}
where $\bar\rho \equiv |\vec \rho - \vec \rho'|=\sqrt{\rho^2+\rho'^2-2\rho\rho'\cos(\phi-\phi')}$ is the distance between the point $(\rho,\phi)$ and the center of the original perturbation at $(\rho',\phi')$.
Thus the physical picture of the above transverse wave becomes transparent: it is a cylindrically symmetric wave  propagating away from the center of the initial perturbation.  The corresponding sound wave velocity field is given by
\begin{eqnarray}\label{eq_BTT}
u_1^\rho &=&
-\frac{\xi\tau'^2}{8\pi c_s}\Big( \frac{\tau}{\tau'}\Big)^{\frac{2}{3}} \int_0^\infty
\frac{J_{{\frac{1-c_s^2}{2}}}(c_s\omega\tau')J_{-{\frac{1-c_s^2}{2}}}(c_s\omega\tau)-J_{-{\frac{1-c_s^2}{2}}}(c_s\omega\tau')J_{{\frac{1-c_s^2}{2}}}(c_s\omega\tau)}
{J_{{\frac{1-c_s^2}{2}}}(c_s\omega\tau')J_{{\frac{1+c_s^2}{2}}}(c_s\omega\tau')+J_{-{\frac{1-c_s^2}{2}}}(c_s\omega\tau')J_{-{\frac{1+c_s^2}{2}}}(c_s\omega\tau')} \nonumber \\
&&\qquad \qquad \qquad \qquad \times e^{-\frac{\sigma^2\omega^2\tau'^2}{2}}  J_{1}(\omega \bar \rho) \frac{\rho-\rho'\cos(\phi-\phi')}{\bar \rho}  \omega  \mathrm{d}\omega,\nonumber\\
u_1^\phi &=& -\frac{\xi\tau'^2}{8\pi c_s}\Big( \frac{\tau}{\tau'}\Big)^{\frac{2}{3}} \int_0^\infty
\frac{J_{{\frac{1-c_s^2}{2}}}(c_s\omega\tau')J_{-{\frac{1-c_s^2}{2}}}(c_s\omega\tau)-J_{-{\frac{1-c_s^2}{2}}}(c_s\omega\tau')J_{{\frac{1-c_s^2}{2}}}(c_s\omega\tau)}
{J_{{\frac{1-c_s^2}{2}}}(c_s\omega\tau')J_{{\frac{1+c_s^2}{2}}}(c_s\omega\tau')+J_{-{\frac{1-c_s^2}{2}}}(c_s\omega\tau')J_{-{\frac{1+c_s^2}{2}}}(c_s\omega\tau')}  \nonumber \\
&&\qquad \qquad \qquad \qquad \times   e^{-\frac{\sigma^2\omega^2\tau'^2}{2}}  J_{1}(\omega \bar \rho) \frac{ \rho'\sin(\phi-\phi')}{\rho \bar \rho}  \omega  \mathrm{d}\omega .
\end{eqnarray}

Finally we consider the case of Dirac-delta form initial perturbation $p_1(\tau') =p_0(\tau')\cdot\tau'^2\delta^{(2)}(\vec\rho - \vec \rho') =p_0(\tau') \cdot \tau'^2\frac{\delta(\rho)}{\pi\rho}$ which can be obtained as a proper limit $\sigma\to0$ of the Guassian case. A careful calculation reveals that the transverse wave solution from such a completely localized initial perturbation can be obtained by simply putting $\sigma=0$ in the above solution given by Eqs.(\ref{eq_BT}) and (\ref{eq_BTT}).

\subsection{Longitudinal sound wave solutions}

Let us then consider solutions for longitudinal sound wave that travels in parallel to the background flow's expansion direction. To do that, we find solutions with vanishing transverse velocity i.e. $u_1^\rho=u_1^\phi=0$. This condition greatly simplifies the Eqs.(\ref{eq_bj_linear}),
\begin{eqnarray}
0&=&\frac{p_1}{\tau} +{\frac{1}{1+c_s^2}}p_{1,\tau} +p_0 u^\eta_{1,\eta}, \nonumber\\
0&=& p_0 u^\eta_{1,\tau}+{\frac{2-c_s^2}{\tau}}p_0 u_1^\eta + {\frac{c_s^2}{1+c_s^2}}\frac{p_{1,\eta}}{\tau^2}.
\end{eqnarray}
It is straightforward to find the solution,
\begin{eqnarray}\label{eq_bj_L}
p_1 &=& p_0 \Big(\frac{\tau}{\tau'}\Big)^{-\frac{1-c_s^2}{2}} \int_{-\infty}^{\infty}dk~
\Big[a_k e^{i[k\eta-c_s\sqrt{k^2-\frac{(1-c_s^2)^2}{4c_s^2}}\ln(\frac{\tau}{\tau'})]}+
b_k e^{i[k\eta+c_s\sqrt{k^2-\frac{(1-c_s^2)^2}{4c_s^2}}\ln(\frac{\tau}{\tau'})]}\Big],\nonumber\\
u_1^\eta &=& \Big(\frac{\tau}{\tau'}\Big)^{-\frac{3-c_s^2}{2}} \int_{-\infty}^{\infty}dk~
\Big[\frac{-i(1-c_s^2)-\sqrt{4c_s^2k^2-(1-c_s^2)^2}}{2k(1+c_s^2)}a_k e^{i[k\eta-c_s\sqrt{k^2-\frac{(1-c_s^2)^2}{4c_s^2}}\ln(\frac{\tau}{\tau'})]}  \nonumber \\&&
+\frac{-i(1-c_s^2)+\sqrt{4c_s^2k^2-(1-c_s^2)^2}}{2k(1+c_s^2)}b_k e^{i[k\eta + c_s\sqrt{k^2-\frac{(1-c_s^2)^2}{4c_s^2}}\ln(\frac{\tau}{\tau'})]}\Big]
\end{eqnarray}
with $k$ the dimensionless longitudinal wave number (in ``conjugation'' to spatial rapidity $\eta$). We note that this particular solution, i.e. longitudinal wave on top of Bjorken background flow has been studied in \cite{Kapusta:2011gt}, and the above solution agrees precisely with that found in \cite{Kapusta:2011gt}.
The small coefficients $a_k,b_k \ll 1$ are determined by matching with initial perturbation at time $\tau'$. They should also satisfy the following constraints  to ensure all the above physical quantities to be real numbers, $a^*_{-k}=b_k$ (thus $b^*_{-k}=a_k$) for $|k| >  \frac{1-c_s^2}{2c_s} $ and $a^*_{-k}=a_k$ and $b^*_{-k}=b_k$ for $|k|< \frac{1-c_s^2}{2c_s} $ and  either of the two conditions for $|k|= \frac{1-c_s^2}{2c_s} $.

Note that the delicate structure $\sqrt{k^2-\frac{(1-c_s^2)^2}{4c_s^2}} $ can be either a real number or an imaginary number. For $|k| > \frac{1-c_s^2}{2c_s}$ (corresponding to short ``wavelength'' or well localized modes), the solution has an oscillating time dependence which resembles a propagating wave. The case of $|k|=  \frac{1-c_s^2}{2c_s} $ is trivial, without any oscillating phase in time. For $0<|k|< \frac{1-c_s^2}{2c_s} $, the behavior of the sound wave becomes quite interesting. Naively it looks like exponentially growing or decaying in time and one may worry about possible instability due to the exponential growth, but that is not true due to the $ \ln(\tau/\tau') $ structure in the exponential term. In fact for  $0<|k|< \frac{1-c_s^2}{2c_s} $ we can write  $\sqrt{k^2-\frac{(1-c_s^2)^2}{4c_s^2}}=i \sqrt{\frac{(1-c_s^2)^2}{4c_s^2}-k^2}$  and then have $e^{i [\pm i c_s \sqrt{ {\frac{(1-c_s^2)^2}{4c_s^2}}-k^2} { \ln(\tau/\tau')]}}\sim (\tau/\tau')^{\pm c_s \sqrt{ {\frac{(1-c_s^2)^2}{4c_s^2}}-k^2}}$. For $c_s \sqrt{ {\frac{(1-c_s^2)^2}{4c_s^2}} - k^2} <  {\frac{1-c_s^2}{2}}$, even with the positive power, neither the pressure nor the velocity field of the sound wave would grow in time, instead they both decrease in time by a power law dependence on time.    Physically this type of behavior may be understood as follows. For the small $|k|$ modes their wavelength becomes so large that  the background Bjorken flow will stretch the different parts inside the same wavelength away from each other significantly (recalling that the Bjorken flow is just a 1D Hubble flow in which each local fluid cell ``sees'' all other cells expanding away) and thus render the usual time oscillation no longer possible. The $k=0$ case requires some special discussion. In this case the pressure is finite, but the velocity field integration bears a logarithmic singularity $\sim \int dk / k$ at small $k$. However the leading order of the integrand is an odd function  for $k\to 0$ and thus the divergence will be canceled.

One may be curious about the behavior of the solutions under longitudinal boost. One may boost to a reference frame with a velocity $\tilde v$ along longitudinal direction relative to the original flow by  the following coordinate transformation:
\begin{eqnarray}
\tau&\to&\tau ,\nonumber\\
\eta&\to&\eta+\frac{1}{2}\mathrm{ln}\frac{1+\tilde v}{1- \tilde v} = \eta + \delta \eta .
\end{eqnarray}
Note that in the solutions (\ref{eq_bj_L}) the $\eta$ dependence is entirely in $e^{ik\eta}$, and upon the above transformation one gets  extra factors $e^{i k \delta \eta}$ which can all be absorbed into a redefinition of the coefficients $a_k$ and $b_k$ (and it is not difficult to see that such redefinition satisfies the constraints for these coefficients). One therefore sees that the longitudinal sound wave solution has a boost-invariant form.

Let us also look at the dispersion relation for the propagating modes with $|k|>  {\frac{1-c_s^2}{2c_s}}$. From the oscillating phase $e^{i[k\eta\mp c_s \sqrt{k^2- {\frac{(1-c_s^2)^2}{4c_s^2}}}\ln(\frac{\tau}{\tau'})]}$, one may identify a phase velocity $\frac{\tau \delta \eta}{\delta \tau}=\pm c_s \sqrt{1- {\frac{(1-c_s^2)^2}{4c_s^2k^2}}}$ which approaches $c_s$ for large wavenumber $k\to \infty$. It is instructive to also look at the dispersion in the flat coordinate $\frac{\delta z}{\delta t} = \frac{\tanh\eta\pm c_s}{1\pm c_s \tanh\eta } = \frac{v_z \pm c_s }{1 \pm v_z c_s}$  where $v_z=z/t$ is the local background flow velocity. Clearly the phase velocity in flat coordinate has the interpretation of red/blue shifted wave propagation by the local background flow.

As an example, let us examine the evolution of a static Gaussian  perturbation introduced at $\tau=\tau'$ and $\eta=\eta'$,
\begin{eqnarray}
\label{gaussing2}
p_1(\tau') &=&   p_0(\tau')\frac{\xi}{\sqrt{2\pi}\sigma} e^{-\frac{(\eta-\eta')^2}{2\sigma^2}},\nonumber\\
u_1^\eta(\tau') &=& 0.
\end{eqnarray}
The corresponding sound wave resulting from such perturbation can be written as
\begin{eqnarray} \label{eq_bj_L_G}
\frac{p_1(\tau,\eta)}{p_0} &=& \frac{\xi}{2\pi} \Big(\frac{\tau}{\tau'}\Big)^{-\frac{1-c_s^2}{2}} \int_{-\infty}^{\infty}dk~ e^{-\frac{\sigma^2 k^2}{2}}
\cos[k(\eta-\eta')]\cos[c_s\sqrt{k^2-\frac{(1-c_s^2)^2}{4c_s^2}}\ln(\tau/\tau')] ,  \nonumber\\
u_1^\eta &=& \frac{\xi}{4\pi} \Big(\frac{\tau}{\tau'}\Big)^{-\frac{3-c_s^2}{2}} \int_{-\infty}^{\infty}dk~ e^{-\frac{\sigma^2 k^2}{2}}\frac{\sin[k(\eta-\eta')]}{k}
\bigg (\frac{1-c_s^2}{1+c_s^2}\cos\left[c_s\sqrt{k^2-\frac{(1-c_s^2)^2}{4c_s^2}}\ln(\tau/\tau')\right] \nonumber\\ &&
- \frac{\sqrt{4c_s^2 k^2-(1-c_s^2)^2}}{1+c_s^2}\sin\left[c_s\sqrt{k^2-\frac{(1-c_s^2)^2}{4c_s^2}}\ln(\tau/\tau')\right ] \bigg) .
\end{eqnarray}
Note that there is no divergence of  the integrand in the above velocity field in the $k\to 0$ region by virtue of the fact $\sin[k(\eta-\eta')]\sim k$.

Similarly, by taking the limit $\sigma\to0$ in the above equations, one obtains the solution with Dirac-delta initial perturbation. An interesting point, as discussed in \cite{Kapusta:2011gt}, is that the sound wave solution originating from a delta-function perturbation shall bear a singular ``sound front'' if viscosity is neglected. Indeed, our sound wave solution from Guassian perturbation in the $\sigma \to 0$ limit shows the existence of such singularity. To see that, let us take the limit $\sigma \to 0$ and examine the behavior of $p_1/p_0$ in Eq.(\ref{eq_bj_L_G}), for which the $k$-integration part can be rewritten as 
\begin{eqnarray}
&&	\int_{-\infty}^{\infty} dk \cos[k(\eta-\eta')] \cos\left[c_s\sqrt{k^2-\frac{(1-c_s^2)^2}{4c_s^2}}\ln(\tau/\tau')\right]  \nonumber \\
&=&	\int_{-\infty}^{\infty} dk \cos[k(\eta-\eta')] \cos[c_s k \ln(\tau / \tau')]  \nonumber\\
&&        +	\int_{-\infty}^{\infty} dk \cos[k(\eta-\eta')]  \left[ \cos[c_s k \sqrt{1-\frac{(1-c_s^2)^2}{4c_s^2k^2}}\ln(\tau / \tau')] - \cos[c_s k \ln(\tau / \tau')] \right]   \nonumber \\
&=&	\frac{1}{2} \int_{-\infty}^{\infty} dk e^{i [ k(\eta-\eta') + c_s k \ln(\tau / \tau')]}
               +	\frac{1}{2}  \int_{-\infty}^{\infty} dk e^{i [ k(\eta-\eta')  -  c_s k \ln(\tau / \tau')]} \nonumber\\
&&        +	\int_{-\infty}^{\infty} dk \cos[k(\eta-\eta')]  \left[ \cos[c_s k \sqrt{1-\frac{(1-c_s^2)^2}{4c_s^2 k^2}}\ln(\tau / \tau')] - \cos[c_s k \ln(\tau / \tau')] \right]
\end{eqnarray}
Clearly the first two terms in the last step give rise to singularities precisely  at the two``sound front'' positions due to propagation in both longitudinal directions.  The third term $\sim \sin(k c_s \ln(\tau/\tau'))/6k$ at $k\to \infty$ and thus is regular everywhere. This analysis shows that our result is in consistency with that in \cite{Kapusta:2011gt}.

\subsection{General sound wave solutions}

Finally let us find the general sound wave solutions for the background Bjorken flow. We start by rewriting the linearized equations (\ref{eq_bj_linear}) as
\begin{eqnarray}
0&=&\frac{1}{1+c_s^2}\frac{\rho}{\tau^{c_s^2}}\Big(\frac{p_1}{p_0}\Big)_{,\tau}
+ (\tau^{-c_s^2}\rho u^\eta_{1})_{,\eta}
+ (\tau^{-c_s^2}\rho u^\rho_{1})_{,\rho}
+ (\tau^{-c_s^2}\rho u^\phi_{1})_{,\phi}, \label{eq.B1} \\ 
0&=& (\tau^{2-c_s^2} u^\eta_{1})_{,\tau}
+\frac{c_s^2}{1+c_s^2}\frac{1}{\tau^{c_s^2}} \Big(\frac{p_1}{p_0}\Big)_{,\eta}, \label{eq.B2}\\ 
0&=&(\tau^{-c_s^2}\rho  u^\rho_{1})_{,\tau}
+\frac{c_s^2}{1+c_s^2}\frac{\rho}{\tau^{c_s^2}} \Big(\frac{p_1}{p_0}\Big)_{,\rho}, \label{eq.B3}\\  
0&=&(\tau^{-c_s^2}\rho  u^\phi_{1})_{,\tau}
+\frac{c_s^2}{1+c_s^2}\frac{1}{\tau^{c_s^2}\rho} \Big(\frac{p_1}{p_0}\Big)_{,\phi}. \label{eq.B4}
\end{eqnarray}
Let us introduce the notation $\delta\equiv p_1/p_0$.  By the manipulation $(1+c_s^{-2}) \frac{\tau^{c_s^2}}{\rho}[(\ref{eq.B1})_{,\tau}-(\ref{eq.B3})_{,\rho}-(\ref{eq.B4})_{,\phi}]$ we obtain
\begin{eqnarray}
(1+c_s^{-2})\tau^{c_s^2}(\frac{u_1^\eta}{\tau^{c_s^2}})_{,\tau\eta} &=&
\frac{1}{\tau}\delta_{,\tau} -c_s^{-2}\delta_{,\tau\tau}
+\delta_{,\rho\rho} +\frac{1}{\rho}\delta_{,\rho} +\frac{1}{\rho^2}\delta_{,\phi\phi}.
 \label{eq.B5}
\end{eqnarray}
Further using  Eqs.(\ref{eq.B2}) and (\ref{eq.B5}) we can eliminate the variable $u_1^\eta$ and obtain an equation only for variable $\delta$. With the assumption of making variable separation
$\delta(\tau,\eta,\rho,\phi) = \delta_\parallel(\tau,\eta) \delta_\perp(\rho,\phi)$, the procedure leads to separate equations for the longitudinal and transverse fluctuations,
\begin{eqnarray}
&& \delta_{\perp,\rho\rho} +\frac{1}{\rho}\delta_{\perp,\rho} +\frac{1}{\rho^2}\delta_{\perp,\phi\phi} = -\omega^2 \delta_{\perp},\nonumber\\
&& \tau^{1+c_s^2}(\tau^{-1-c_s^2}\delta_{\parallel,\eta\eta})_{,\tau} =
(3-c_s^2)\omega^2\tau\delta_{\parallel}+(\omega^2\tau^2-2+c_s^2)\delta_{\parallel,\tau} + (3c_s^{-2}-2)\tau\delta_{\parallel,\tau\tau} +c_s^{-2}\tau^2\delta_{\parallel,\tau\tau\tau}.
\end{eqnarray}
One can then find the eigen modes for  the above two equations
\begin{eqnarray}
\delta_\perp(\rho,\phi) &\sim & J_m(\omega \rho) \, e^{i m \phi},\nonumber\\
\delta_\parallel(\tau,\eta) &\sim & e^{ik\eta} \times W(\tau).
\end{eqnarray}
Based on these, one can construct the following general solution for the pressure as well as  velocity components,
\begin{eqnarray}\label{eq_bj_s}
\delta(\tau,\eta,\rho,\phi) &=& \sum_m \int_0^\infty d\omega \int_{-\infty}^\infty dk ~
 e^{ik\eta}  J_m(\omega \rho) e^{i m \phi} W(\tau),\nonumber  \\
u_1^\eta (\tau,\eta,\rho,\phi) &=& (\frac{\tau}{\tau'})^{c_s^2-2} u_1^\eta (\tau',\eta,\rho,\phi)
+  \frac{c_s^2}{1+c_s^2}\tau^{c_s^2-2} u_{\perp,\eta} (\tau,\eta,\rho,\phi),\nonumber \\
u_1^\rho (\tau,\eta,\rho,\phi) &=&  (\frac{\tau}{\tau'})^{c_s^2} u_1^\rho (\tau',\eta,\rho,\phi)
+ \frac{c_s^2}{1+c_s^2}\tau^{c_s^2} u_{\perp,\rho} (\tau,\eta,\rho,\phi),\nonumber\\
u_1^\phi (\tau,\eta,\rho,\phi) &=& (\frac{\tau}{\tau'})^{c_s^2} u_1^\phi (\tau',\eta,\rho,\phi)
+ \frac{c_s^2}{1+c_s^2} \frac{\tau^{c_s^2}}{\rho^2} u_{\perp,\phi} (\tau,\eta,\rho,\phi),
\end{eqnarray}
where we have introduced the auxiliary field
\begin{eqnarray}
u_\perp (\tau,\eta,\rho,\phi)  = \sum_m \int_{0}^\infty d\omega \int_{-\infty}^\infty dk ~
  e^{ik\eta}  J_m(\omega \rho) e^{i m \phi}\int_{\tau'}^{\tau} d{\tilde \tau}~ {\tilde \tau}^{-c_s^2} \, W(\tilde \tau),
\end{eqnarray}
and the function $ {W(\tau)}$ including three independent components
\begin{eqnarray} \label{eq_bj_W}
W(\tau) &\equiv& A_{k,\omega,m}  W_1(\tau) + B_{k,\omega,m} W_2(\tau) + C_{k,\omega,m} W_3(\tau),\nonumber\\
W_1(\tau) &=&
(\frac{\tau}{\tau'})^{-\frac{1-c_s^2}{2}-\alpha_k} (\omega\tau)^{\alpha_k}
\Big[ \frac{1-c_s^2-2\alpha_k}{2} J_{-\alpha_k}(c_s\omega\tau)
 -  (c_s \omega\tau) J_{1-\alpha_k}(c_s \omega\tau) \Big],\nonumber\\
W_2(\tau) &=&
(\frac{\tau}{\tau'})^{-\frac{1-c_s^2}{2}+\alpha_k} (\omega\tau)^{-\alpha_k}
\Big[\frac{1-c_s^2-2\alpha_k}{2}J_{\alpha_k}(c_s \omega\tau)
 + (c_s \omega\tau) J_{\alpha_k-1}(c_s \omega\tau) \Big],\nonumber\\
W_3(\tau) &=&
(\omega\tau)^{1+c_s^2}~
_1F_2 [ 2;\frac{7+c_s^2+2\alpha_k}{4},\frac{7+c_s^2-2\alpha_k}{4};-c_s^2\omega^2\tau^2/4 ]
\end{eqnarray}
with $\alpha_k=\sqrt{(1-c_s^2)^2/4-c_s^2k^2}$. In the above solutions, there are a set of coefficients $A_{k,\omega,m}, B_{k,\omega,m}$ and $C_{k,\omega,m}$ depending on wave mode parameters. These parameters are determined by the initial fluctuations, as we will discuss in the next subsection. Note that the hypergeometric function here $_1F_2[a_1;b_1,b_2;x]$ is symmetric with respect to the exchange $b_1 \leftrightarrow b_2$, and one can show that $W_3$ is always real for any $k$.

Clearly the solution maintains a general structure of factorized longitudinal and transverse dynamics. The dependence on $\eta,\rho,\phi$ is basically a superposition of those eigen functions in each of the spatial coordinates. More nontrivial is the time dependence, which is a combination of  the three functions $W_{1,2,3}$ --- loosely speaking this appears in correspondence with three independent sound modes (for given $\omega,k$ parameters), expected in three spatial dimensions.

Let us firstly discuss some general behavior of the functions $W$. 1) It is not difficult to show that for $k^2 \le (1-c_s^2)^2/(4c_s^2)$ all the three functions $W_{1,2,3}$ are real. For $k^2>(1-c_s^2)^2/(4c_s^2)$, one has $W_3=W_3^*$ and $W_1^*=W_2$.
Therefore to ensure the solution to take real values, the set of coefficients should satisfy the constraints  $(-1)^mC_{-k,\omega,-m}^*=C_{k,\omega,m}$ and  $(-1)^m\{A,B\}_{-k,\omega,-m}^*=\{A,B\}_{k,\omega,m}$  for $k^2\le (1-c_s^2)^2/(4c_s^2)$ and $(-1)^m\{B,A\}_{-k,\omega,-m}^*=\{A,B\}_{k,\omega,m}$ for $k^2>(1-c_s^2)^2/(4c_s^2)$.    2) One can show $W_{1,2,3}(\tau) \sim \tau^{1/6}$ at $\tau\to\infty$, which means the same asymptotical  behavior   as the transverse sound wave. (Note that $p_1\sim p_0 W $ would still decrease in time.)
It is also interesting to examine these functions   in the two  limits  of $\omega\to0$ (longitudinal limit) and $k\to0$ (transverse limit),
\begin{eqnarray}
W_1(\tau) &\propto& \left\{ \begin{array}{ll}
 \tau^{-\frac{1-c_s^2}{2}-\alpha_k} & \omega \to 0 \\
 \tau^{\frac{1+c_s^2}{2}}J_{\frac{1+c_s^2}{2}}(c_s\omega\tau) & k \to 0 \\
 \end{array} \right. , \nonumber\\
W_2(\tau) &\propto& \left\{ \begin{array}{ll}
 \tau^{-\frac{1-c_s^2}{2}+\alpha_k} & \omega \to 0 \\
 \tau^{\frac{1+c_s^2}{2}}J_{-\frac{1+c_s^2}{2}}(c_s\omega\tau) & k \to 0 \\
 \end{array} \right. ,\nonumber\\
 W_3(\tau) &\propto& \left\{ \begin{array}{ll}
 0 & \omega \to 0 \\
 \tau^{\frac{1+c_s^2}{2}}J_{\frac{1+c_s^2}{2}}(c_s\omega\tau)& k \to 0 \\
 \end{array} \right. .
\end{eqnarray}
These limit cases appear in close correspondence with the previously found longitudinal and transverse wave solutions.
3)   In the limit $\omega>>1$ and $\omega>>k$, it can be shown that the function $W(\tau)$ reveals a symptomatic phase factor $\sim e^{\pm ic_s\omega\tau}$ which together with the asymptotic form of $J_m(\omega\rho)$ at large $\rho$ reveals a propagating transverse wave with phase velocity $c_s$.
On the other hand, if we take  $k>>1$ and $k>>\omega$ limit, we will find a similar structure $\sim e^{\pm ic_sk\tau}$, which together with the factor $e^{i k \eta}$ reveals a propagating longitudinal wave in correspondence to our previous analysis.

\subsection{Determining the coefficients from initial perturbation}

With the general solution in Eqs.(\ref{eq_bj_s}), one needs to determine the coefficients $\{A,B,C\}_{k,\omega,m}$ in the function $W$ by matching to the initial condition $\delta(\tau')$ and $u_1(\tau')$ when such perturbation occurs. One needs three conditions, and the strategy is as follows. A first constraint is provided by the initial pressure perturbation $\delta(\tau')$. One uses the initial perturbation together with the hydro equations to determine the first and second order time derivatives of the pressure field i.e. $\partial_\tau \delta$ and $\partial^2_\tau \delta$. Then one can use the three matching conditions for $\delta$, $\partial_\tau \delta$, as well as $\partial^2_\tau \delta$ to completely fix the coefficients. Let us see how that works. Using the linearized hydro equations at time $\tau'$, it is not difficult to obtain
\begin{eqnarray}\label{eq_bj_deldt}
\partial_\tau \delta (\tau') &=& -(1+c_s^2)\Big[ u^\eta_{1,\eta}(\tau')
+\frac{(\rho u^\rho_1)_{,\rho}(\tau')}{\rho}+u^\phi_{1,\phi}(\tau') \Big],\nonumber\\
\partial^2_\tau \delta(\tau') &=&
c_s^2 \Big[ \frac{\delta_{,\eta\eta}(\tau')}{\tau'^2} + \frac{(\rho\delta_{,\rho}(\tau'))_{,\rho}}{\rho} + \frac{\delta_{,\phi\phi}(\tau')}{\rho^2} \Big]
 \nonumber \\
 && \qquad  + \frac{c_s^2(1+c_s^2)}{\tau'} \Big[ \frac{(1+c_s^{-2})(2-c_s^2)}{1+c_s^2} u^\eta_{1,\eta}(\tau') - \frac{(\rho u_1^\rho(\tau'))_{,\rho}}{\rho} - u^\phi_{1,\phi}(\tau') \Big]
\end{eqnarray}
which can completely fix $\partial_\tau \delta$ and $\partial^2_\tau \delta$ at the time moment $\tau'$ from the initial perturbation $\delta$ and $u_1$ at $\tau'$.
Now we focus on $\delta$ in Eq.(\ref{eq_bj_s}). By integrating the two sides of the equation over spatial coordinates multiplied by proper basis functions (similar to the inverse of Fourier transformation), we have
\begin{eqnarray} \label{eq_Wt}
W(\tau')   &=&
 \frac{\omega}{4\pi^2}\int_0^\infty \rho d\rho \int_{-\infty}^{\infty} d\eta \int_0^{2\pi} d\phi \,
\delta(\tau',\eta,\rho,\phi)  e^{-ik\eta} e^{-im\phi} J_m(\omega\rho).
\end{eqnarray}
We further take the first and second order time derivatives of the above equation,
\begin{eqnarray} \label{eq_Wdt}
\partial_\tau W (\tau')    &=&
 \frac{\omega}{4\pi^2}\int_0^\infty \rho d\rho \int_{-\infty}^{\infty} d\eta \int_0^{2\pi} d\phi \,
[\partial_\tau \delta (\tau',\eta,\rho,\phi)]  e^{-ik\eta} e^{-im\phi} J_m(\omega\rho),\nonumber\\
\partial^2_\tau W (\tau')  &=&
 \frac{\omega}{4\pi^2}\int_0^\infty \rho d\rho \int_{-\infty}^{\infty} d\eta \int_0^{2\pi} d\phi \,
[\partial_\tau^2 \delta (\tau',\eta,\rho,\phi)]  e^{-ik\eta} e^{-im\phi} J_m(\omega\rho) \, .
\end{eqnarray}
Now the quantities $W(\tau')$, $\partial_\tau W (\tau')$ and $\partial^2_\tau W (\tau')$ are all fixed by initial perturbations.

From Eq.(\ref{eq_bj_W}), $W(\tau')$, $\partial_\tau W (\tau')$, and $\partial^2_\tau W (\tau')$ can also be determined by the three coefficients $\{A,B,C\}_{k,\omega,m}$. This provides the way to fix $\{A,B,C\}_{k,\omega,m}$ from $W(\tau')$, $\partial_\tau W (\tau')$, and $\partial^2_\tau W (\tau')$,
\begin{eqnarray}
\left(\begin{array}{c}
A_{k,\omega,m}  \\ B_{k,\omega,m} \\ C_{k,\omega,m} \\
\end{array}\right) &=&
\left(\begin{array}{ccc}
W_1(\tau') & W_2(\tau') & W_3(\tau') \\
\partial_\tau W_1(\tau') & \partial_\tau W_2(\tau') & \partial_\tau W_3(\tau') \\
\partial_\tau^2 W_1(\tau') & \partial_\tau^2 W_2(\tau') & \partial_\tau^2 W_3(\tau') \\
\end{array}\right)^{-1} \cdot
\left(\begin{array}{c}
W(\tau') \\ \partial_\tau W(\tau') \\ \partial_\tau^2 W(\tau') \\
\end{array}\right).
\end{eqnarray}
In the above, the matrix elements $\partial_\tau^{\{0,1,2\}} W_{1,2,3}$ can all be directly computed from Eqs.(\ref{eq_bj_W}). In this way we've shown the method to determine the sound wave solutions completely for any arbitrarily given initial perturbation.

To make the procedure more transparent, let us give  an explicit example, by considering a static initial Gaussian perturbation  with cylindrical symmetry,
\begin{eqnarray}
\delta(\tau') &=& \frac{\xi}{(2\pi\sigma^2)^{3/2}}e^{-\frac{(\eta-\eta')^2+\rho^2/\tau'^2}{2\sigma^2}},\nonumber\\
u_1(\tau') &=& 0.
\end{eqnarray}
In this case one can obtain from Eqs.(\ref{eq_bj_deldt}),
\begin{eqnarray}
\frac{d\delta}{d\tau}|_{\tau'} &=& 0  , \nonumber\\
\frac{d^2\delta}{d\tau^2}|_{\tau'} &=& - \frac{ {3c_s^2}}{\sigma^2\tau'^2}\delta(\tau')+ {c_s^2}\frac{\tau'^2(\eta-\eta')^2+\rho^2}{\sigma^4\tau'^4}\delta(\tau').
\end{eqnarray}
We can then use the Eqs.(\ref{eq_Wt}) and (\ref{eq_Wdt}) to derive
\begin{eqnarray}
W(\tau') &=& \delta_{m,0} \frac{\xi \tau'^2}{4\pi^2}\omega e^{-\frac{\sigma^2(\omega^2\tau'^2+k^2)}{2}}
e^{-ik\eta'},\nonumber\\
\partial_\tau W(\tau') &=& 0,\nonumber\\
\partial_\tau^2 W(\tau') &=&   \delta_{m,0} \left(- {c_s^2}\frac{\omega^2\tau'^2 + k^2}{\tau'^2} \right) W(\tau')
\end{eqnarray}
and in turn the coefficients
\begin{eqnarray} \label{eq_bj_G_3D}
\left(\begin{array}{c}
A_{k,\omega,m} \\ B_{k,\omega,m} \\  C_{k,\omega,m}\\
\end{array}\right) &=&  \delta_{m,0} \frac{\xi \tau'^2}{4\pi^2}\omega e^{-\frac{\sigma^2(\omega^2\tau'^2+k^2)}{2}}
e^{-ik\eta'}
\left(\begin{array}{ccc}
W_1(\tau') & W_2(\tau') & W_3(\tau') \\
\partial_\tau W_1(\tau') & \partial_\tau W_2(\tau') & \partial_\tau W_3(\tau') \\
\partial_\tau^2 W_1(\tau') & \partial_\tau^2 W_2(\tau') & \partial_\tau^2 W_3(\tau') \\
\end{array}\right)^{-1}
\left(\begin{array}{c}
1 \\ 0 \\- {c_s^2}\frac{\omega^2\tau'^2 + k^2}{\tau'^2} \\
\end{array}\right),
\end{eqnarray}
where again $\partial_\tau^{\{0,1,2\}} W_{1,2,3}$ are computed from Eqs.(\ref{eq_bj_W}).

Clearly to obtain the sound wave solutions arising form an initial delta-function perturbation, one simply takes $\sigma\to 0$ in the above calculations.

\section{Sound waves on top of 3D Hubble flow}
\label{s4}
In this Section, we study another example of sound waves on expanding background and present general solutions to   the linearized hydrodynamic equations on the background 3D Hubble flow. The 3D Hubble flow, mostly studied for   the Universe expansion, also provides an approximate description of  the relatively late time expansion (when transverse flow becomes significant) of hot QCD fluid in a  heavy ion collision.

\subsection{The linearized hydrodynamic equations}

A 3D Hubble flow expands radially with a velocity field $\vec v = \vec r/t$. As is well known, from the point of view of any local rest frame in the fluid, the whole system expands in the same  rotationally  symmetric way. For describing such a flow background, it's most convenient to use the following coordinates
\begin{eqnarray}
\tau &=& \sqrt{t^2-r^2},~~~~~~~~~~~~~~~~~~~
\eta ~=~ \frac{1}{2}\mathrm{ln}\frac{t+r}{t-r},~~~\nonumber\\
\theta &=& \frac{1}{2i}\mathrm{ln}\frac{z+i\sqrt{x^2+y^2}}{z-i\sqrt{x^2+y^2}},~~~~~
\phi ~=~ \frac{1}{2i}\mathrm{ln}\frac{x+i~y}{x-i~y}    \label{eq_hb_coordinate}
\end{eqnarray}
with $r=\sqrt{x^2+y^2+z^2}$. Note that the $\eta$ here has a different definition from the 1D Bjorken case, while $\theta$ and $\phi$ are the  polar and azimuthal angles in usual spherical frame. To keep the main line of our discussions clear, we leave many of the details regarding the coordinate system, the metric and connections, as well as the full hydrodynamic equations in this coordinate system in the Appendix \ref{app_1}.

The 3D Hubble flow, in the above coordinates, is conveniently described by
\begin{eqnarray}  \label{eq_hb_flow}
p_0 = \frac{p_0(\tau_0)\tau_0^{3(1+c_s^2)}}{\tau^{3(1+c_s^2)}}   \,\,  , \,\, u_0^\mu&=&(1,0,0,0).
\end{eqnarray}
We now consider the sound wave on top of this background, $p=p_0 + p_1$ and $u^\mu=u_0^\mu + u_1^\mu$. By substituting these into the hydrodynamic equations and keeping the leading order in perturbation, we obtain the following linearized equations on top of the 3D Hubble background,
\begin{eqnarray}
0&=&{\frac{1}{1+c_s^2}}p_{1,\tau}+\frac{3}{\tau}p_1
+p_0 u^\eta_{1,\eta}+p_0u^\theta_{1,\theta}+p_0u^\phi_{1,\phi}
+2\frac{\cosh\eta}{\sinh\eta}p_0u_1^\eta
+\frac{\cos\theta}{\sin\theta}p_0u^\theta_{1}  \label{eq_hb_lin1}  \\ 
0&=&p_0 u^\eta_{1,\tau}+{\frac{2-3c_s^2}{\tau}} p_0 u_1^\eta
+{\frac{c_s^2}{1+c_s^2}}\frac{p_{1,\eta}}{\tau^2}      \label{eq_hb_lin2}    \\
0&=&p_0u^\theta_{1,\tau}+{\frac{2-3c_s^2}{\tau}} p_0u^\theta_{1}
+{\frac{c_s^2}{1+c_s^2}}\frac{1}{\tau^2\sinh^2\eta}p_{1,\theta}         \label{eq_hb_lin3}     \\
0&=&p_0u^\phi_{1,\tau}+{\frac{2-3c_s^2}{\tau}} p_0u^\phi_{1}
+{\frac{c_s^2}{1+c_s^2}}\frac{1}{\tau^2\sinh^2\eta~\sin^2\theta}p_{1,\phi}         \label{eq_hb_lin4}  \,\, .
\end{eqnarray}
Note also that the four-velocity constraint $u^\mu u_\mu=1$ requires (at linear order of perturbation) $u_1^\tau=0$.

\subsection{General sound wave solutions}
\label{4b}

Now we proceed to solve the linearized equations, using a similar strategy as before i.e. to combine them into a higher-order differential equation for the pressure perturbation with a form allowing variable separations. That can be done by  (\ref{eq_hb_lin1})$_{,\tau}-$(\ref{eq_hb_lin2})$_{,\eta}-$(\ref{eq_hb_lin3})$_{,\theta}-$(\ref{eq_hb_lin4})$_{,\phi}+\frac{5}{\tau}\times$ (\ref{eq_hb_lin1}) $-2\mathrm{coth}\eta\times$ (\ref{eq_hb_lin2})$-\mathrm{cot}\theta\times$ (\ref{eq_hb_lin3}) which leads to the equation
\begin{eqnarray}
{c_s^{-2}}\tau^2p_{1,\tau\tau} +{(3+8c_s^{-2})} \tau p_{1,\tau}+{12(1+c_s^{-2})}p_1=p_{1,\eta\eta}+2 \frac{\cosh\eta}{\sinh\eta} p_{1,\eta}+\frac{1}{\sinh^2\eta}\big( p_{1,\theta\theta}+\frac{\cos\theta}{\sin\theta}p_{1,\theta}+\frac{1}{\sin^2\theta}p_{1,\phi\phi} \big).
\end{eqnarray}
Deferring the detailed derivation into the Appendices \ref{app_2}\&\ref{app_3}, here we present the final results for the general sound wave solutions  on top of the 3D Hubble flow,
\begin{eqnarray}\label{eq_hb_s}
\frac{p_1(\tau,\eta,\theta,\phi)}{p_0}
&=&~~ {(\frac{\tau}{\tau'})^{\frac{3c_s^2-1}{2}}}
\sum_{l,m}\int_{-\infty}^\infty
a_{l,m}(k) \cos\left[{\beta_k}\ln(\tau/\tau')\right]
R_l (k,\eta) Y_l^m(\theta,\phi) ~\mathrm{d}k \no
&&+ {(\frac{\tau}{\tau'})^{\frac{3c_s^2-1}{2}}}
\sum_{l,m}\int_{-\infty}^\infty
b_{l,m}(k) \sin\left[{\beta_k}\ln(\tau/\tau')\right]
R_l (k,\eta) Y_l^m(\theta,\phi) ~\mathrm{d}k,\nonumber\\
u_1^\eta (\tau,\eta,\theta,\phi) &=&
{(\frac{\tau'}{\tau})^{2-3c_s^2}}u_1^\eta (\tau',\eta,\theta,\phi)  +
{\frac{c_s^2}{1+c_s^2}}u_{\perp,\eta} (\tau',\eta,\theta,\phi),\nonumber\\
u_1^\theta (\tau,\eta,\theta,\phi) &=&
{(\frac{\tau'}{\tau})^{2-3c_s^2}}u_1^\theta (\tau',\eta,\theta,\phi)  +
{\frac{c_s^2}{1+c_s^2}}\frac{1}{\sinh^{2}\eta}u_{\perp,\theta} (\tau',\eta,\theta,\phi),\nonumber\\
u_1^\phi (\tau,\eta,\theta,\phi) &=&
{(\frac{\tau'}{\tau})^{2-3c_s^2}}u_1^\phi (\tau',\eta,\theta,\phi) +
{\frac{c_s^2}{1+c_s^2}}\frac{1}{\sinh^{2}\eta~\sin^2\theta} u_{\perp,\phi} (\tau',\eta,\theta,\phi)
\end{eqnarray}
with  $\beta_k\equiv c_s \sqrt{k^2+(1-c_s^2)(9c_s^2-1)/4c_s^2}$. In the above solution, we've introduced two auxiliary functions $R_l(k,x)$ and $u_\perp(\tau,\eta,\theta,\phi)$ defined as
\begin{eqnarray}
R_l(k,\eta) &=& \sqrt{\frac{\Gamma(l+1+ik)\Gamma(l+1-ik)}{\pi2^{2l+2}\Gamma(l+3/2)^2}}
\sinh^l\eta~ _2F_1(\frac{l+1+ik}{2},\frac{l+1-ik}{2},l+3/2,-\sinh^2\eta)\nonumber\\
&=& \frac{\Gamma(l+1)}{\sqrt{\Gamma(l+1-ik)\Gamma(l+1+ik)}}
\frac{2^l}{\sinh(\pi k)}\sinh^{l}\eta C^{(l+1)}_{ik-l-1}(\cosh\eta),\nonumber\\
u_\perp(\tau,\eta,\theta,\phi) &=&
\frac{1}{\tau}\sum_{l,m}\int_{-\infty}^\infty
\frac{\beta_k b_{l,m}(k)+(3c_s^2-1)a_{l,m}(k) }{(3c_s^2-1)^2+\beta_k^2}
\left[\cos\left(\beta_k\ln\frac{\tau}{\tau'}\right) -1\right]
R_l (k,\eta) Y_l^m(\theta,\phi)~\mathrm{d}k \nonumber\\
&&
+\frac{1}{\tau}\sum_{l,m}\int_{-\infty}^\infty
\frac{-\beta_k a_{l,m}(k)+(3c_s^2-1)b_{l,m}(k) }{(3c_s^2-1)^2+\beta_k^2}
\sin\left(\beta_k\ln\frac{\tau}{\tau'}\right)
R_l (k,\eta) Y_l^m(\theta,\phi)~\mathrm{d}k,  \label{eq_hb_up}
\end{eqnarray}
where
 $C_\mu^{(\nu)}$ are Gegenbauer functions. The coefficients $a_{l,m}$ and $b_{l,m}$ satisfy the constraints $a_{l,m}^*=(-1)^m a_{l,-m}$, $b_{l,m}^*=(-1)^m b_{l,-m}$, $a_{l,m}(-k)=a_{l,m}(k)$ and $b_{l,m}(-k)=b_{l,m}(k)$, they are determined by the initial  perturbation at time $\tau'$.

It is interesting to take a look at the simplest sound wave mode here, i.e. the spherically symmetric wave with $l=m=0$ and  $R_0(k,\eta) = \frac{1}{ \sqrt{\pi k \sinh(\pi k)} } \frac{\sin k\eta}{\sinh\eta}$ . In this case the solution can be simplified as
\begin{eqnarray}
p_1&=&p_0\int_{-\infty}^\infty a_{0,0}(k) e^{\pm i  \beta_k \ln(\tau/\tau')} \frac{\sin(k\eta)}{k \sinh\eta}~\mathrm{d}k.
\end{eqnarray}
By rewriting $ {\sin(k\eta)} = (e^{ik\eta} - e^{-i k \eta})/(2i)$, one may literally extract a phase velocity of the wave propagation
\begin{eqnarray}
\frac{\tau \delta \eta}{\delta \tau} =  \pm  \frac{\beta_k}{k} =  \pm  \frac{c_s}{k} \sqrt{k^2+\frac{(1-c_s^2)(9c_s^2-1)}{4c_s^2}}
\end{eqnarray}
which at large $k$ limit approaches the speed of sound on static background. By returning to the original flat coordinates one can obtain the phase velocity
\begin{eqnarray}
\frac{\delta r}{\delta t}
= \frac{\tanh\eta \pm {c_s}}{1 \pm c_s \tanh\eta}
= \frac{v_r \pm{c_s}}{1 \pm v_r {c_s}} \, .
\end{eqnarray}
Its physical meaning becomes transparent: these are two sound wave modes, one traveling inward while the other traveling outward   with the sound speed ${c_s}$ relative to the underlying Hubble flow.

\subsection{Determining the coefficients from initial perturbation}

In general, for a given initial perturbation $\delta(\tau')\equiv p_1(\tau') / p_0 (\tau')$ and $u_1(\tau')$, we need to determine the coefficients $a_{l,m}(k)$ and $b_{l,m}(k)$ from Eqs.(\ref{eq_hb_s}). As the expressions with symbolic $c_s$ become too lengthy and complicated to display, we will use the speed of sound  $c_s=1/\sqrt3$ in this part. Note that by expanding the initial pressure field in terms of eigen-functions can only provide one set of constraints. In addition, we can get the first time derivative of the pressure at $\tau'$ through Eq.(\ref{eq_hb_lin1}), namely
\begin{eqnarray} \label{eq_hb_pdpt}
\frac{\partial\delta}{\partial\tau}{\bigg |}_{\tau'} &=& - {\frac{4}{3}} \left[
\frac{[\sinh^2\eta u_1^\eta(\tau')]_{,\eta}}{\sinh^2\eta}
+ \frac{[\sin\theta u_1^\theta(\tau')]_{,\theta}}{\sin\theta}
+ u_{1,\phi}^{\phi}(\tau')\right].
\end{eqnarray}

With the two constraints from $\delta(\tau')$ and $\frac{\partial\delta}{\partial\tau}|_{\tau'}$ we can then completely fix the two sets of coefficients
\begin{eqnarray}\label{eq_hb_ab}
a_{l,m}(k) &=& k\sinh(\pi k)
 \int_{4\pi} d\Omega \int_0^\infty \sinh^2\eta d\eta ~
\delta(\tau',\eta,\theta,\phi) R_l(k,\eta) Y_l^m(\theta,\phi)^* ,\nonumber\\
b_{l,m}(k) &=&\frac{\tau'}{\beta_k} k\sinh(\pi k)
\int_{4\pi} d\Omega \int_0^\infty \sinh^2\eta d\eta ~
\frac{\partial\delta}{\partial\tau}(\tau' {,\eta,\theta,\phi}) R_l(k,\eta) Y_l^m(\theta,\phi)^*
\end{eqnarray}
with $\beta_k=\sqrt{\frac{k^2+1}{3}}$ when we use $c_s^2=1/3$. Note that the technical details for fixing the normalization coefficients are given in the Appendix C.  With these coefficients determined, one can substitute them back to the solution in Eq.(\ref{eq_hb_s}), and simplify the solution by using the following summation identity:
\begin{eqnarray}  \label{eq_hb_sum}
R_0(k,\bar\eta)R_0(k,0)|Y_0^0|^2 = \sum_{l,m}
R_l(k,\eta')R_l(k,\eta)Y_l^m(\theta,\phi)Y_l^m(\theta',\phi')^*
\end{eqnarray}
in which the parameter $\bar \eta$ is defined via
\begin{eqnarray}
\cosh\bar \eta = \cosh\eta'\cosh\eta-\sinh\eta'\sinh\eta
[\cos\theta'\cos\theta+\sin\theta'\sin\theta\cos(\phi-\phi')]  \,\, .
\end{eqnarray}
The ultimate origin of the above identity lies in a 3D boost-invariance of the Hubble flow (i.e. one sees the same Hubble flow when changes to any fluid cell's local rest frame). A strict mathematical proof of the above identity is provided in Appendix \ref{app_4}. The space-time evolution of the pressure perturbation is then given as follows,
\begin{eqnarray} \label{eq_hb_final}
\delta = \frac{p_1}{p_0}
&=&~~
\int_{-\infty}^\infty \frac{k^2}{4\pi^2}
\mathrm{d}k \int_{4\pi} d\Omega' \int_0^\infty \sinh^2\eta' d\eta'~
  {\cos\left[\beta_k\ln(\tau/\tau')\right]
\frac{\sin(k\bar\eta)}{k\sinh\bar\eta}}\, \delta(\tau',\eta',\theta',\phi') \no
&&+
\int_{-\infty}^\infty \frac{k^2}{4\pi^2}
\mathrm{d}k \int_{4\pi} d\Omega' \int_0^\infty \sinh^2\eta' d\eta'~
\frac{\tau' \sin\left[\beta_k\ln(\tau/\tau')\right] }{\beta_k}
 {\frac{\sin(k\bar\eta)}{k\sinh\bar\eta}} \, \left[\frac{\partial\delta}{\partial\tau}(\tau' {,\eta,\theta,\phi}) \right].
\end{eqnarray}
The above form could be physically understood as follows:  the first term is the superposition of the wave generated from each localized delta-function source in pressure convoluted with the initial pressure perturbation field $\delta(\tau')$, while the second term is the superposition of the wave generated from each localized  delta-function source in velocity field convoluted with the initial velocity perturbation field $\frac{\partial\delta}{\partial\tau}(\tau')$ (via Eq.(\ref{eq_hb_pdpt})).

As a concrete example let us consider again a static Gaussian-like perturbation,
\begin{eqnarray}
p_1(\tau') &=&  p_0(\tau')\cdot\frac{\xi}{(2\pi\sigma^2)^{3/2}} e^{-\frac{\bar\eta^2}{2\sigma^2}},\nonumber\\
u_1(\tau') &=& 0.
\end{eqnarray}
After fixing the coefficients from  Eqs.(\ref{eq_hb_ab}), we obtain the following   solution
\begin{eqnarray}
\frac{p_1(\tau,\eta,\theta,\phi)}{p_0}
&=&{\xi}\sum_{l,m}
\int_{-\infty}^\infty
k\sinh(\pi k) \frac{\sin(k\sigma^2)}{k\sigma^2e^{-\sigma^2/2}}e^{-\frac{\sigma^2 k^2}{2}}
\no&&~~~~~~~~~~~~~\times
\cos[\beta_k \ln(\tau/\tau')]
 R_l (k,\eta') Y_l^m(\theta',\phi')^* R_l (k,\eta)  Y_l^m(\theta,\phi) ~\mathrm{d}k   \,\, .
 \end{eqnarray}
The physical picture of the above sound wave is best manifested after using the summation identity (\ref{eq_hb_sum}) to obtain:
\begin{eqnarray}
\frac{p_1(\tau,\eta,\theta,\phi)}{p_0}
&=&\frac{\xi}{4\pi^2}
\int_{-\infty}^\infty \frac{\sin(k\sigma^2)}{k\sigma^2e^{-\sigma^2/2}}e^{-\frac{\sigma^2 k^2}{2}}
\frac{\sin(k\bar\eta)}{\sinh(\bar\eta)} \cos[\beta_k \ln(\tau/\tau')] k~dk  \,\, . \label{eq_hb_ga}
\end{eqnarray}
In fact this can be directly obtained by starting from Eq.(\ref{eq_hb_final}).
 Intuitively the solution  is a spherically symmetric sound wave triggered from the center of the initial Gaussian perturbation, which becomes apparent when one makes a boost of coordinates  to the local rest frame at the center of the initial Gaussian perturbation. The corresponding velocity field of the sound wave is given by
\begin{eqnarray}
u_\perp(\tau,\eta,\theta,\phi) &=&
-\frac{ \xi}{4\pi^2\tau}
\int_{-\infty}^\infty \frac{\sin(k\sigma^2)}{k\sigma^2e^{-\sigma^2/2}}
\frac{e^{-\frac{\sigma^2 k^2}{2}}}{\beta_k}
\frac{\sin(k\bar\eta)}{\sinh(\bar\eta)} \sin\left[\beta_k \ln(\tau/\tau')\right] k~dk,\nonumber\\
\frac{\partial u_\perp}{\partial\bar\eta}   &=&
-\frac{ \xi}{4\pi^2\tau}
\int_{-\infty}^\infty \frac{\sin(k\sigma^2)}{k\sigma^2e^{-\sigma^2/2}}
\frac{e^{-\frac{\sigma^2 k^2}{2}}}{\beta_k}
\left[ \frac{k\cos(k\bar\eta)}{\sinh(\bar\eta)}
-\frac{\sin(k\bar\eta)\cosh(\bar\eta)}{\sinh^2(\bar\eta)} \right]
\sin\left[\beta_k \ln(\tau/\tau')\right] k~dk,\nonumber\\
u_1^\eta (\tau,\eta,\theta,\phi) &=& \frac{u_{\perp,\eta}}{4}
= \frac{1}{4\sinh\bar\eta} \left(\frac{\partial u_\perp }{\partial\bar\eta} \right)
[\cosh\eta'\sinh\eta-
\sinh\eta'\cosh\eta (\cos\theta'\cos\theta+\sin\theta'\sin\theta\cos(\phi-\phi'))],\nonumber \\
u_1^\theta (\tau,\eta,\theta,\phi) &=&
\frac{u_{\perp,\theta}}{4\sinh^{2}\eta}
= \frac{1}{4\sinh\bar\eta}  \left( \frac{\partial u_\perp }{\partial\bar\eta}  \right)
\frac{\sinh\eta'}{\sinh\eta} (\cos\theta'\sin\theta-\sin\theta'\cos\theta\cos(\phi-\phi')),\nonumber\\
u_1^\phi (\tau,\eta,\theta,\phi) &=&
\frac{u_{\perp,\phi}}{4\sinh^{2}\eta~\sin^2\theta}
= \frac{1}{4\sinh\bar\eta}  \left(\frac{\partial u_\perp }{\partial\bar\eta}  \right)
\frac{\sinh\eta'\sin\theta'}{\sinh\eta\sin\theta}\sin(\phi-\phi') .
\end{eqnarray}

Finally one can get the sound wave solution from the delta-function initial perturbation by taking the limit  $\sigma\to0$,
i.e. $ p_1(\tau') = p_0(\tau')\cdot\delta^{(3)}(\bar\eta) =
p_0(\tau')\cdot\frac{\delta(\bar\eta)}{2\pi\bar\eta^2}$. This amounts to reduce the factor  $\frac{\sin(k\sigma^2)}{k\sigma^2e^{-\sigma^2/2}}e^{-\frac{\sigma^2 k^2}{2}}$ in Eq.(\ref{eq_hb_ga})  to be one, from which we obtain:
\begin{eqnarray}
\frac{p_1(\tau,\eta,\theta,\phi)}{p_0}
&=&\frac{\xi}{4\pi^2}
\int_{-\infty}^\infty
\frac{\sin(k\bar\eta)}{\sinh(\bar\eta)} \cos[\beta_k \ln(\tau/\tau')] \, k~dk.
\end{eqnarray}

\section{Numerical Results and Discussions}
\label{s5}
In the previous two sections, we have found the analytic solutions describing how a general form of perturbation, once introduced, propagates on top of the Bjorken flow as well as the Hubble flow. As already discussed in the Introduction, there can be many applications of these solutions in order to understand the manifestations of various types of fluctuations through observables like rapidity and azimuthal angle correlations in high energy heavy-ion (as well as pp or pA/dA) collisions. As also mentioned, the primary purpose of the present paper is to find these solutions, while a detailed investigation of various phenomenological applications is underway and will be reported elsewhere. In this Section, we show only a number of simple examples. We will focus on the rapidity distribution of pressure fluctuations $\delta(\tau,\eta)$ from a static Guassian initial perturbation occurring at an earlier time $\tau'<\tau$. Note that for a conformal equation of state $p=c_s^2 \epsilon = \epsilon/3$, one has $\delta \equiv \frac{p_1}{p_0}=\frac{\Delta p}{p}
=\frac{\Delta \epsilon}{\epsilon}=\frac{4}{3} \frac{\Delta s}{s}$. We also study  the resulting equal-time rapidity correlation:
\begin{eqnarray}
C(\Delta \eta) = \int d\eta_1  \delta(\tau,\eta_1) \delta(\tau,\eta_1+\Delta\eta) \, ,
\end{eqnarray}
In the following we study and compare these quantities for different types of waves on top of the two types of background flows. For later convenience, we introduce the two types of rapidity variables, the radial rapidity and the longitudinal rapidity, for which we use the notation $\eta_r=1/2\ln(t+r)/(t-r)$ and $\eta_z=1/2\ln(t+z)/(t-z)$ respectively.

\subsection{Longitudinal wave on top of Bjorken flow}

\begin{figure}[!hbt]
\includegraphics[width=0.3\textwidth]{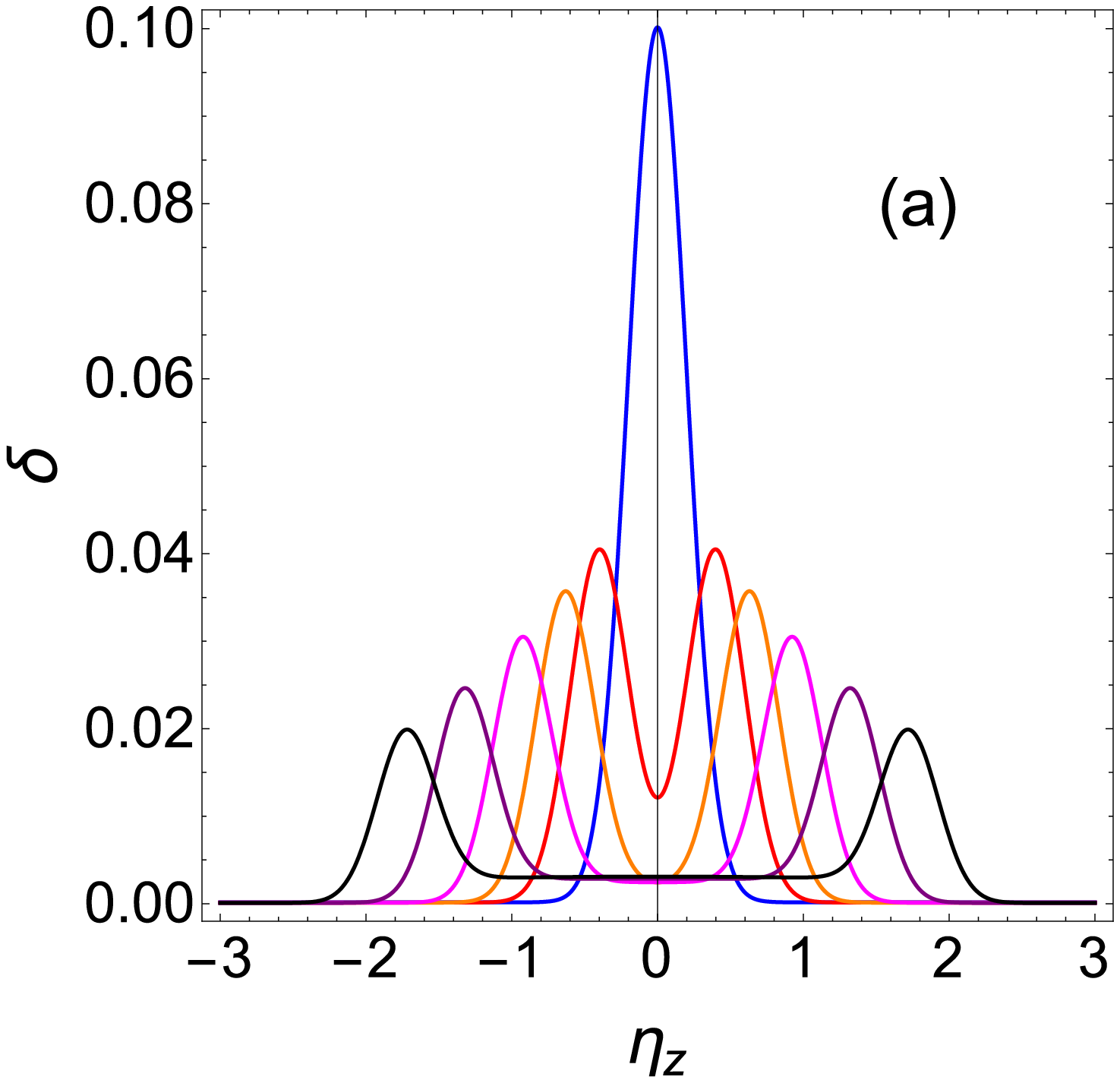} \hspace{0.2in}
\includegraphics[width=0.3\textwidth]{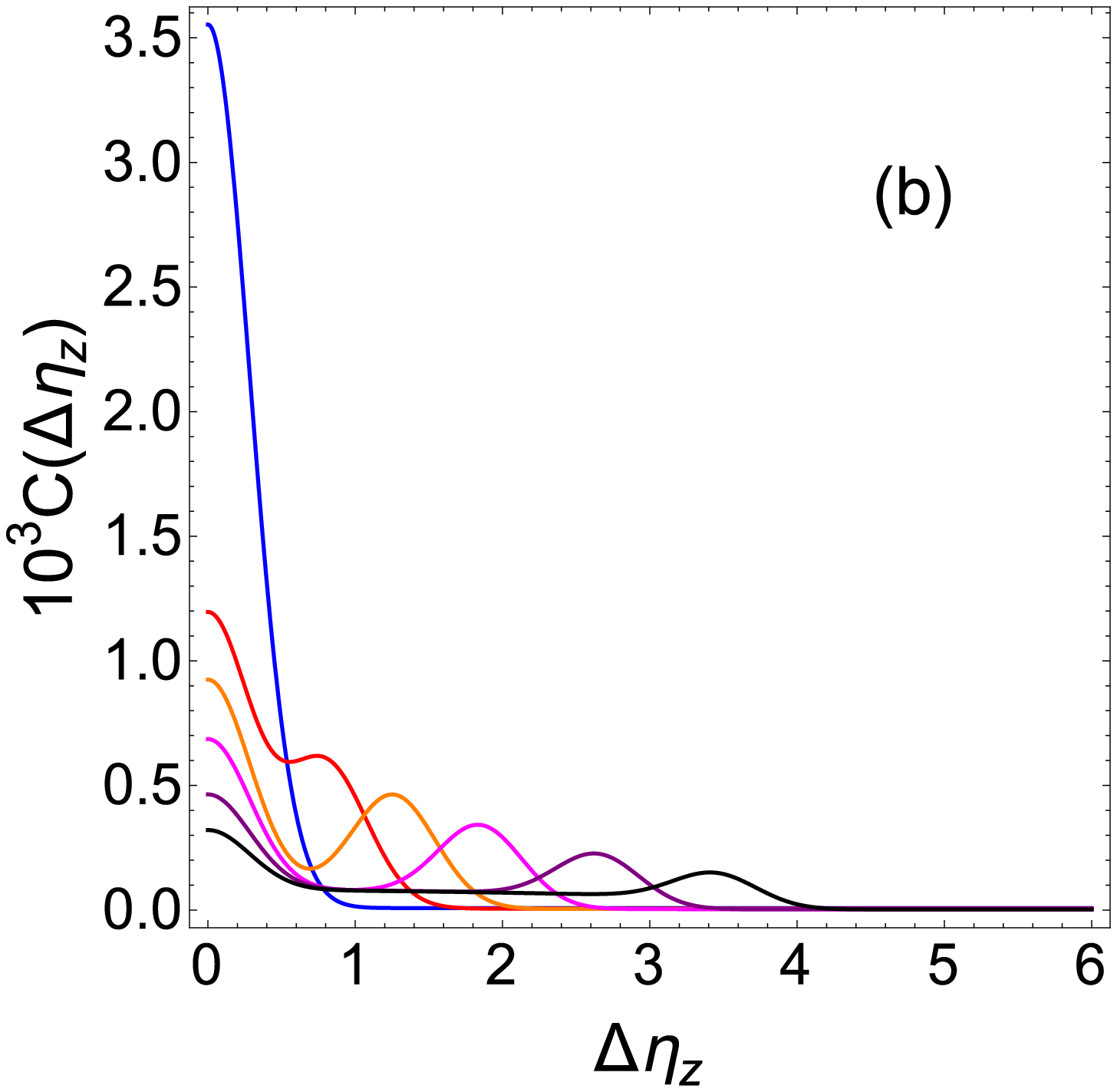}
\caption{(Color online) The longitudinal  sound wave amplitude $\delta=p_1/p_0$ (a) and the resulting pressure-pressure rapidity correlation (b) for an evolution time $\tau/\tau'=$ 1 (blue), 2, (red), 3 (orange), 5 (magenta), 10 (purple) and 20 (black).} \label{fig_BJ_L}
\end{figure}

The first case we consider is the longitudinal sound wave (i.e. homogeneous on transverse plane) on top of the Bjorken flow. The solution for $\delta(\tau,\eta)$ resulting from a static Gaussian perturbation at time $\tau'$ is given in Eq.(\ref{eq_bj_L_G}), and the rapidity correlation $C(\Delta \eta)$ is given in Eq.(\ref{eq_CC_bj_L}) of the Appendix E. To visualize the patterns of the solution, we use the following concrete numbers for plotting the solution: the center of the perturbation at $\eta_0=0$, the Gaussian width $\sigma=0.2$ and the perturbation amplitude parameter $\xi=0.1\sqrt{2\pi}\sigma$. Note that all these parameters introduced in this paper are dimensionless. In Fig.\ref{fig_BJ_L}, we show the wave amplitude $\delta(\eta)=p_1/p_0$ (left) and the resulting pressure-pressure rapidity correlation $C(\Delta \eta)$ (right) for an evolution time $\tau/\tau'=$ 1 (blue), 2, (red), 3 (orange), 5 (magenta), 10 (purple) and 20 (black). One can clearly see the propagation of the wave in rapidity from the original perturbation center toward the two side of large rapidity. While the amplitude of the two peaks is reduced with propagation time, the platform in between the two peaks formed at longer time becomes almost a constant. As a result, the correlation shows a clear pattern. There is always a short-range correlation peaked at zero separation, though  its magnitude and width become smaller and smaller with time. With longer propagation time, the correlation becomes more and more stretched in rapidity separation $\Delta \eta$. The long rapidity structure is associated with and bounded by the ``sound horizon''~\cite{Staig:2010pn}, the maximal distance that a sound wave could travel in a finite time. Note that these features for the longitudinal wave on top of Bjorken flow were previously already studied in \cite{Kapusta:2011gt}.

\begin{figure}[!hbt]
\includegraphics[width=0.3\textwidth]{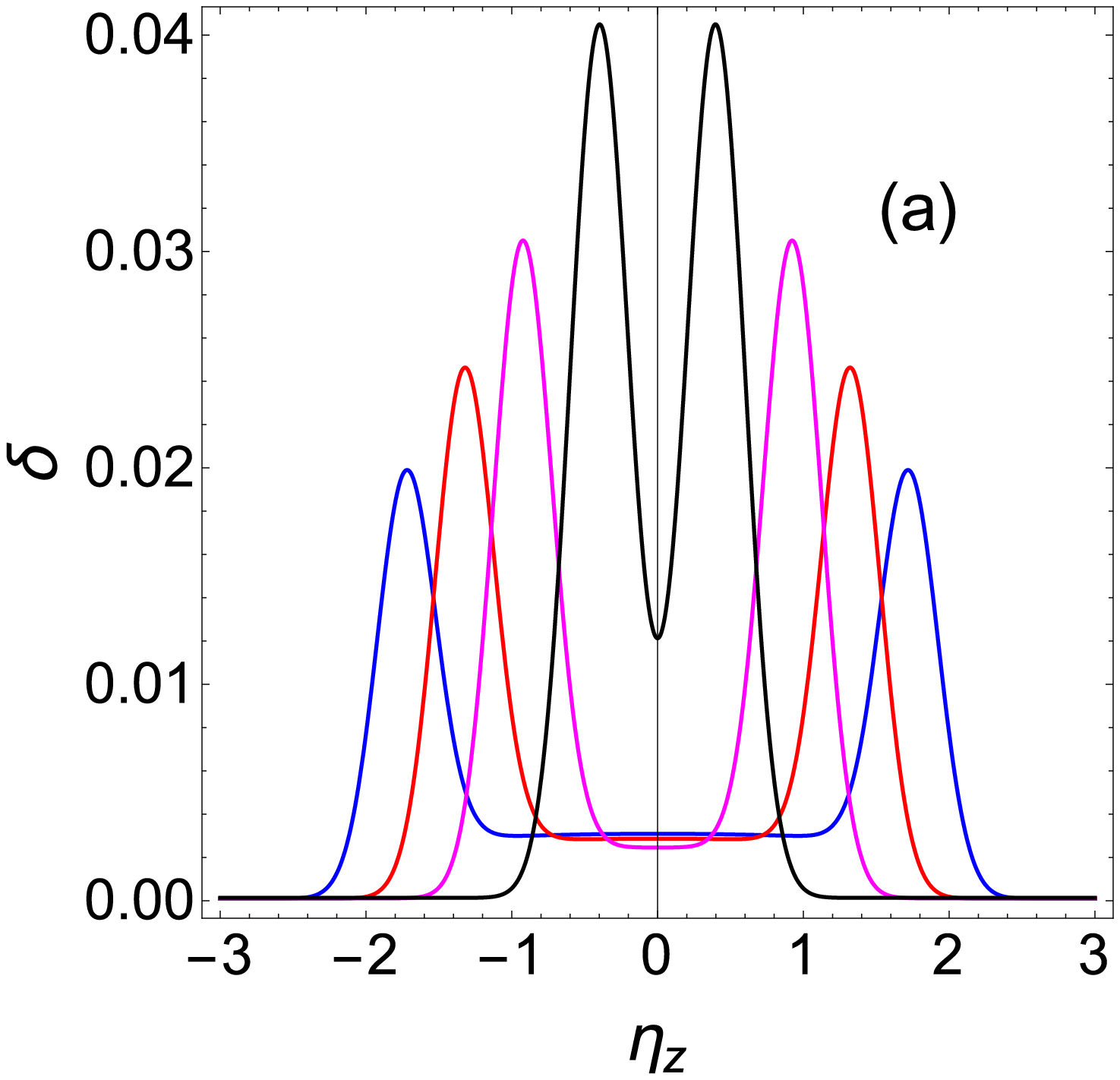} \hspace{0.2in}
\includegraphics[width=0.3\textwidth]{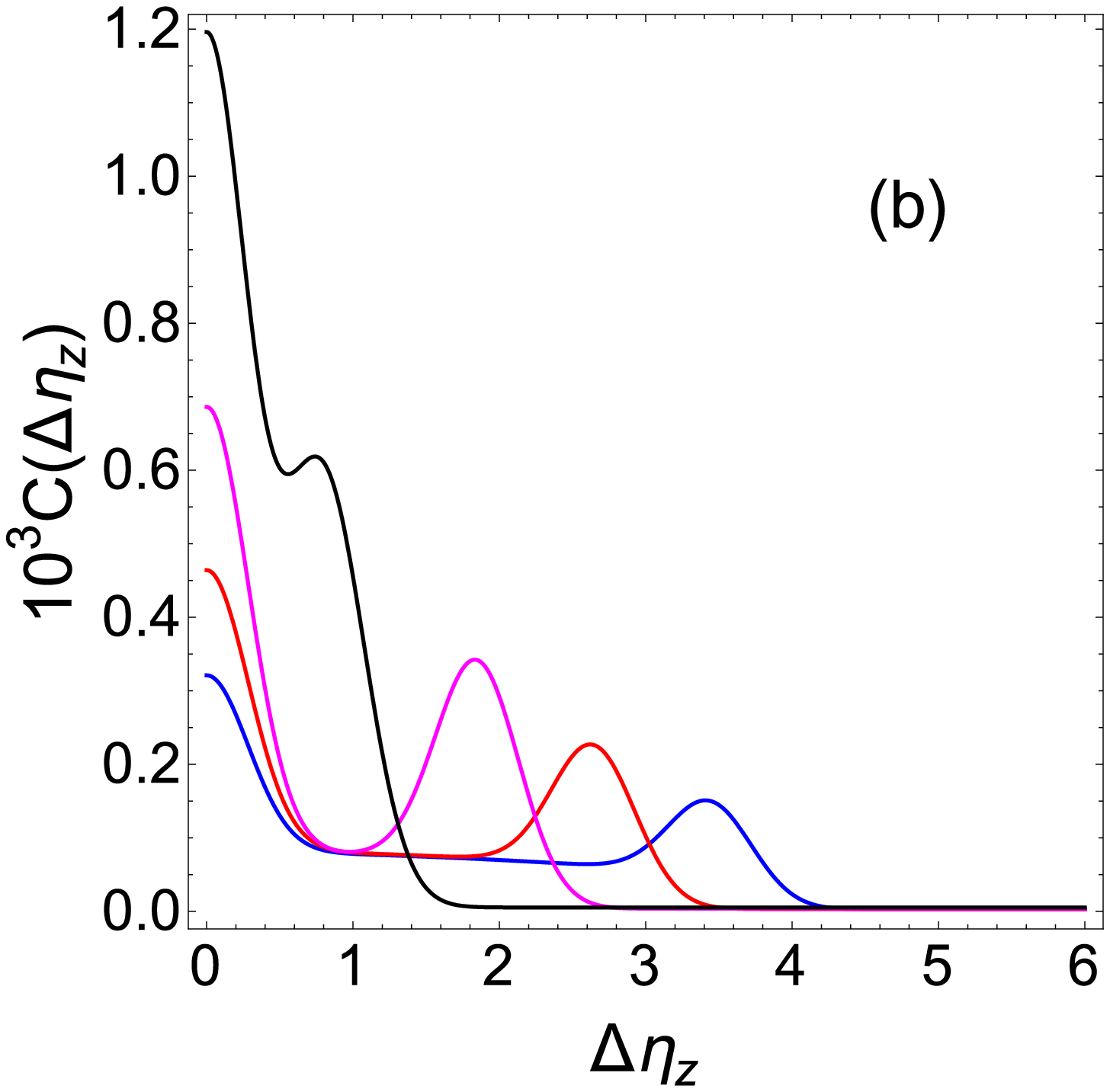}
\caption{(Color online) The longitudinal sound wave amplitude $\delta=p_1/p_0$ (a) and the resulting pressure-pressure rapidity correlation (b) at a fixed freeze-out time $\tau_{f}=10$ fm/c, arising from fluctuations that occur at time $\tau'=$ 0.5 (blue), 1 (red), 2 (magenta) and 5 (black) fm/c. } \label{fig_BJ_L_fo}
\end{figure}

\begin{figure}[!hbt]
\includegraphics[width=0.3\textwidth]{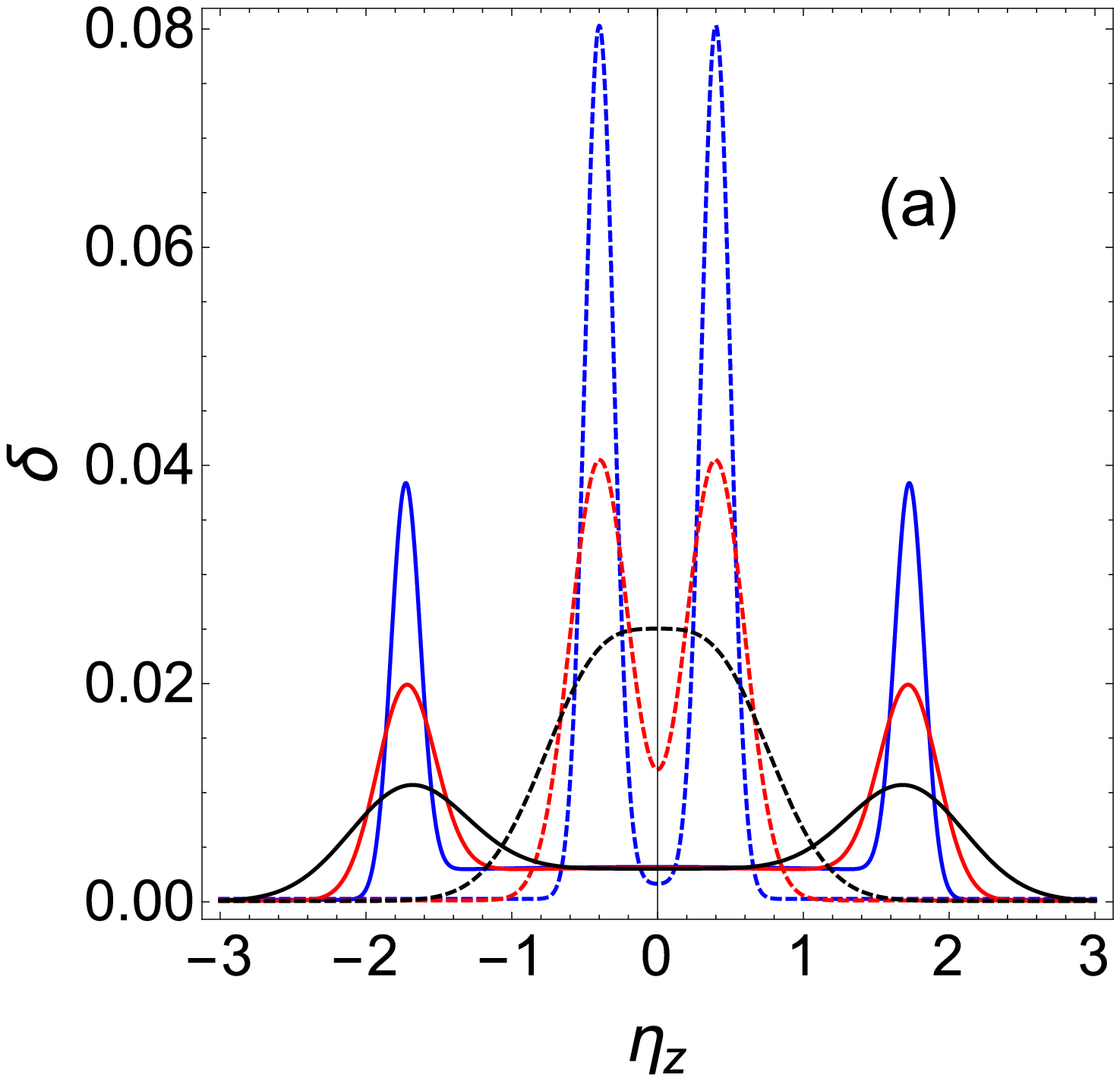} \hspace{0.2in}
\includegraphics[width=0.3\textwidth]{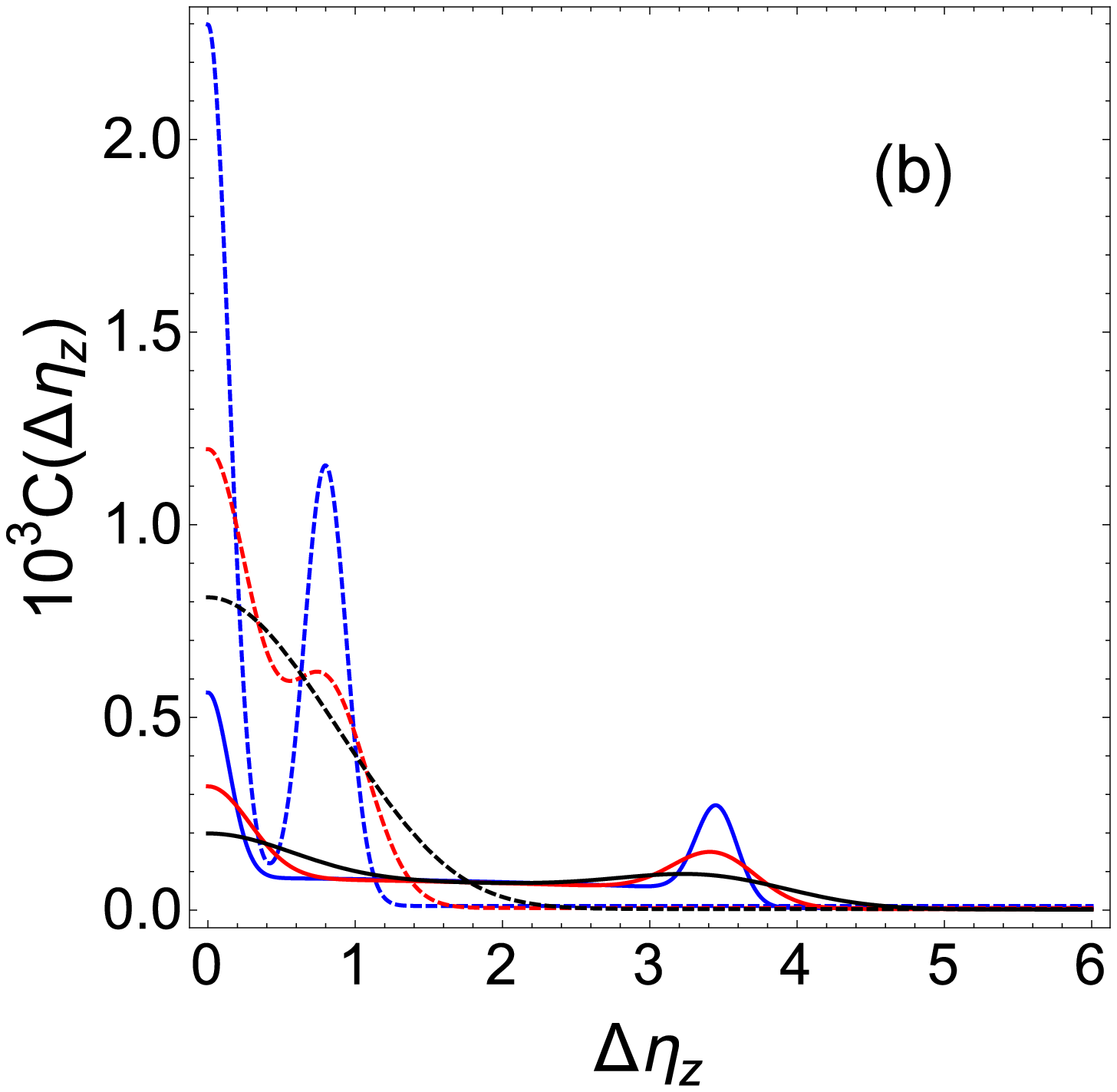}
\caption{(Color online) The longitudinal sound wave amplitude $\delta=p_1/p_0$ (left) and the resulting pressure-pressure rapidity correlation (right) at a fixed freeze-out time $\tau_{f}=10$ fm/c, arising from fluctuations that occur at time $\tau'=$ 0.5 (solid curves) and  5 (dashed curves) fm/c with Guassian fluctuation width $\sigma=$ 0.1 (blue), 0.2 (red) and 0.4 (black).  } \label{fig_BJ_L_sig}
\end{figure}

In view of application to heavy ion collisions, it would be interesting to see the patterns of the sound wave at a fixed final time $\tau_{f}$(i.e. the freeze-out time) from perturbations that occur at any time before. In Fig.\ref{fig_BJ_L_fo}, we show the wave amplitude $\delta(\eta)=p_1/p_0$ (left) and the resulting pressure-pressure rapidity correlation $C(\Delta \eta)$ (right) observed at $\tau_{f}=10$ fm/c,  arising from fluctuations that occur at time $\tau'=$ 0.5 (blue), 1 (red), 2 (magenta) and 5 (black) fm/c. Again, one sees that the perturbation  from early time travels over large rapidity interval and contributes to long-range rapidity correlation. In realistic case, the final correlation should be a time-integrated result by convoluting the fluctuation spectrum at all different times with their respective waves propagating to the freeze-out time. Clearly this contributes to the observed  rapidity correlations. How important this component is (in view of the measured pattern) would require a more quantitative study.

Lastly we study how sensitive these patterns are to the width parameter of the Gaussian perturbation. In Fig.\ref{fig_BJ_L_sig}, we show  sound wave amplitude $\delta=p_1/p_0$ (left) and the resulting pressure-pressure rapidity correlation (right) at a fixed freeze-out time $\tau_{f}=10$ fm/c, arising from fluctuations that occur at time $\tau'=$ 0.5 (solid curves) and  5 (dashed curves) fm/c with width $\sigma=$ 0.1 (blue), 0.2 (red) and 0.4 (black). The results clearly show that the rapidity distribution and correlation patterns are sensitive to the width for short propagation time, but become largely similar after long propagation time despite different width parameters.

\subsection{3D wave on top of Bjorken flow}

We now consider  the 3D sound wave on top of the Bjorken flow. While the background flow is the same as the previous case, now the perturbation induces the sound wave that propagates both in longitudinal and transverse directions. This should be a more realistic case for application to heavy ion collisions, and is not previously studied. The solution for $\delta=p_1/p_0$ resulting from a static (3D) Gaussian perturbation  at time $\tau'$ is given in Eq.(\ref{eq_bj_s},\ref{eq_bj_W},\ref{eq_bj_G_3D}). For comparison with the previous case, here we also focus on the wave amplitude $\delta$  and the resulting pressure-pressure rapidity correlation $C(\Delta \eta)$, and see how the wave propagation in transverse plane may affect the longitudinal patterns.

\begin{figure}[!hbt]
\includegraphics[width=0.3\textwidth]{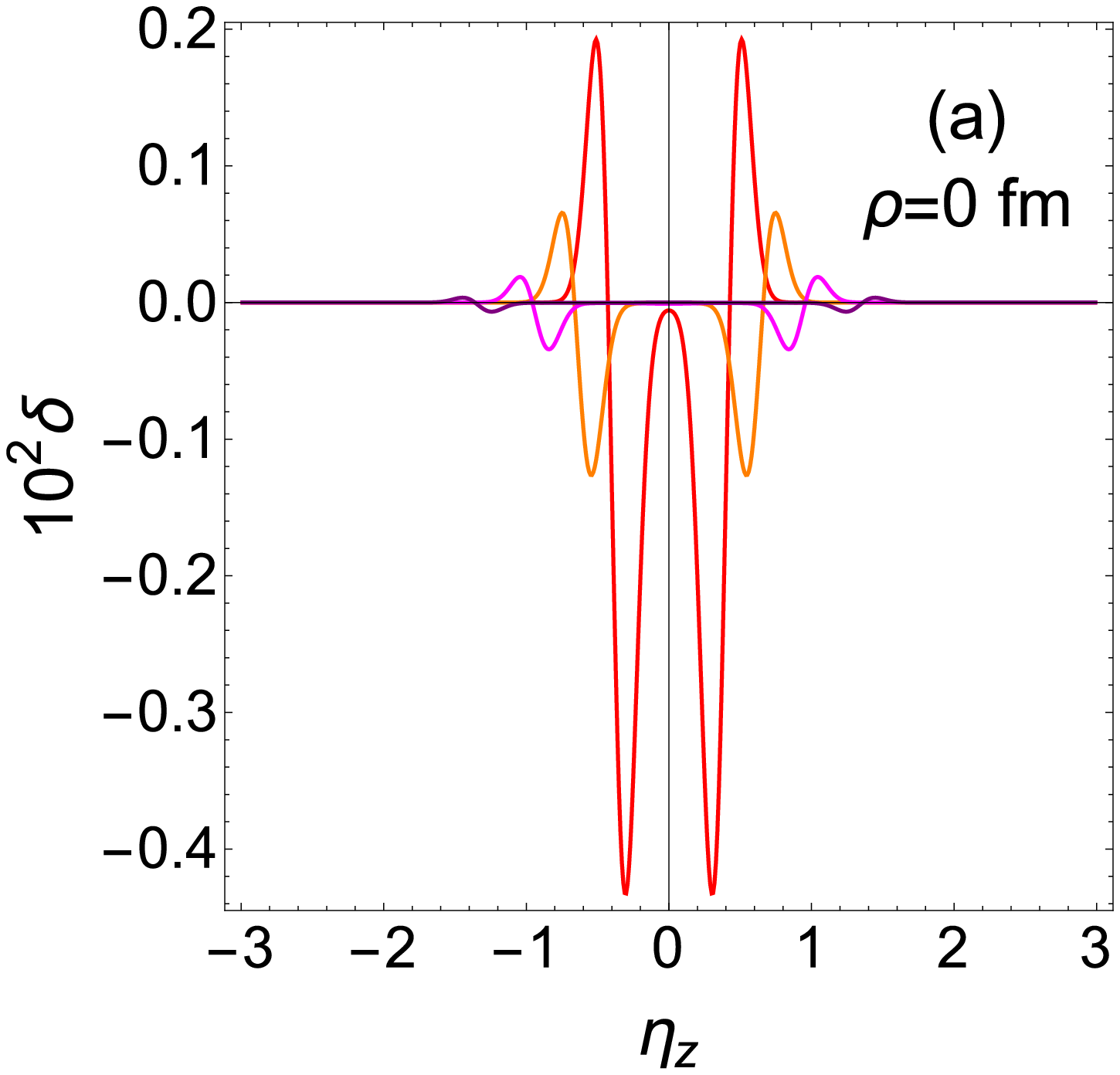}\hspace{0.05in}
\includegraphics[width=0.3\textwidth]{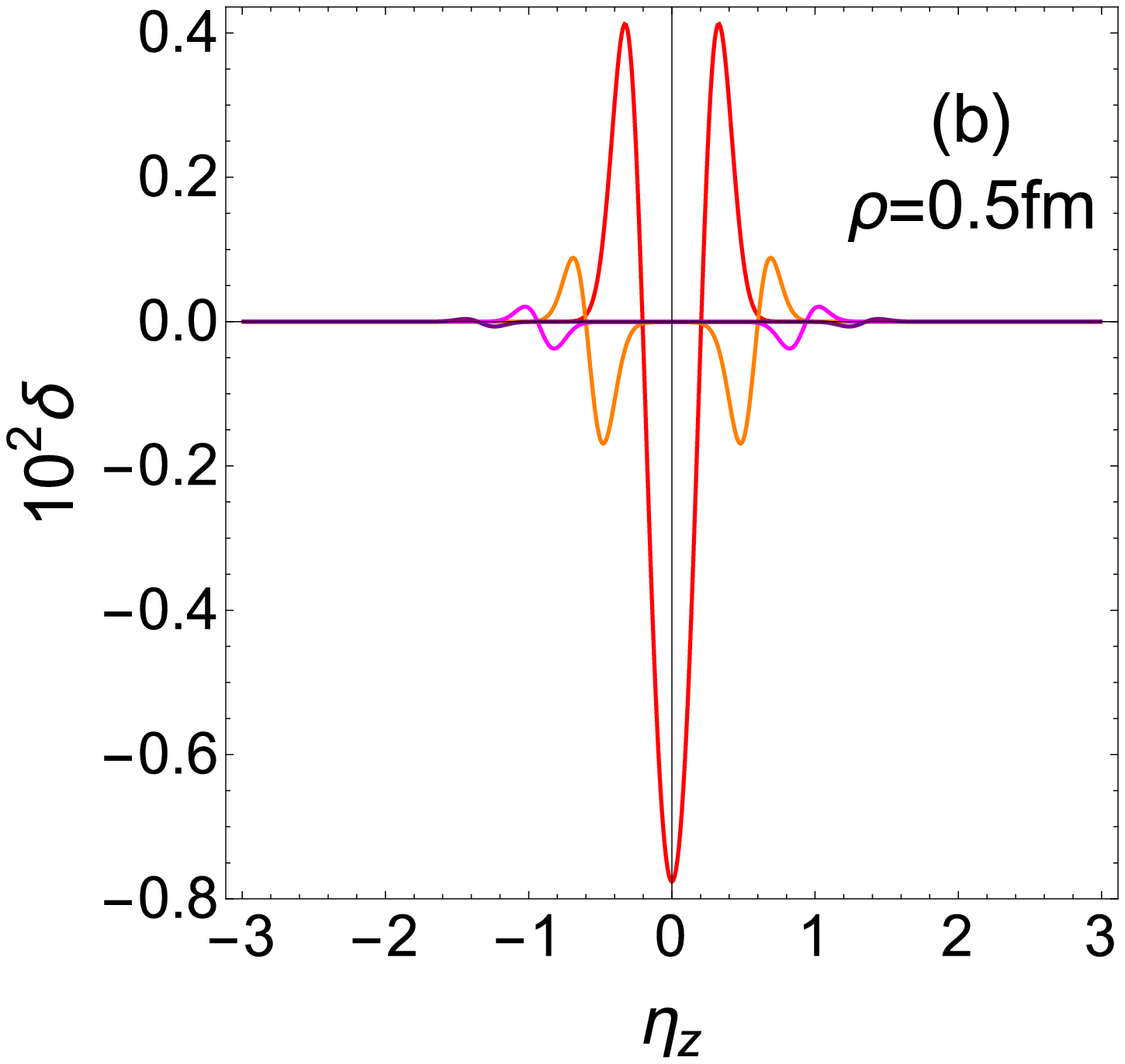}\hspace{0.05in}
\includegraphics[width=0.3\textwidth]{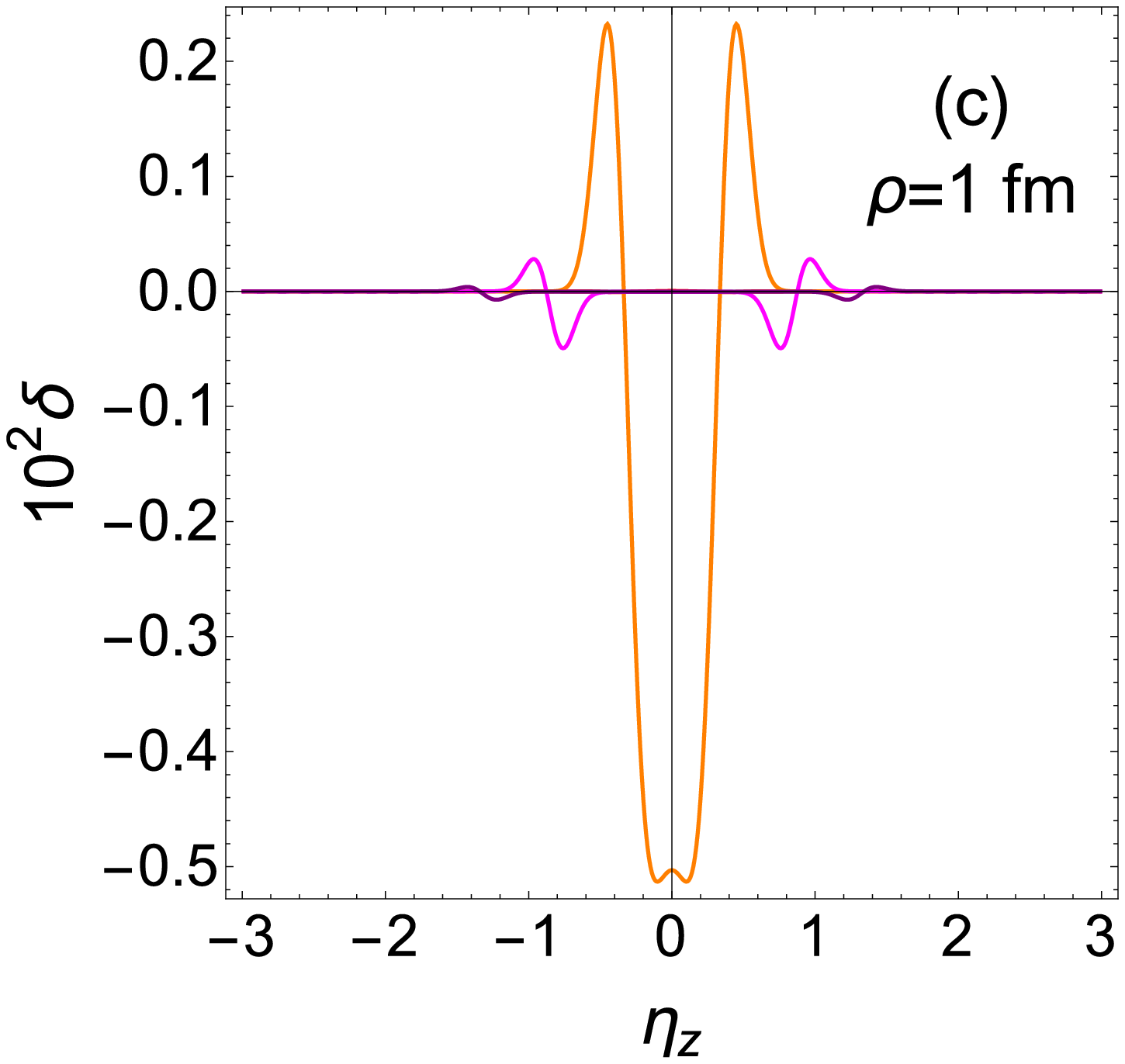}\\
\includegraphics[width=0.3\textwidth]{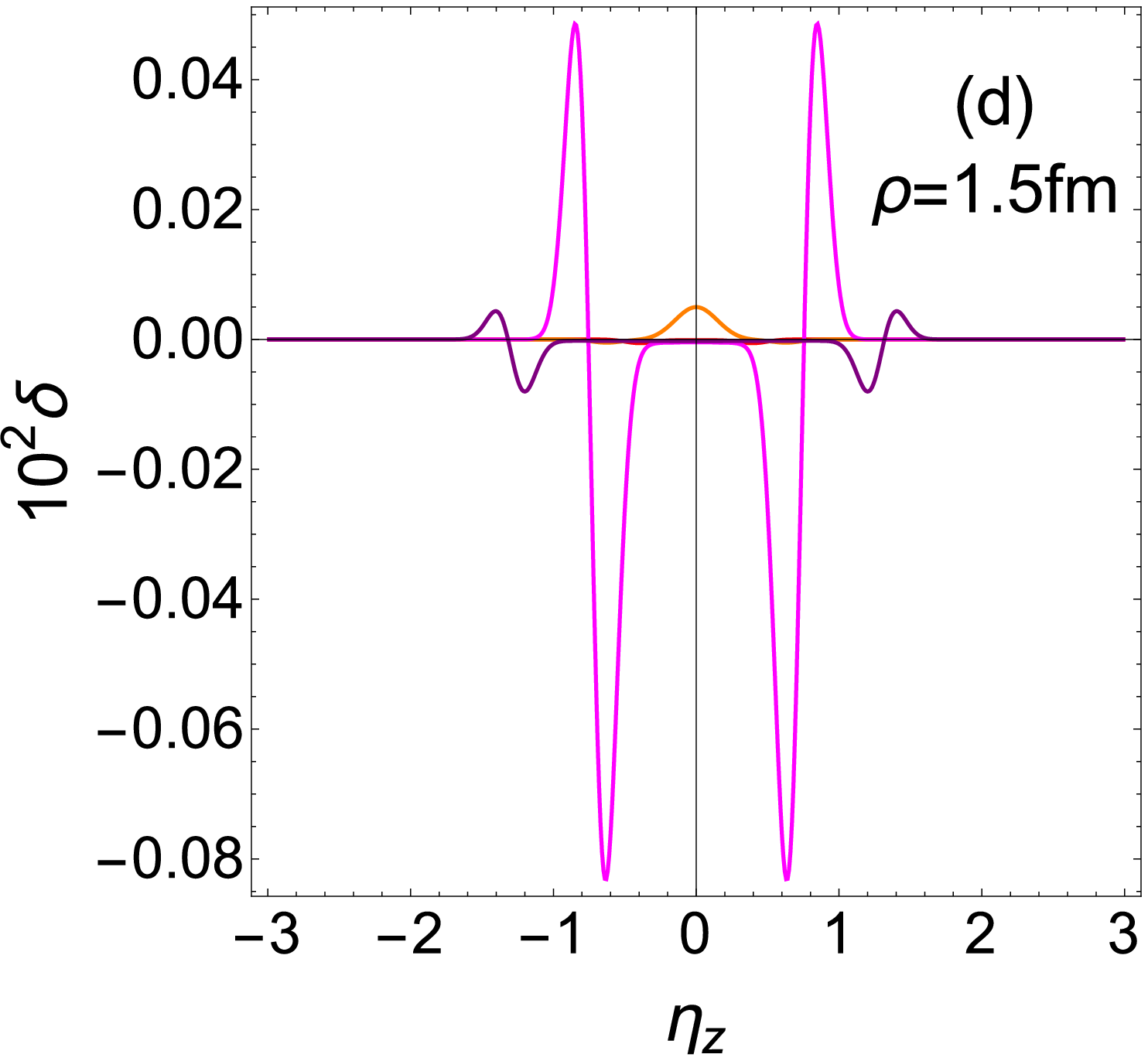}\hspace{0.05in}
\includegraphics[width=0.3\textwidth]{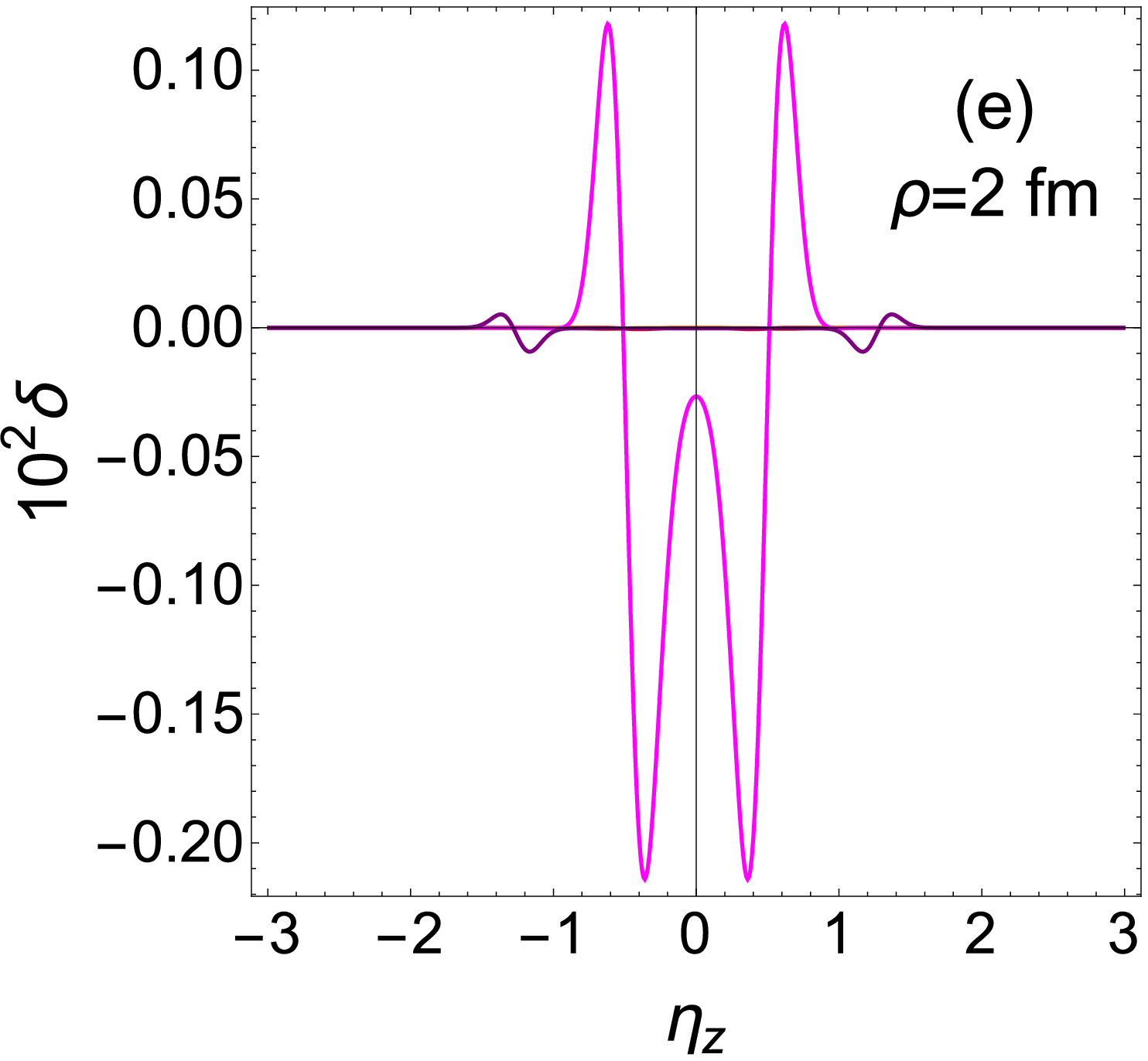}\hspace{0.05in}
\includegraphics[width=0.3\textwidth]{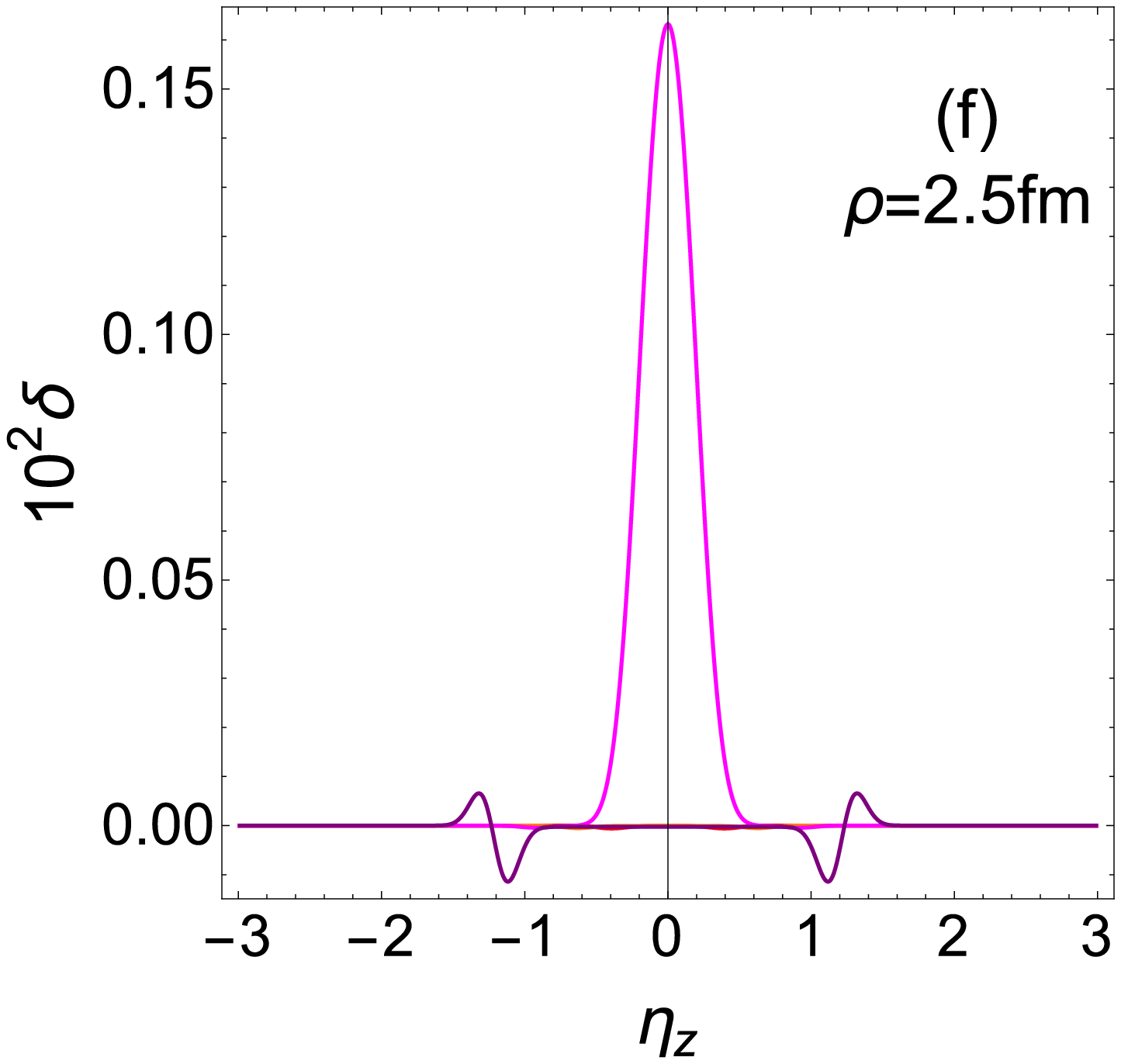}
\caption{(Color online) The sound wave amplitude $\delta=p_1/p_0$ at fixed transverse distance $\rho$ and at evolution time $\tau/\tau'=$ 2 (red), 3 (orange), 5 (magenta) and 10 (purple).} \label{fig_BJ_G_fix}
\end{figure}

\begin{figure}[!hbt]
\includegraphics[width=0.3\textwidth]{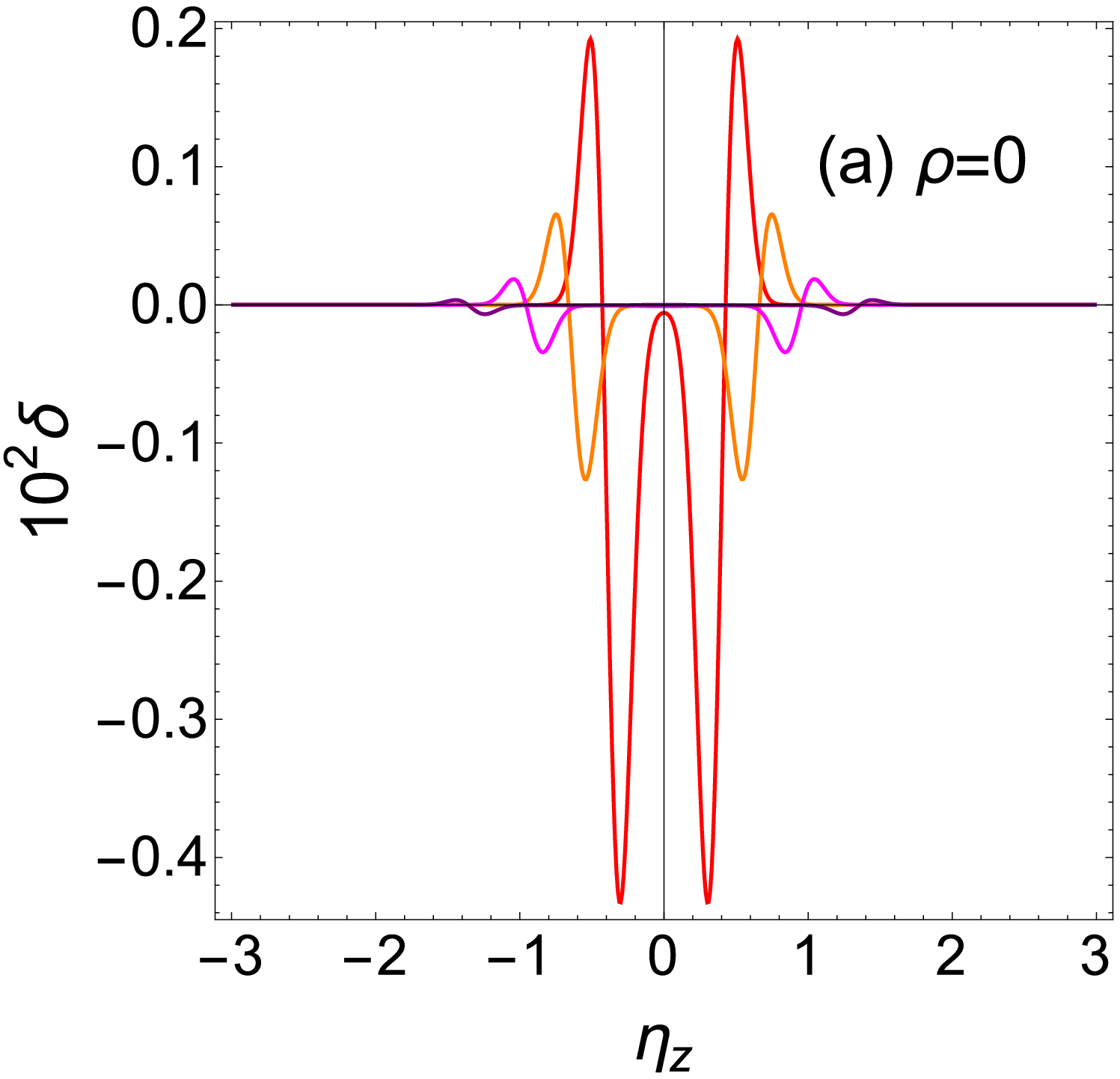} \hspace{0.05in}
\includegraphics[width=0.3\textwidth]{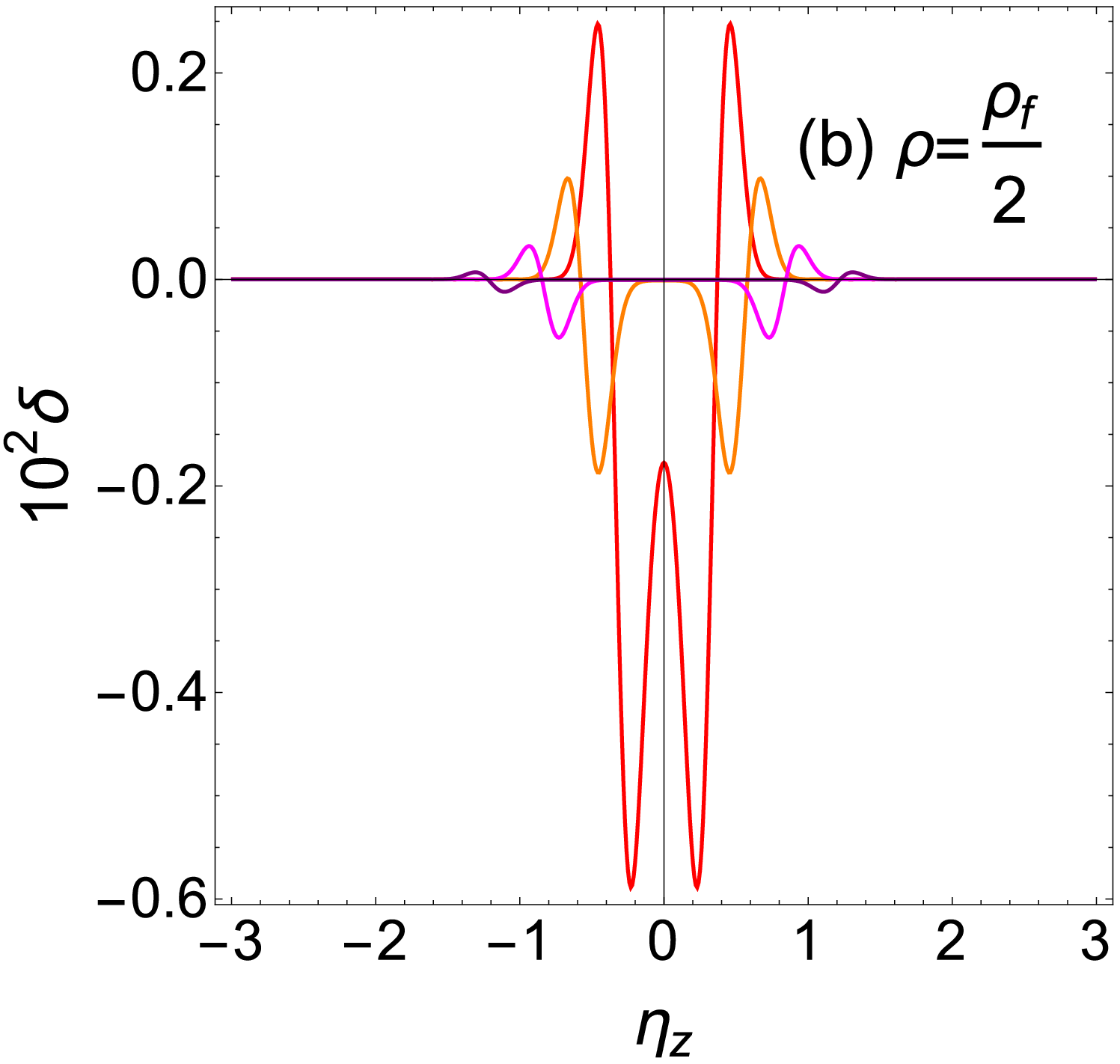} \hspace{0.05in}
\includegraphics[width=0.3\textwidth]{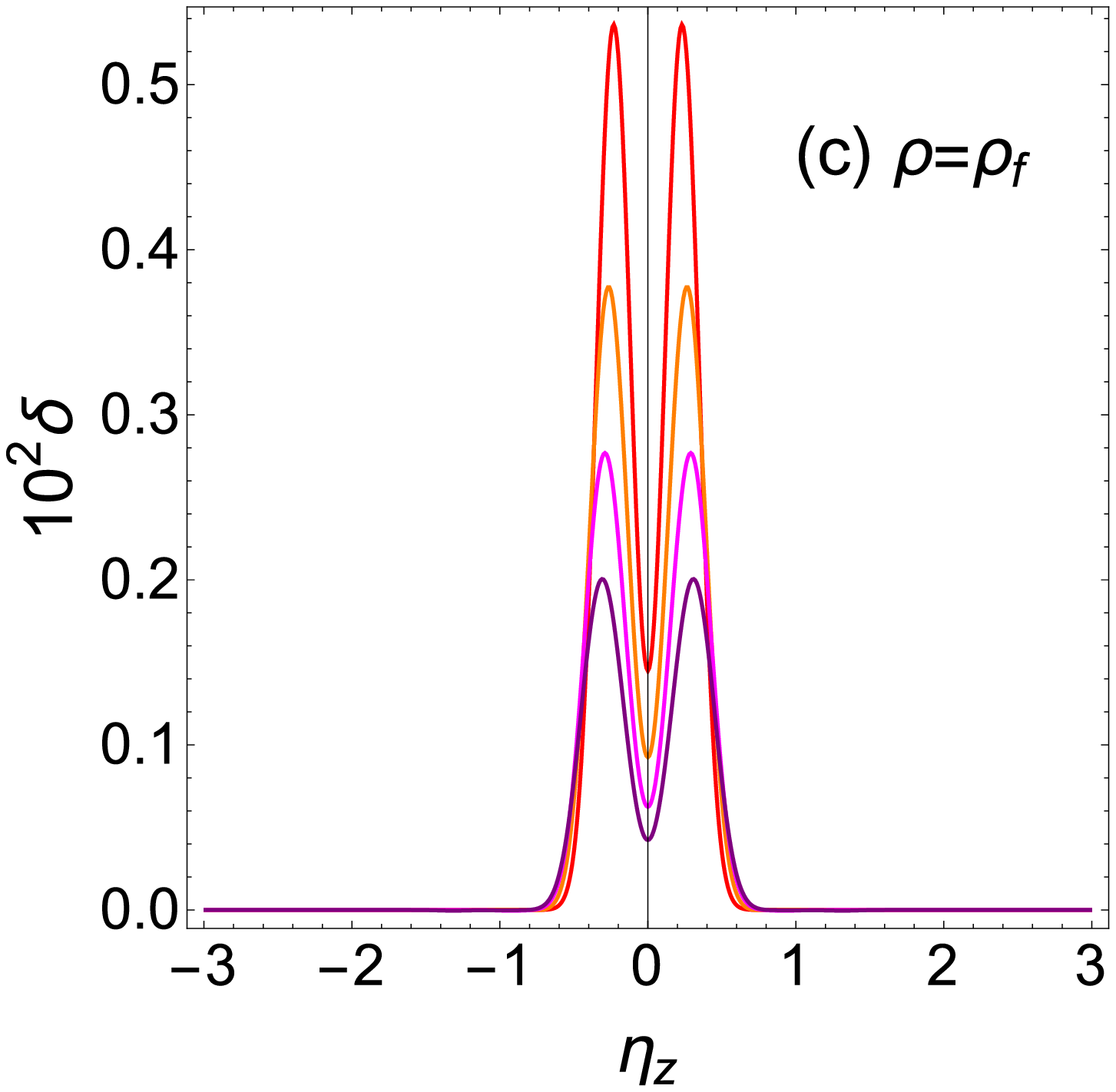}
\caption{(Color online) The sound wave amplitude $\delta=p_1/p_0$ at different transverse distance $\rho$ (measured in unit of the transverse wave-front position $\rho_f=c_s(\tau-\tau')$)  at evolution time $\tau/\tau'=$ 2 (red), 3 (orange), 5 (magenta) and 10 (purple). }  \label{fig_BJ_G}
\end{figure}

\begin{figure}[!hbt]
\includegraphics[width=0.35\textwidth]{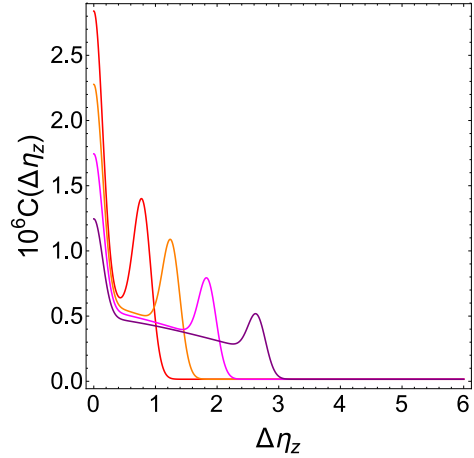}
\caption{(Color online) The pressure-pressure rapidity correlation (after integration over transverse coordinates) at evolution time $\tau/\tau'=$ 2 (red), 3 (orange), 5 (magenta) and 10 (purple).} \label{fig_BJ_G_2}
\end{figure}

We consider the static (3D) Gaussian perturbation located at center $\eta_0=\rho'=0$, with width parameter $\sigma=0.1$ and amplitude parameter $\xi=0.1 \left(2\pi\sigma^2\right)^{3/2}$ . Because of the cylindric symmetry of the system and the initial perturbation, there will be no azimuthal structure in all the quantities. In Figs.\ref{fig_BJ_G_fix} the wave amplitude  at evolution time $\tau/\tau'=$ 2 (red), 3 (orange), 5 (magenta) and 10 (purple) are shown for various transverse positions. One can see the decrease of wave amplitude and its propagation along rapidity direction with time at fixed $\rho$. It is also instructive to plot the wave amplitude for transverse radius $\rho$ measured in unit of the time-dependent transverse wave-front position $\rho_f=c_s(\tau-\tau')$, as shown in Fig.\ref{fig_BJ_G}. For each fixed $\rho$ relative to $\rho_f$, one sees the wave propagation in longitudinal direction with two fronts toward large rapidity. The longitudinal propagation is faster for smaller radius. This could be simply understood as follows: the wave modes propagating to larger transverse radius have larger transverse velocity component while smaller longitudinal velocity component. There is a major   difference between the 3D sound wave amplitude patterns here and the longitudinal sound wave patterns in the previous Subsection. In the present 3D wave case following the propagating crests there are wave trough regions  of  negative pressure perturbation $\delta < 0$, which are absent in the pure longitudinal wave case. This could be understood as a stronger ``push-out'' in the 3D wave case creating ``pressure void'' behind the wavefronts. We will further discuss this behavior in the next Subsection together with the Hubble case.

Finally we examine the pressure-pressure correlation in this case. The correlation now depends on transverse radius $\rho$. Since our focus is the rapidity correlation and for convenience in comparison with the longitudinal wave case, we integrate over $\rho$ to obtain the pressure-pressure rapidity correlation. The detailed calculation is shown in Appendix \ref{app_cor_BG} and the results are plotted in Fig.\ref{fig_BJ_G_2}   at evolution time $\tau/\tau'=$ 2 (red), 3 (orange), 5 (magenta) and 10 (purple). We see that the rapid correlation has a similar pattern to the longitudinal wave case: a peak at very small rapidity separation; a smaller peak at large rapidity separation arising from the ``sound horizon''; and a relatively flat regime in between. The correlation has its strength decrease while extends to large rapidity separation with increasing time for wave propagation.

\subsection{3D wave on top of Hubble flow}

Finally we study the 3D sound wave on top of the Hubble flow. By comparison with the previous case for the 3D wave on top of Bjorken flow, this will allow us to see how a different background flow, in particular how the background transverse expansion, will affect the sound wave propagation from the same perturbation and how the resulting rapidity correlation patterns may change. We emphasize that this may also be interesting for the recent intensive discussions on possible hydrodynamic explosion in high multiplicity pp and pA/dA collisions, since these small colliding systems have a small transverse size (with high pressure gradients) and thus the transverse flow may build up more quickly than in the AA case. The solution for $\delta$ resulting from a static (3D) Gaussian perturbation  at time $\tau'$ is given in Eq.(\ref{eq_hb_ga}).

\begin{figure}[!hbt]
\includegraphics[width=0.35\textwidth]{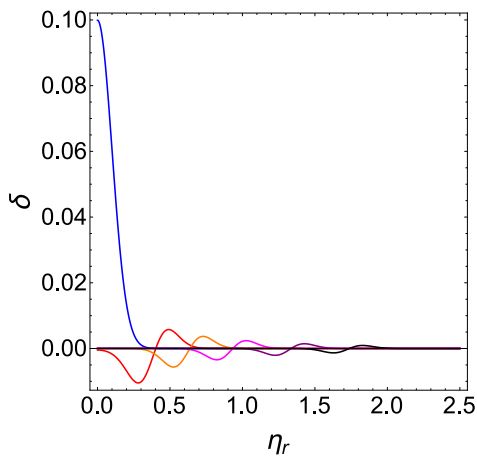}
\caption{(Color online) The wave amplitude $\delta $ at evolution time $\tau/\tau'=$ 1 (blue), 2 (red), 3 (orange), 5 (magenta), 10 (purple) and 20 (black).} \label{fig_Hu_H}
\end{figure}

To be concrete we use  the following parameters for the Guassian perturbation: $\eta'=0$, $\sigma=0.1$, and $\xi=0.1\left(2\pi\sigma^2\right)^{3/2}$ . Fig.\ref{fig_Hu_H} shows the time evolution of the sound wave amplitude in radial rapidity $\eta_r$.  Note that $\eta_r$ is not the observable rapidity, a meaningful correlation is that of the pseudo rapidity, which can be obtain from the angle $\theta$. As the sound wave is spherically symmetric here, there is no angle dependence of the pressure. Consequently, the pressure correlation should be a constant of pseudo rapidity.

\begin{figure}[!hbt]
\includegraphics[width=0.3\textwidth]{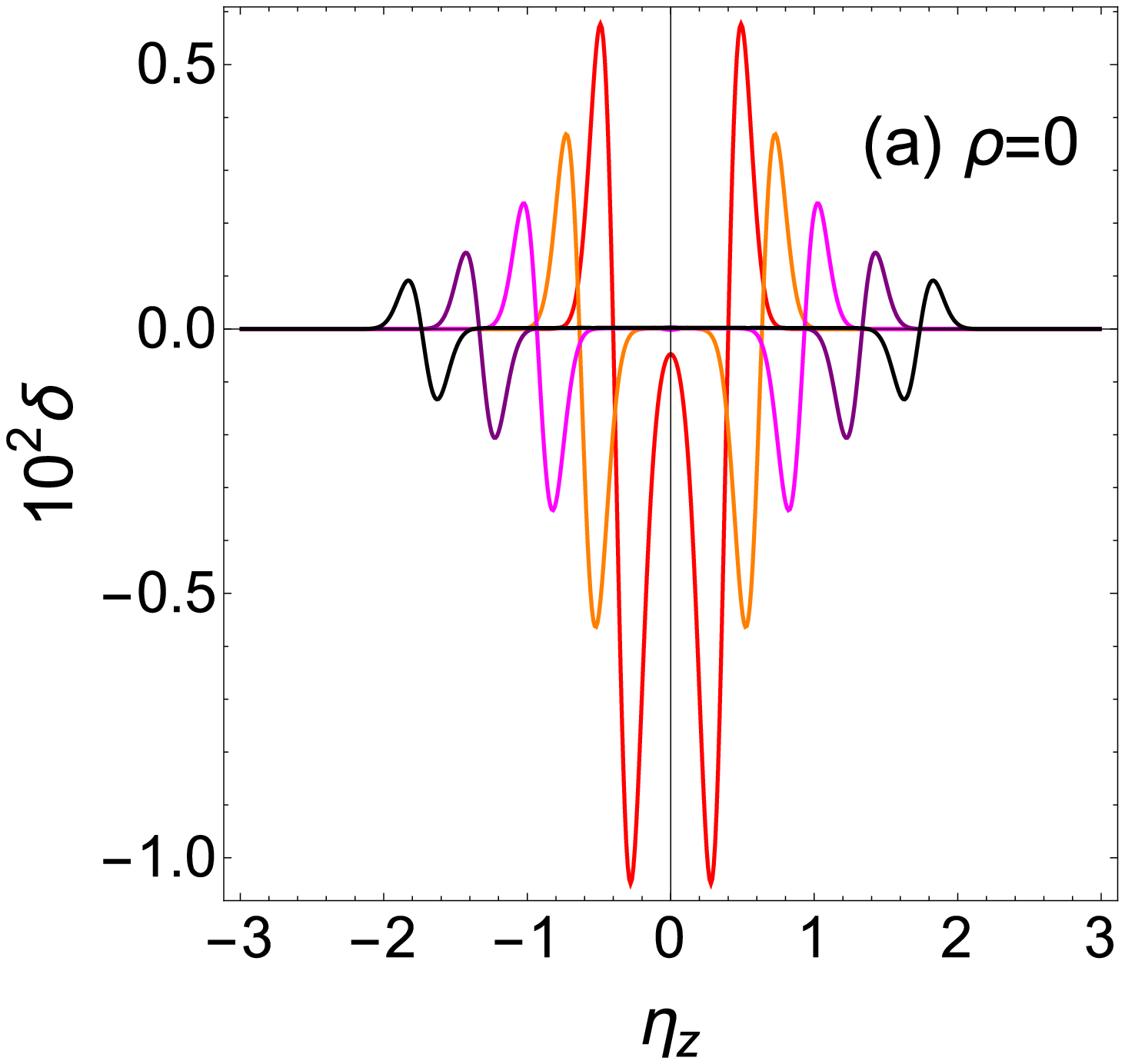} \hspace{0.05in}
\includegraphics[width=0.3\textwidth]{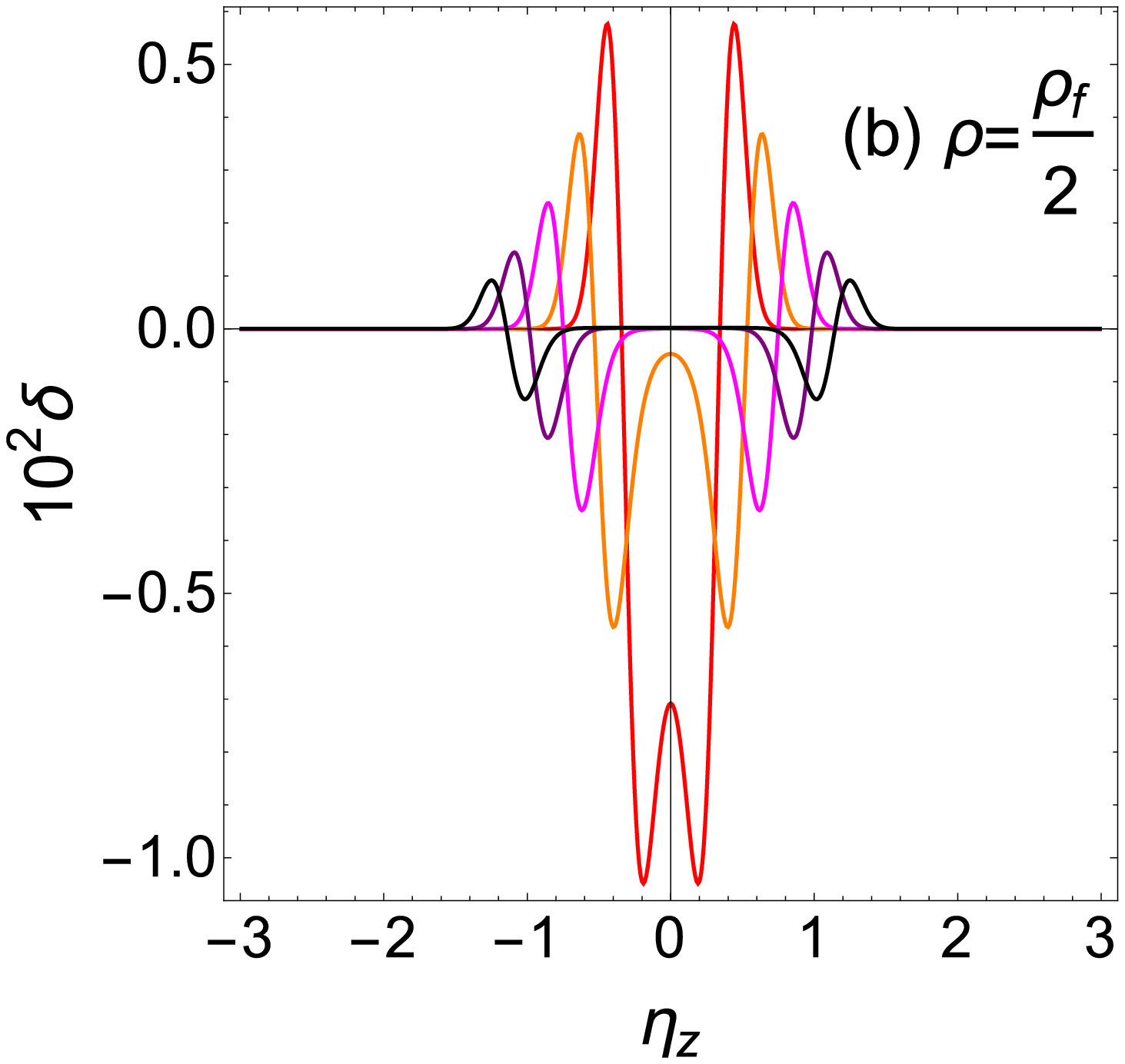} \hspace{0.05in}
\includegraphics[width=0.3\textwidth]{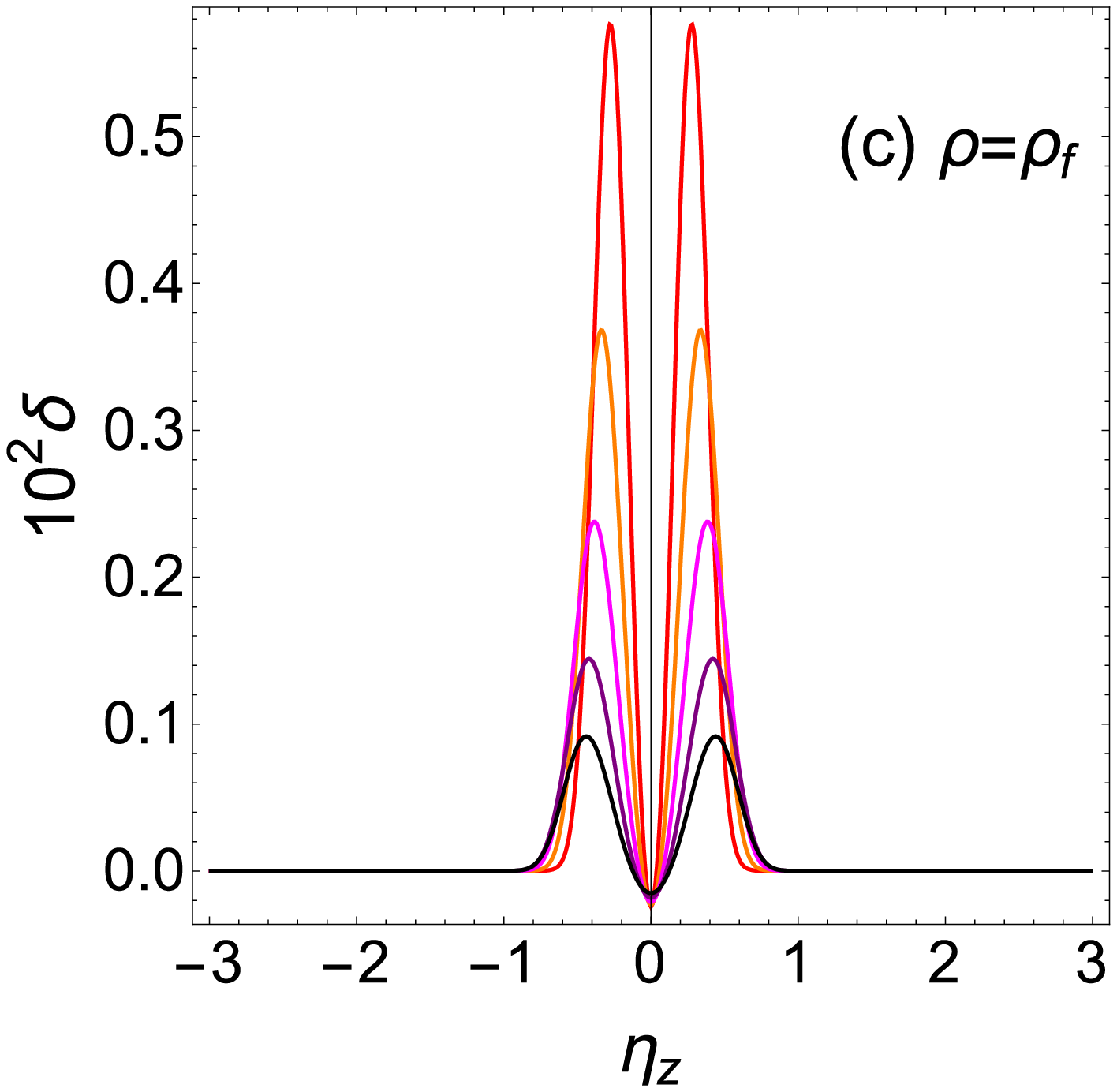} \\
\vspace{-0.15in}
\includegraphics[width=0.3\textwidth]{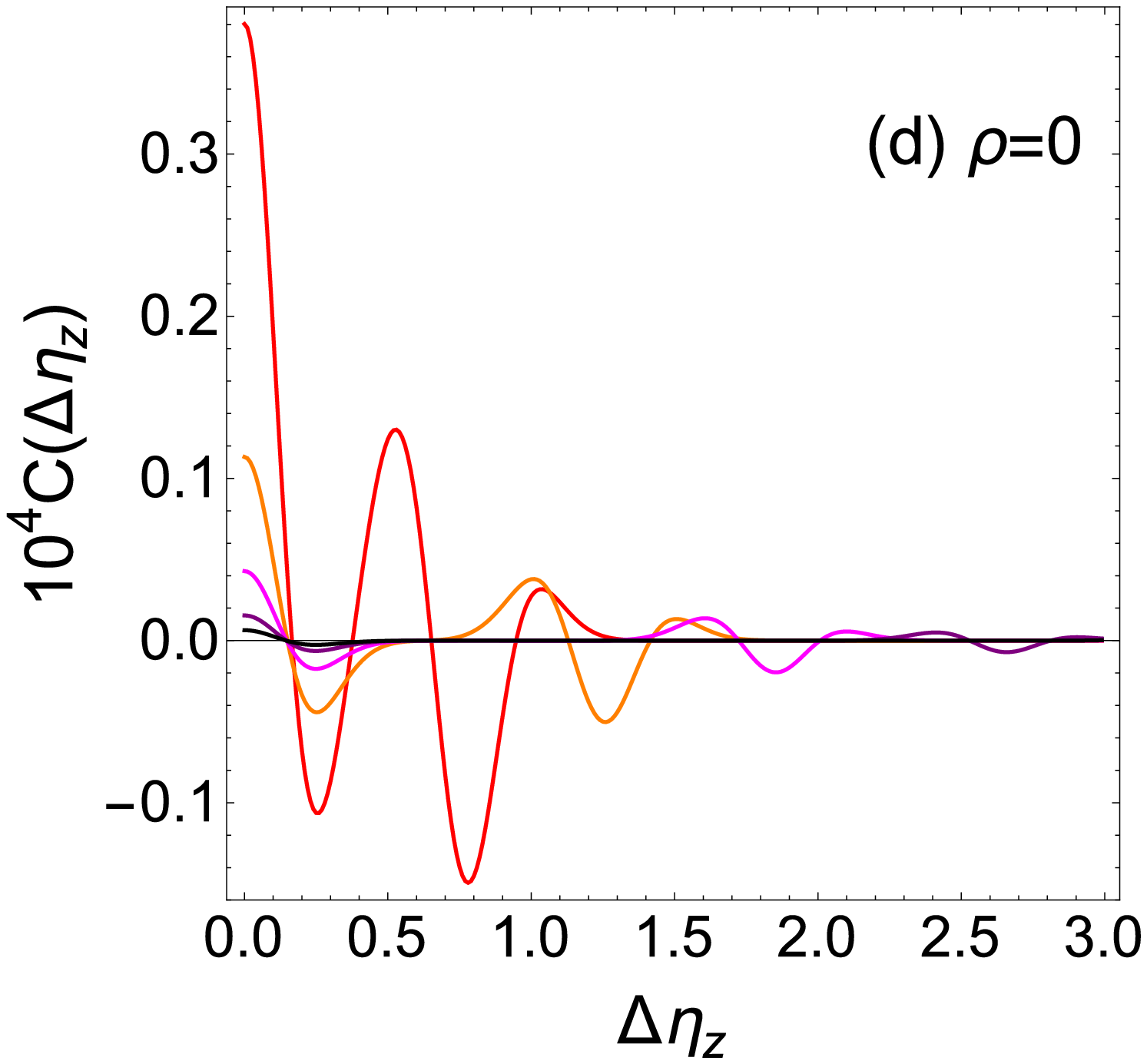} \hspace{0.05in}
\includegraphics[width=0.3\textwidth]{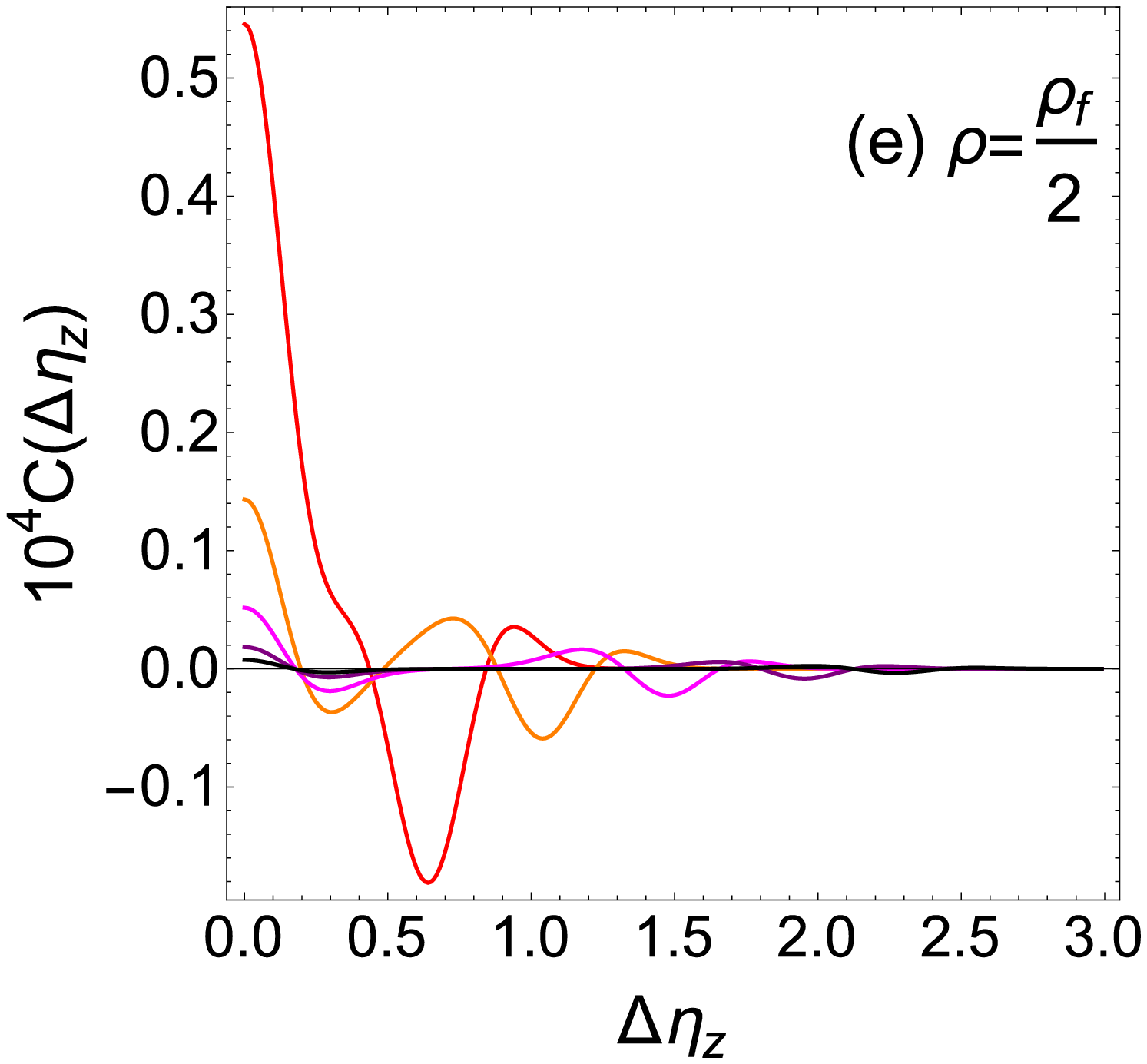} \hspace{0.05in}
\includegraphics[width=0.3\textwidth]{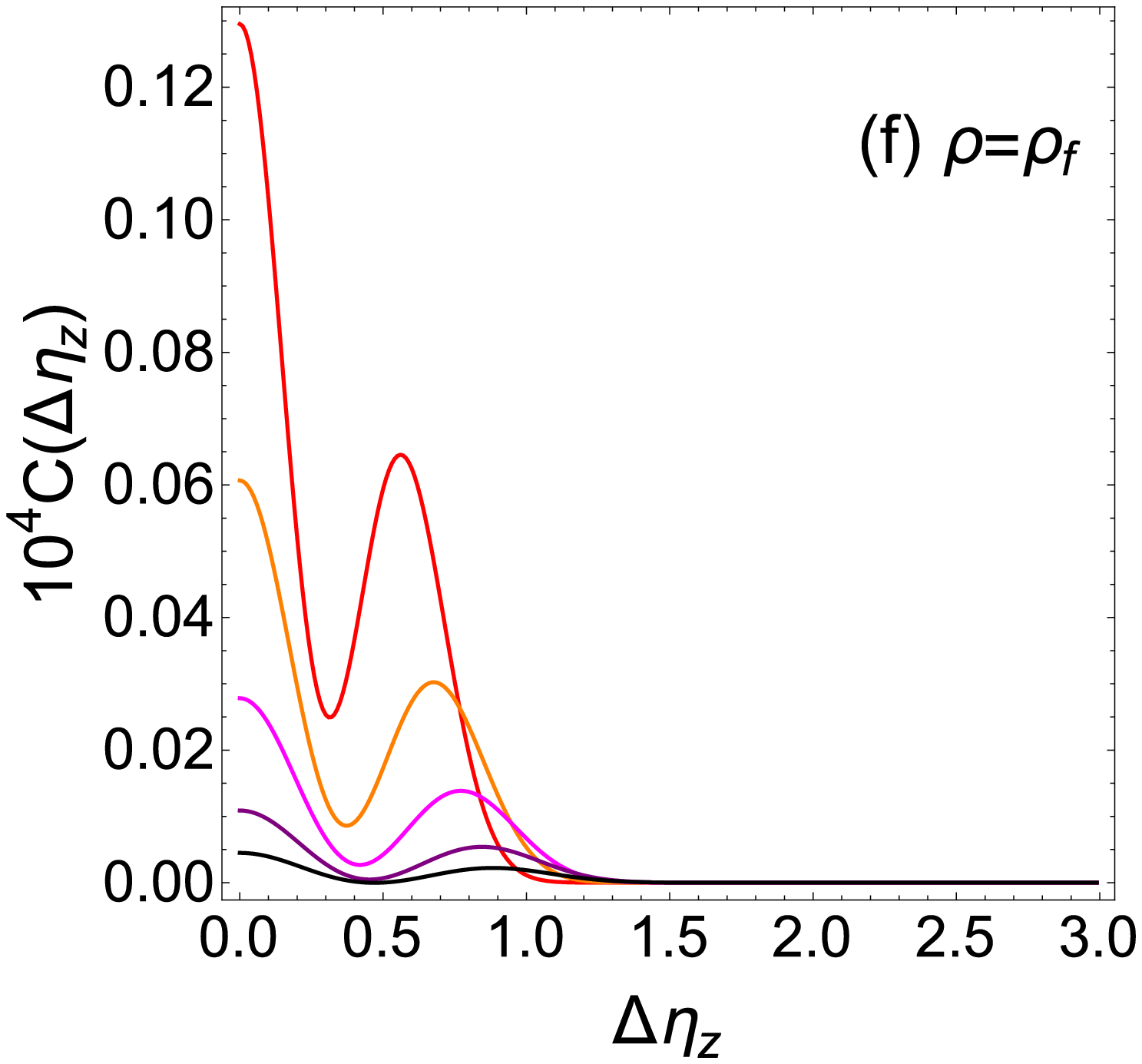}
\vspace{-0.25in}
\caption{(Color online) The sound wave amplitude $\delta=p_1/p_0$ (a-c) and the resulting pressure-pressure rapidity correlation (d-f) at fixed transverse distance $\rho$ and at evolution time $\tau_r/\tau_r'=$ 2 (red), 3 (orange), 5 (magenta), 10 (purple) and 20 (black) fm/c. $\rho_f=\tau_r \sinh[c_s\ln(\tau_r/\tau'_r)]$ is the transverse wave-front at the corresponding time. }\label{fig_HB}
\end{figure}

\begin{figure}[!hbt]
\includegraphics[width=0.3\textwidth]{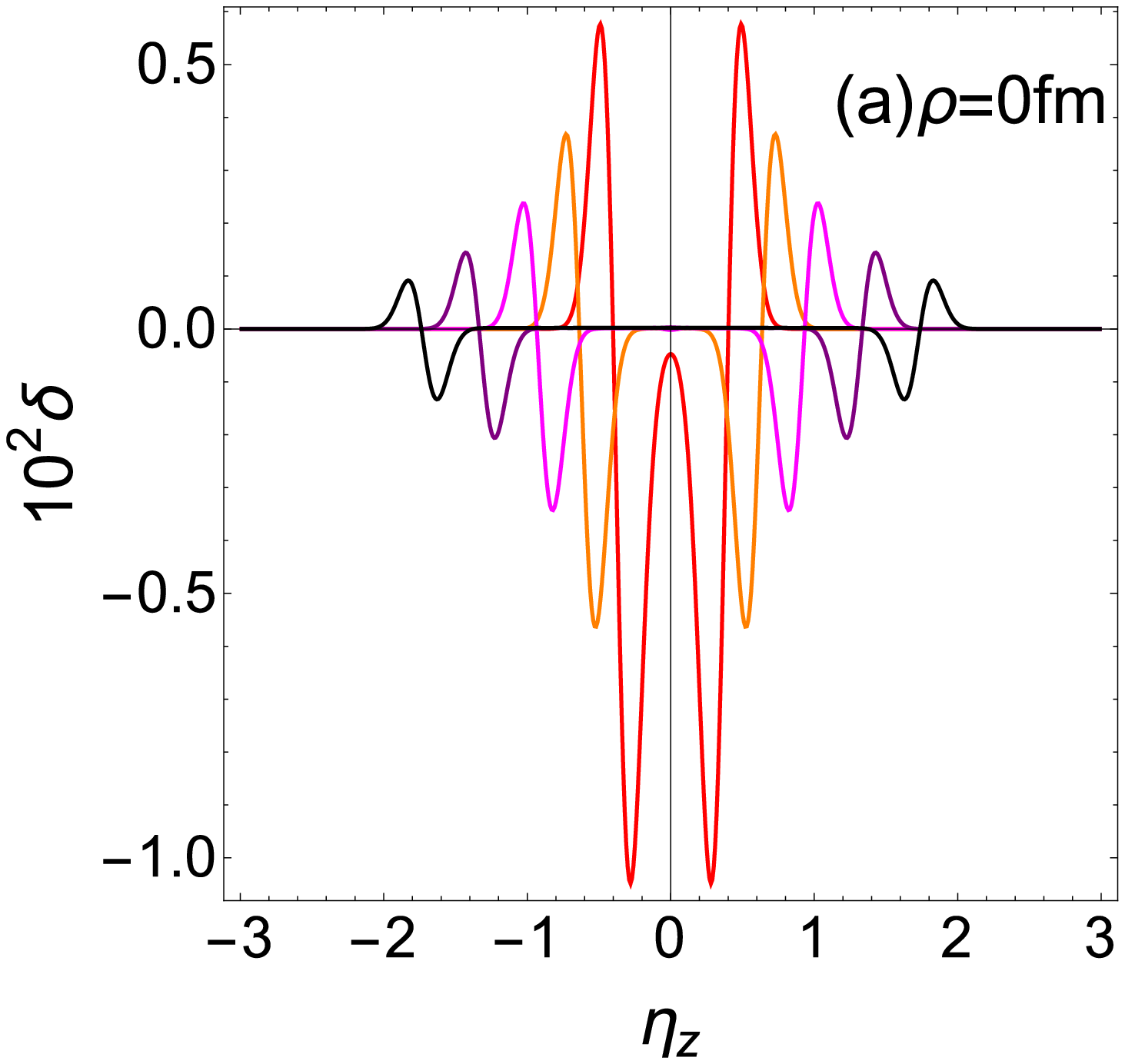} \hspace{0.05in}
\includegraphics[width=0.3\textwidth]{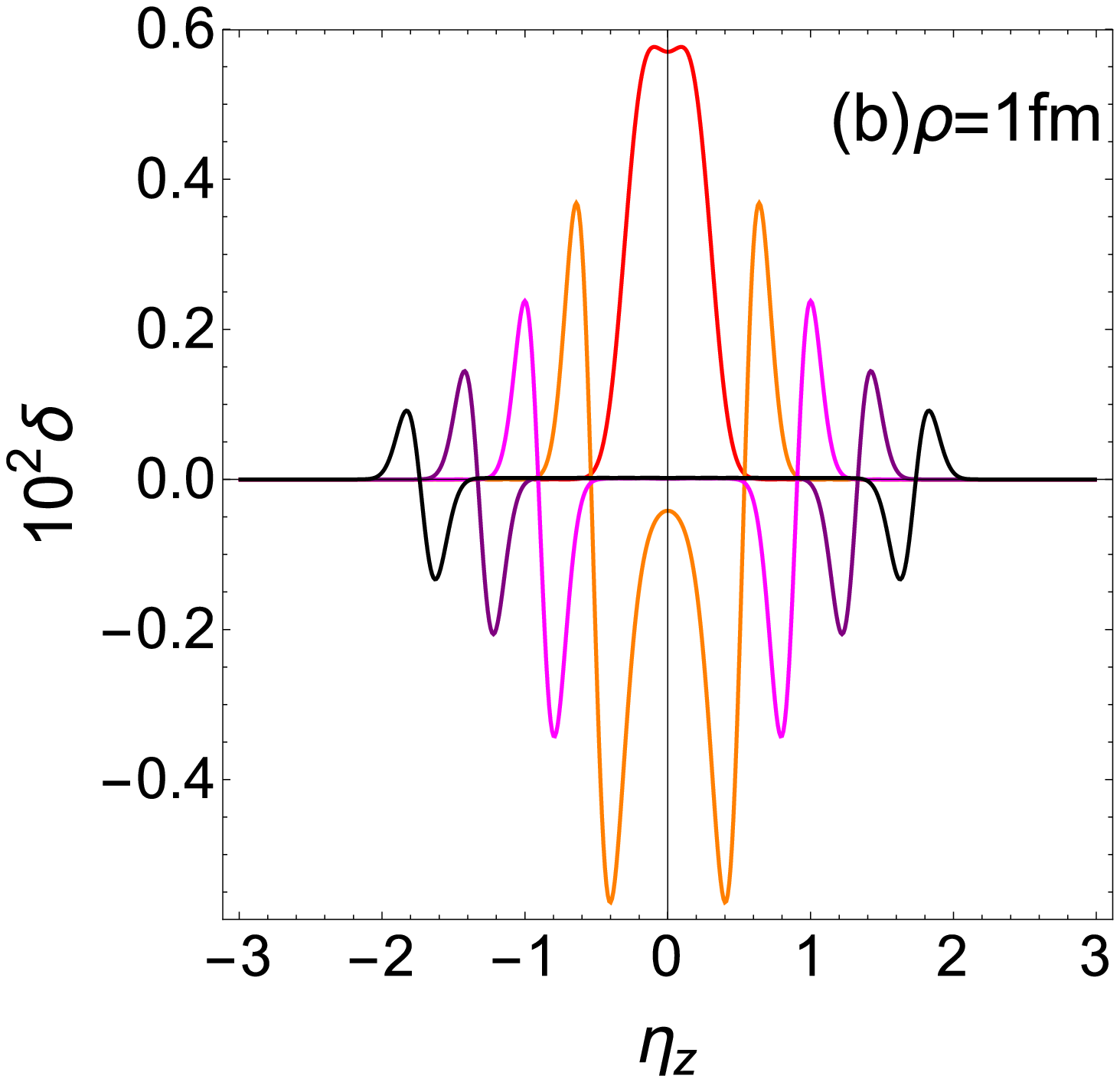} \hspace{0.05in}
\includegraphics[width=0.3\textwidth]{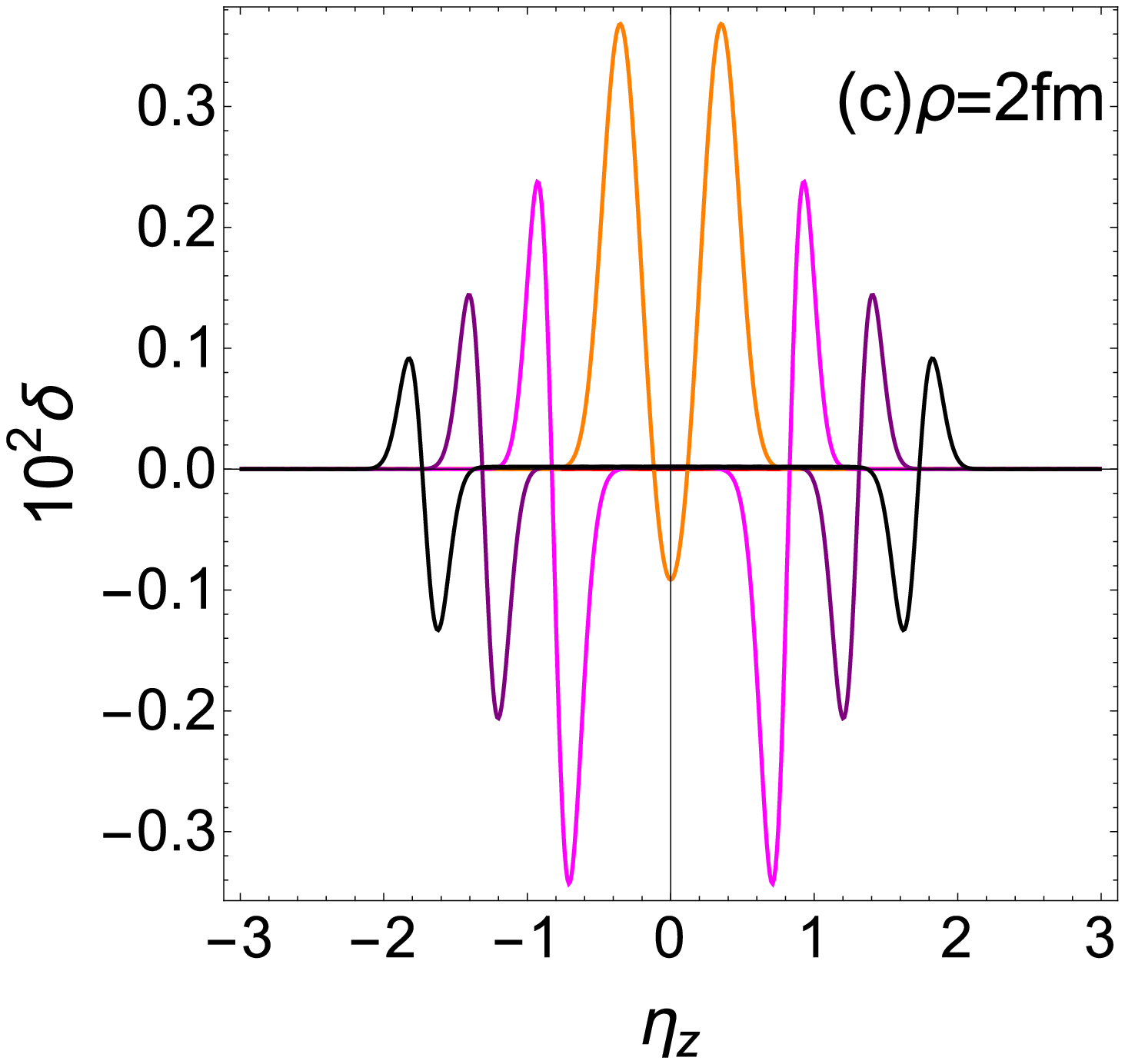} \\
\vspace{-0.15in}
\includegraphics[width=0.3\textwidth]{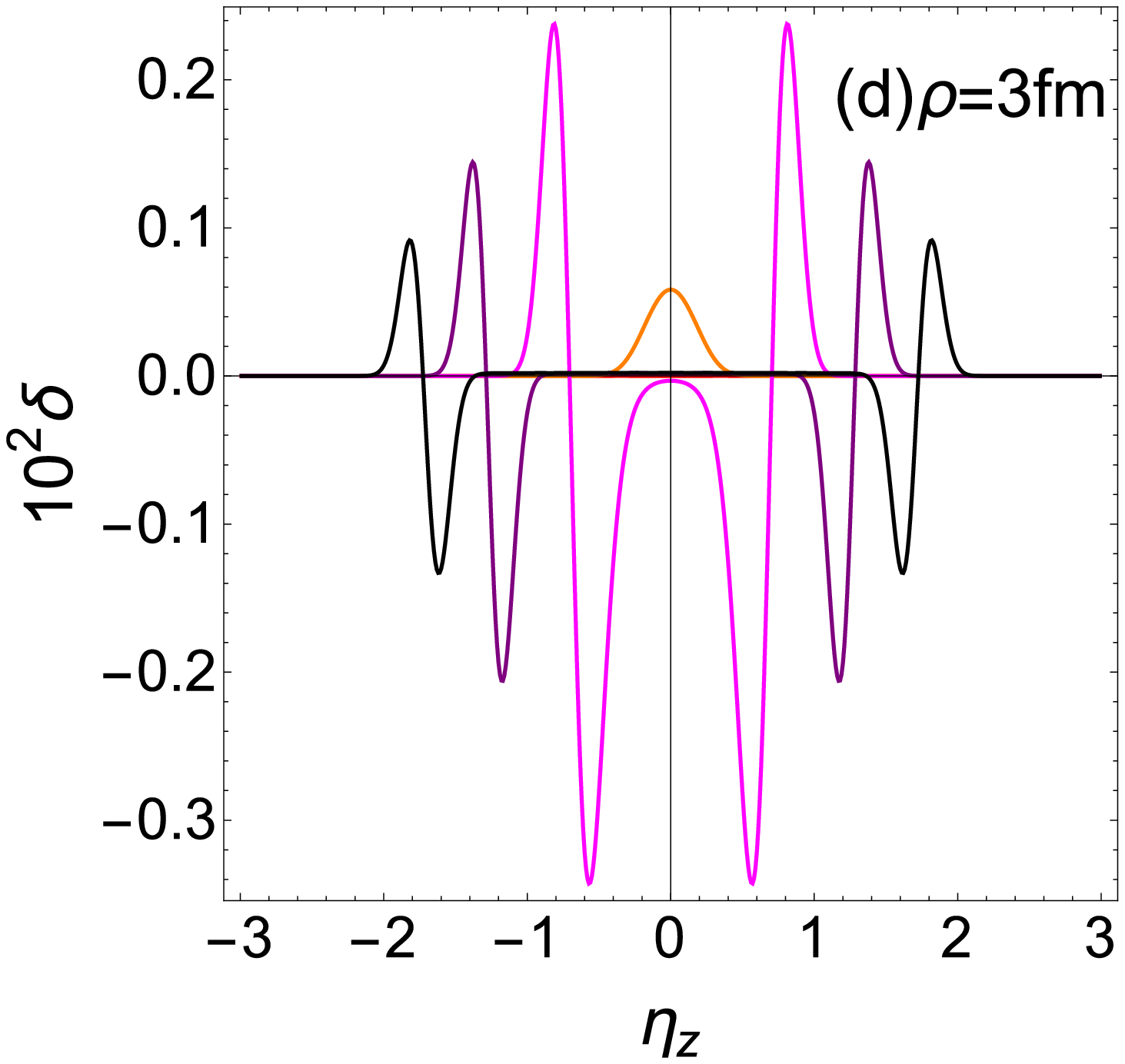} \hspace{0.05in}
\includegraphics[width=0.3\textwidth]{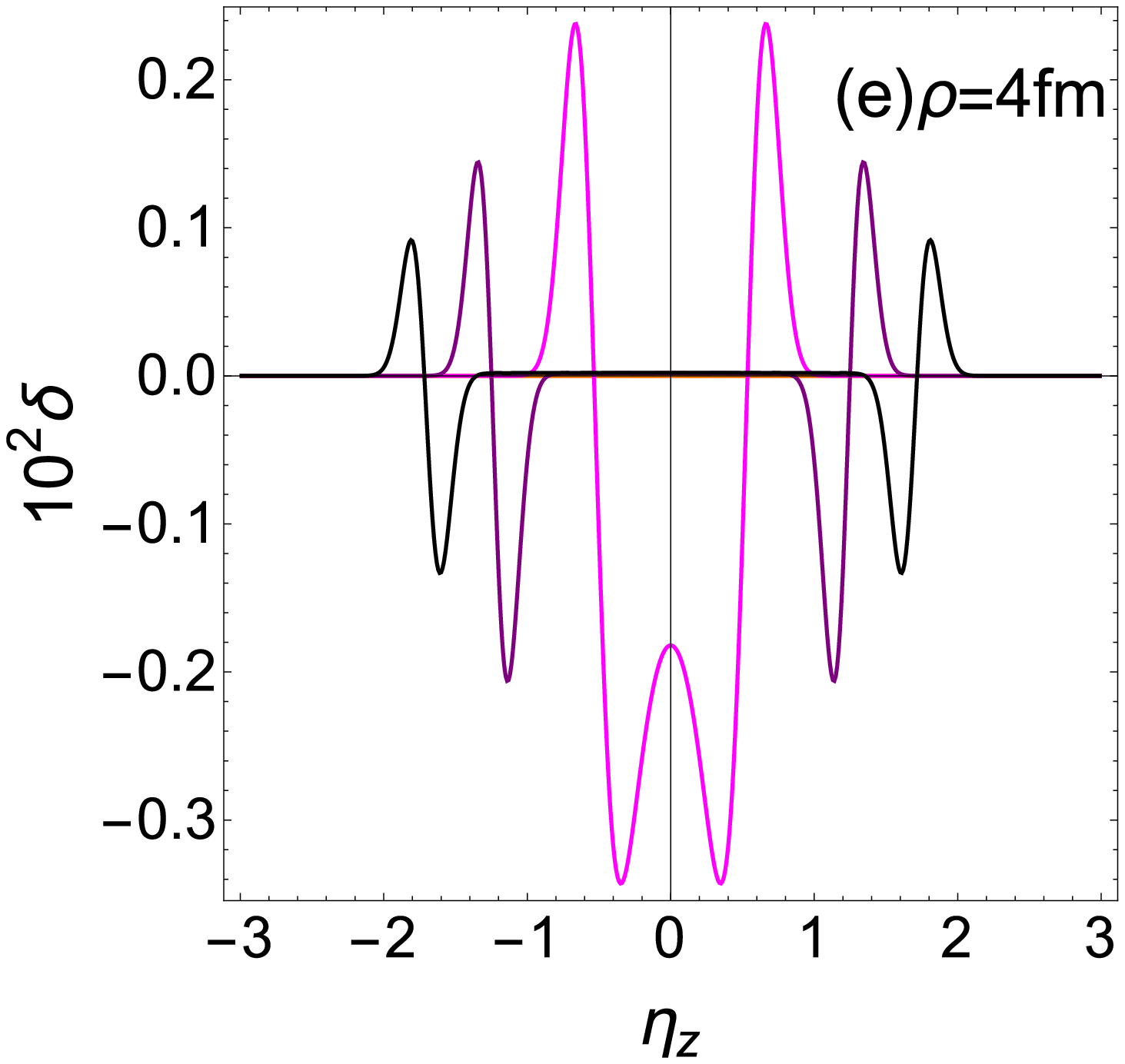} \hspace{0.05in}
\includegraphics[width=0.3\textwidth]{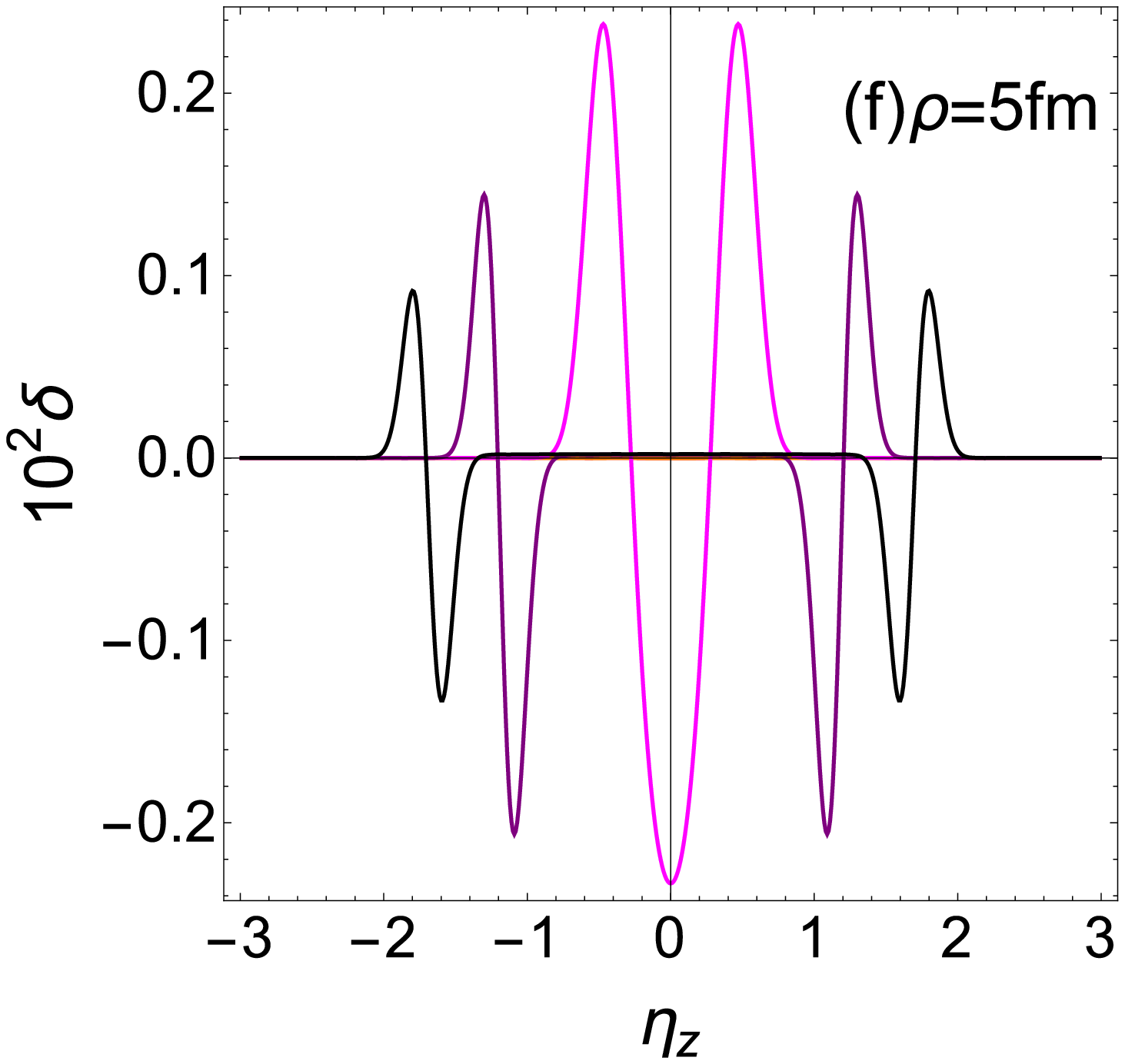}
\vspace{-0.25in}
\caption{(Color online) The sound wave amplitude $\delta=p_1/p_0$ at fixed transverse distance $\rho$ and at evolution time $\tau_r/\tau_r'=$ being 2 (red), 3 (orange), 5 (magenta), 10 (purple) and 20 (black) fm/c. } \label{fig_HB_fix}
\end{figure}

\begin{figure}[!hbt]
\includegraphics[width=0.3\textwidth]{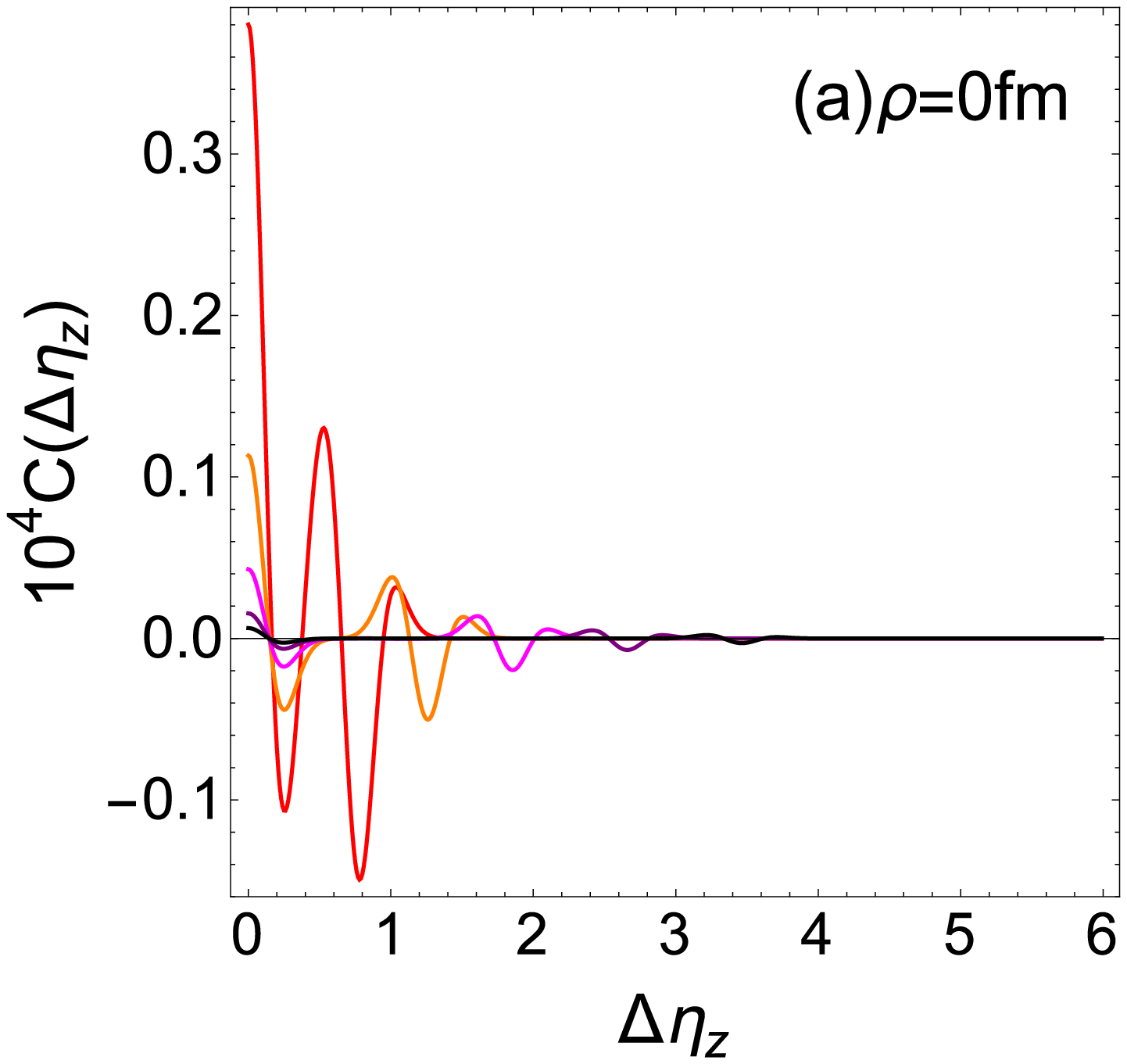} \hspace{0.05in}
\includegraphics[width=0.3\textwidth]{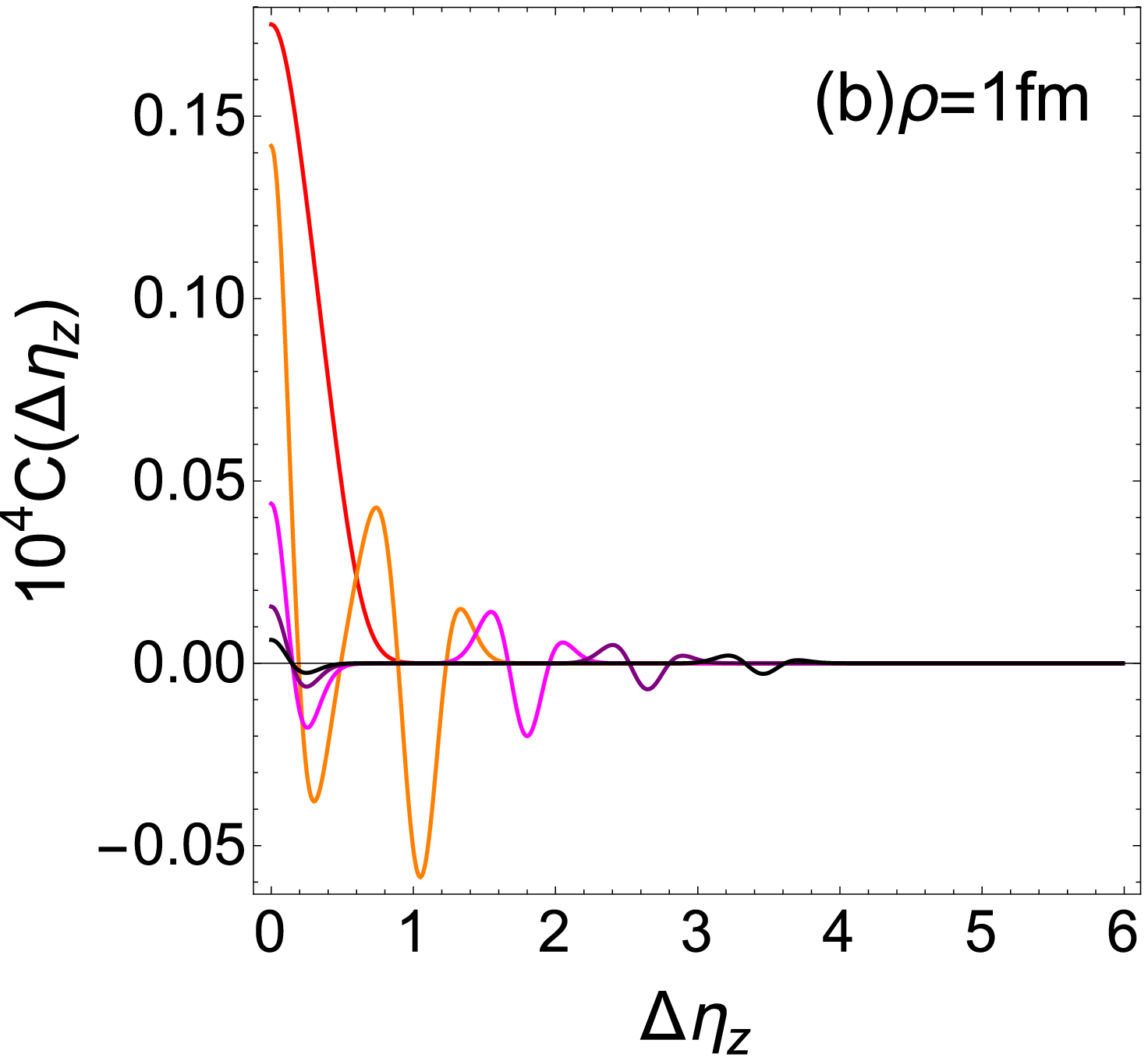} \hspace{0.05in}
\includegraphics[width=0.3\textwidth]{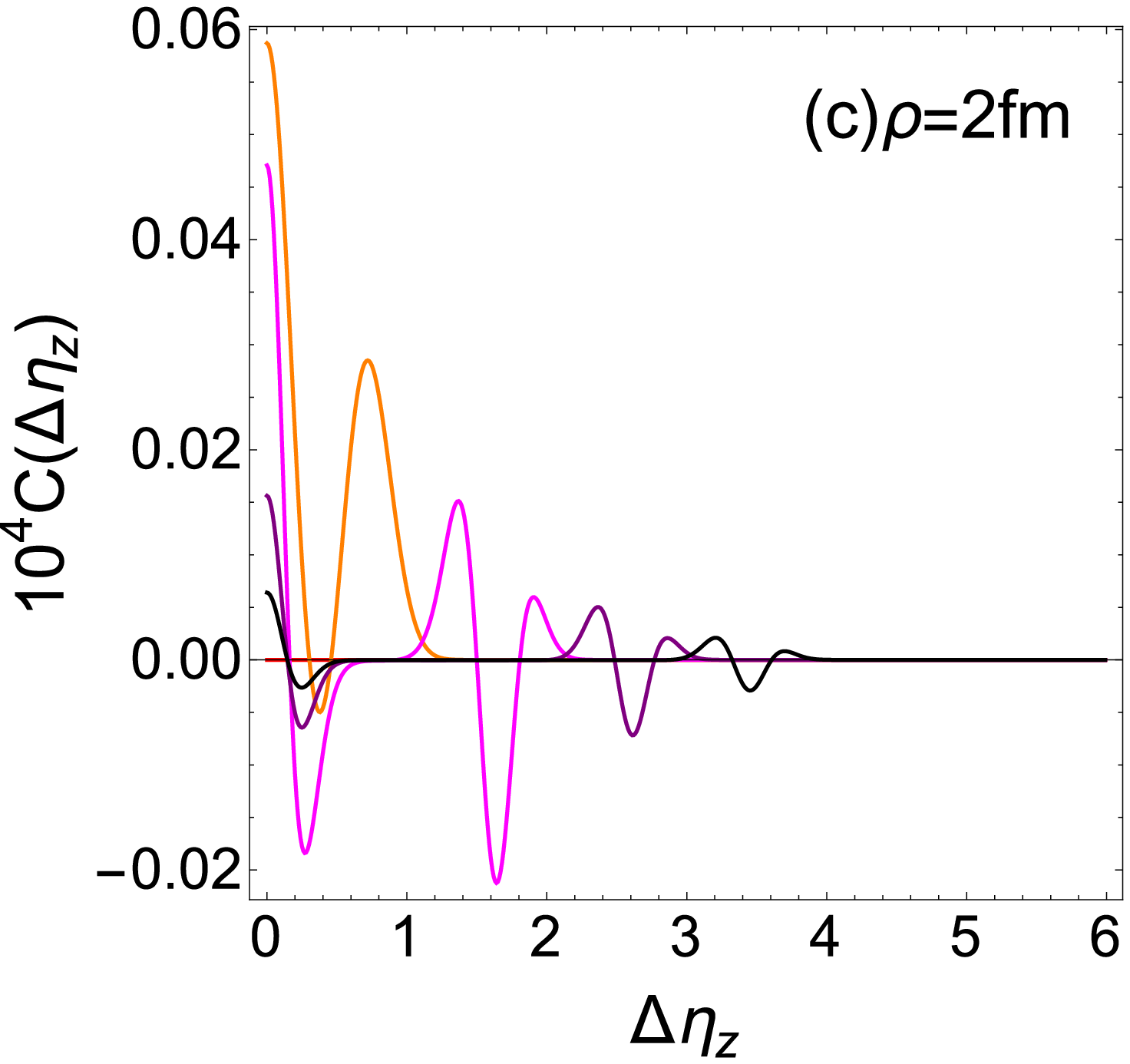} \\
\vspace{-0.15in}
\includegraphics[width=0.3\textwidth]{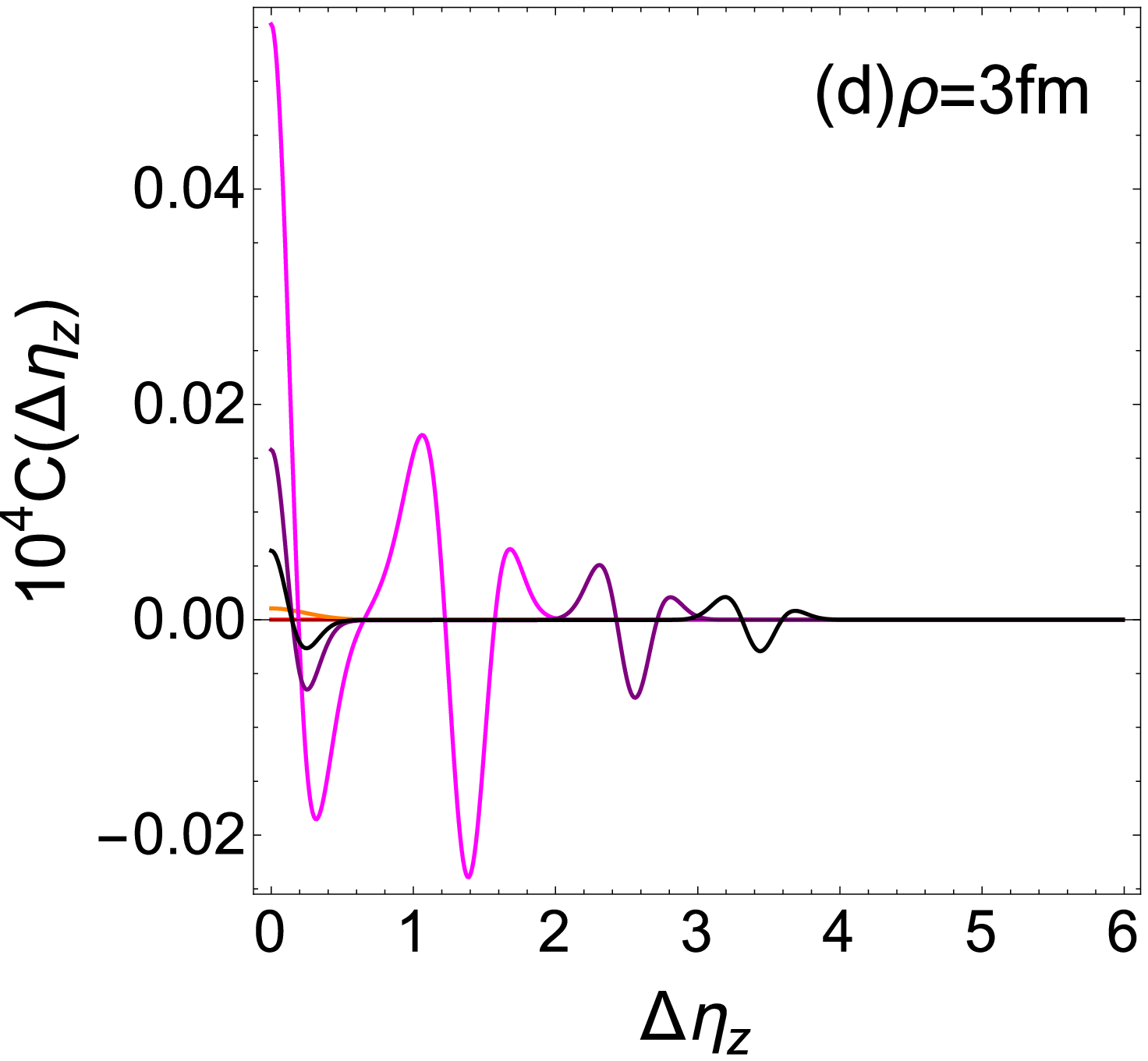} \hspace{0.05in}
\includegraphics[width=0.3\textwidth]{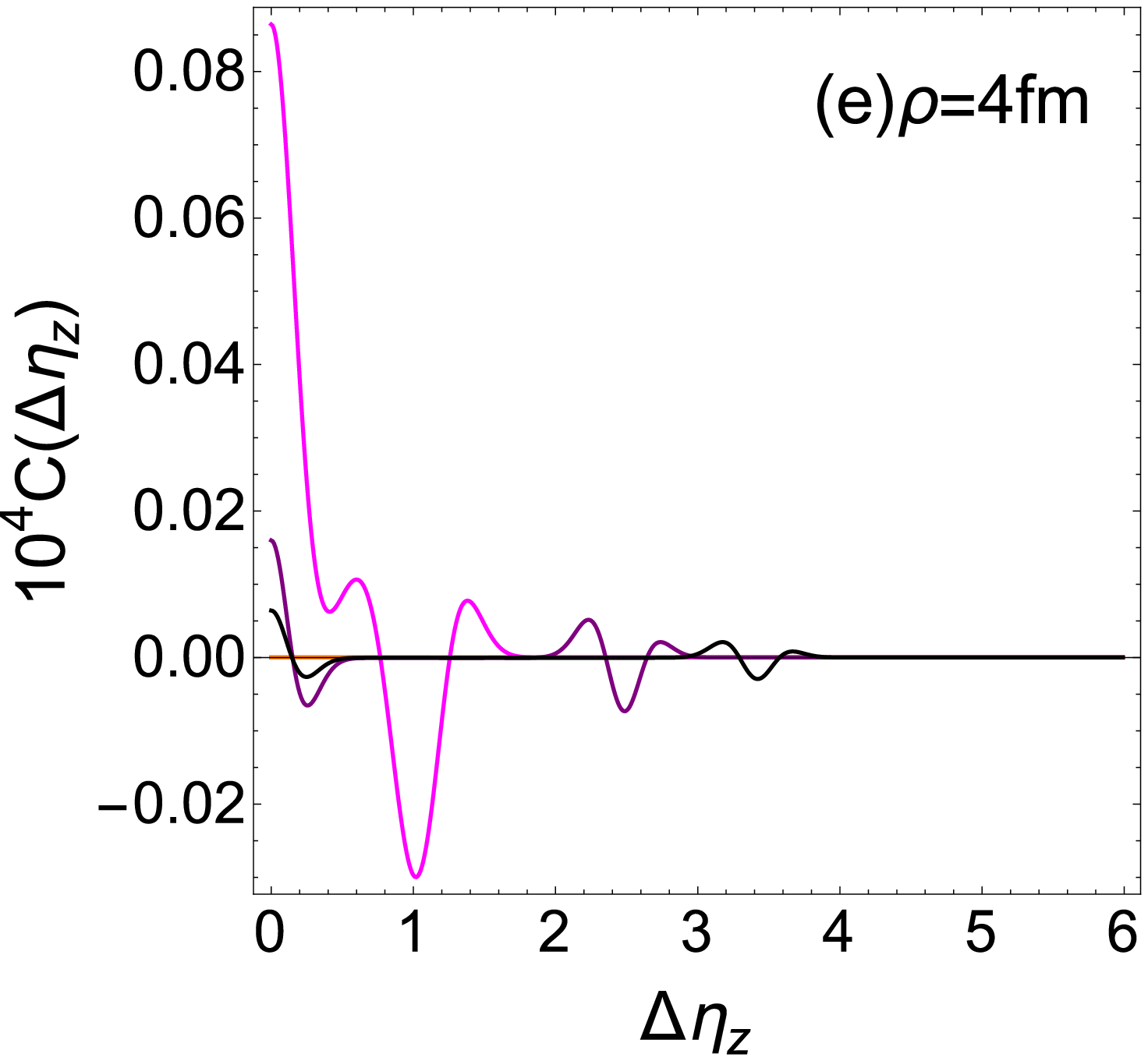} \hspace{0.05in}
\includegraphics[width=0.3\textwidth]{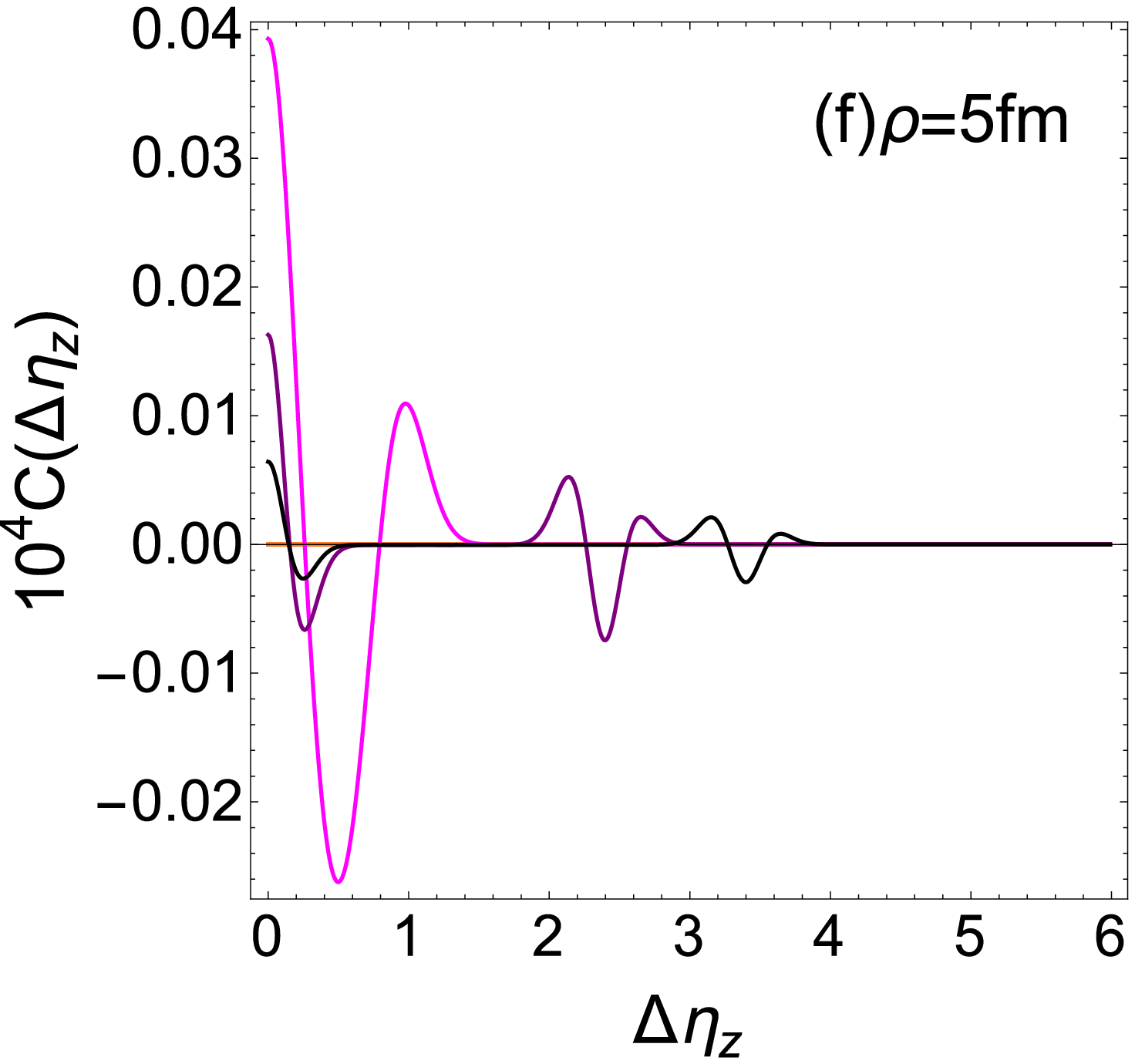}
\vspace{-0.2in}
\caption{(Color online) The pressure-pressure rapidity correlation at fixed transverse distance $\rho$ and at evolution time $\tau_r/\tau_r'=$ 2 (red), 3 (orange), 5 (magenta), 10 (purple) and 20 (black) fm/c. } \label{fig_HB_fix_cor}
\end{figure}

\vspace{-0.2in}
For better comparison with the previous cases from Bjorken flow, we focus on the longitudinal rapidity distribution of the wave amplitude $\delta(\eta_z)$ and the resulting pressure-pressure rapidity correlation $C(\Delta \eta_z)$. Here we make a projection and recast the spatial coordinates in the solution to be $(\tau_r,\eta_z,\rho,\phi)$: note the $\eta_z,\rho,\phi$ are the same as that in the Bjorken case, while we still keep the time coordinate to be $\tau_r$, which is most proper for the background Hubble flow (e.g. with regard to the freeze-out condition). Details about the coordinate projection is given in Appendix \ref{app_cor_Hu}.
Figs.\ref{fig_HB}, \ref{fig_HB_fix}, and \ref{fig_HB_fix_cor} show the evolution of the perturbation as well as the resulting pressure-pressure rapidity correlation at  different radial coordinate $\rho$. As the background is expanding in all directions, the transverse position of the wave-front in this case should be defined as $\rho_f=\tau_r\sinh[c_s\ln(\tau_r/\tau'_r)]$ (see detailed derivation in the Appendix \ref{app_cor_Hu}). We can see that the wave patterns at given $\rho$ relative to the wave-front, are quite similar to the 3D sound waves on top of Bjorken background in Fig.\ref{fig_BJ_G}.  There is however one important difference:  for the 3D wave on top of Bjorken flow, one sees only one wave crest and one trough for each propagating direction; while for the 3D wave on top of Hubble flow, one sees a much stronger oscillatory pattern with multiple crests and troughs along each propagating direction. As a result of such difference, the rapidity correlation in the present Hubble case also develops an oscillating pattern with positive and negative regions in rapidity while in the previous Bjorken case the correlation is always positive without multiple oscillation in rapidity. Clearly, such difference arises from the different background flows in the two cases. A plausible origin of the multiple oscillation pattern in the Hubble case could be the nontrivial interplay between the wave propagation and the transverse expansion of the background flow.

\begin{figure}[!hbt]
\includegraphics[width=0.3\textwidth]{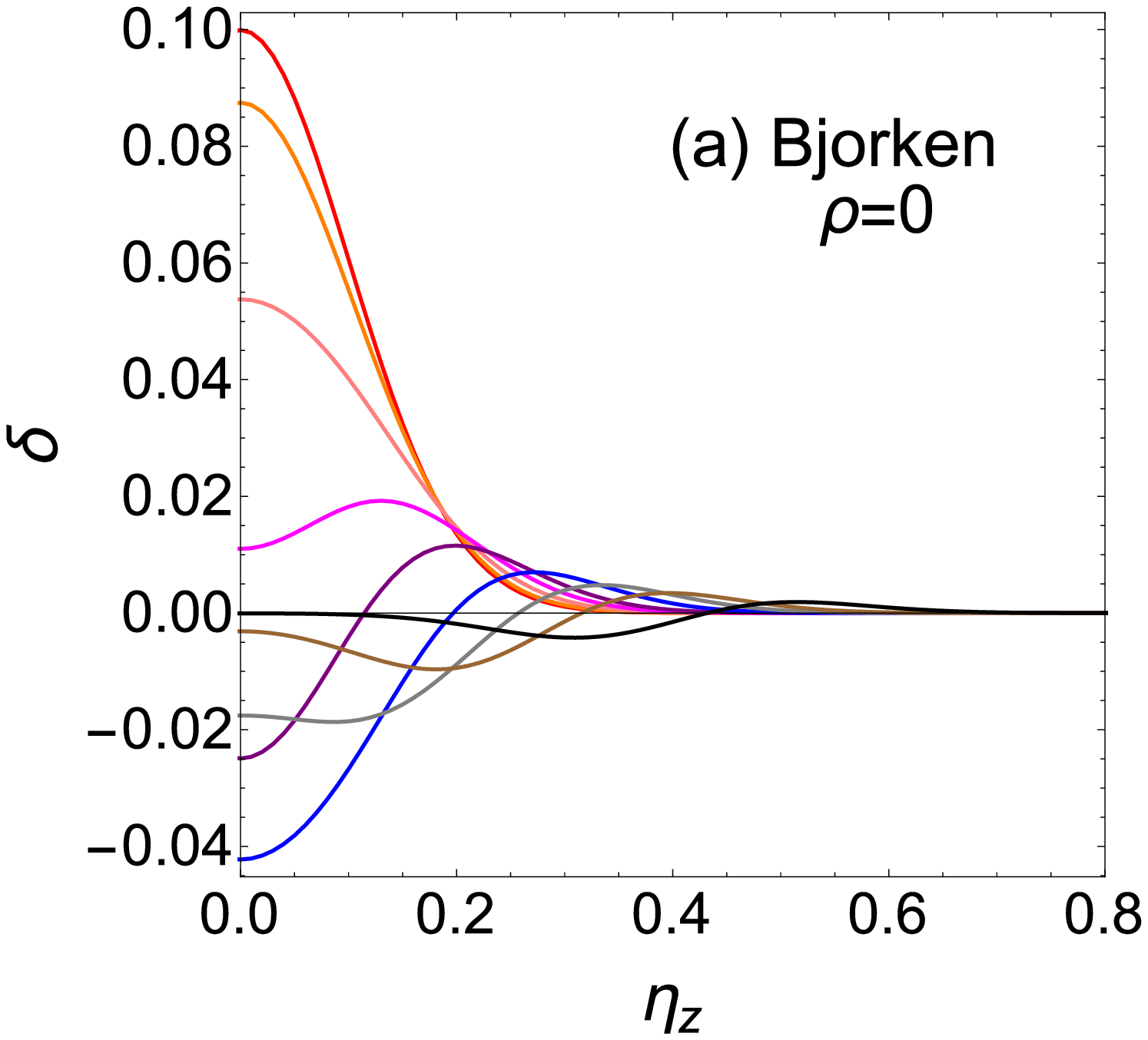}\hspace{0.05in}
\includegraphics[width=0.3\textwidth]{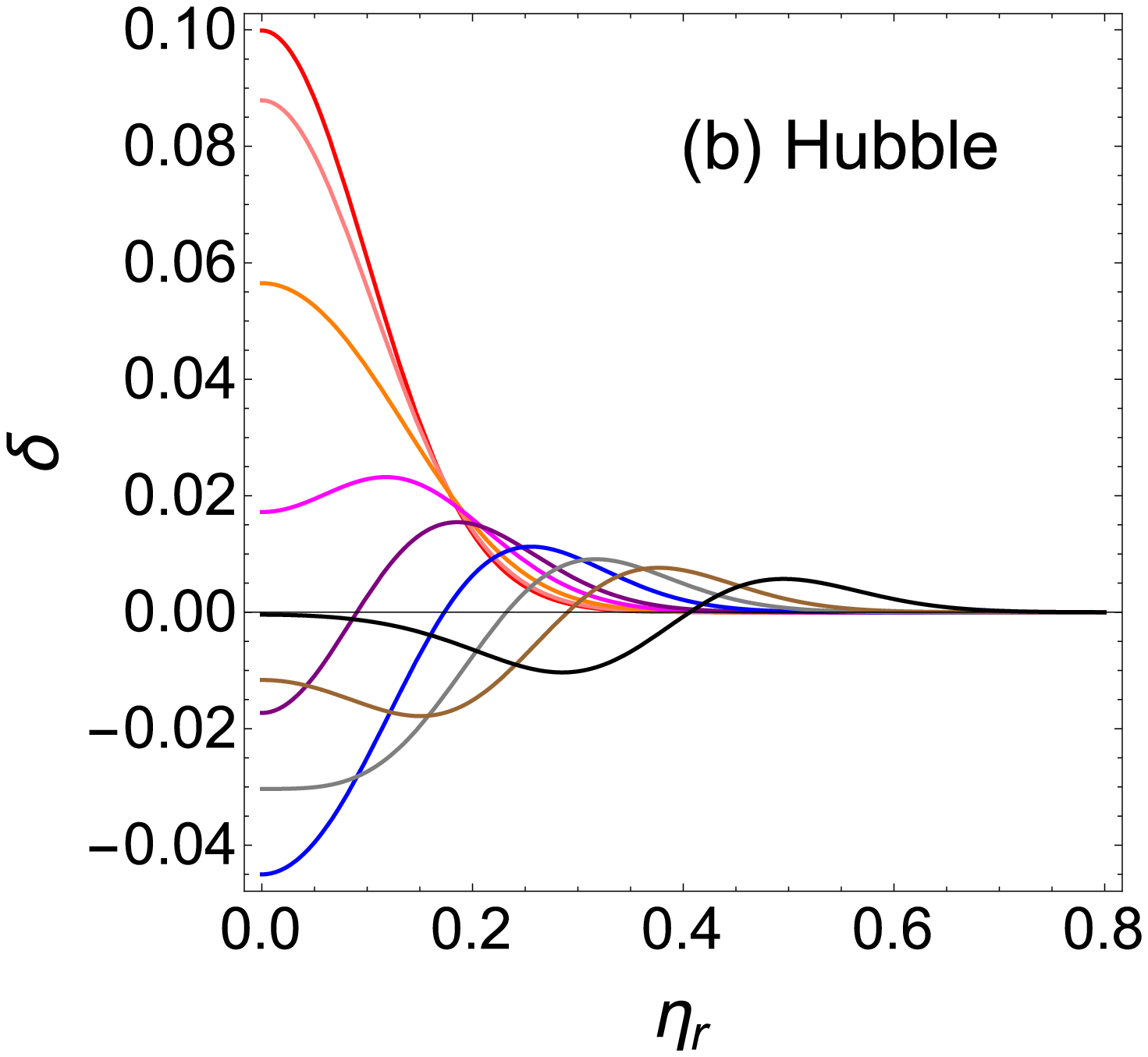}\hspace{0.05in}
\includegraphics[width=0.3\textwidth]{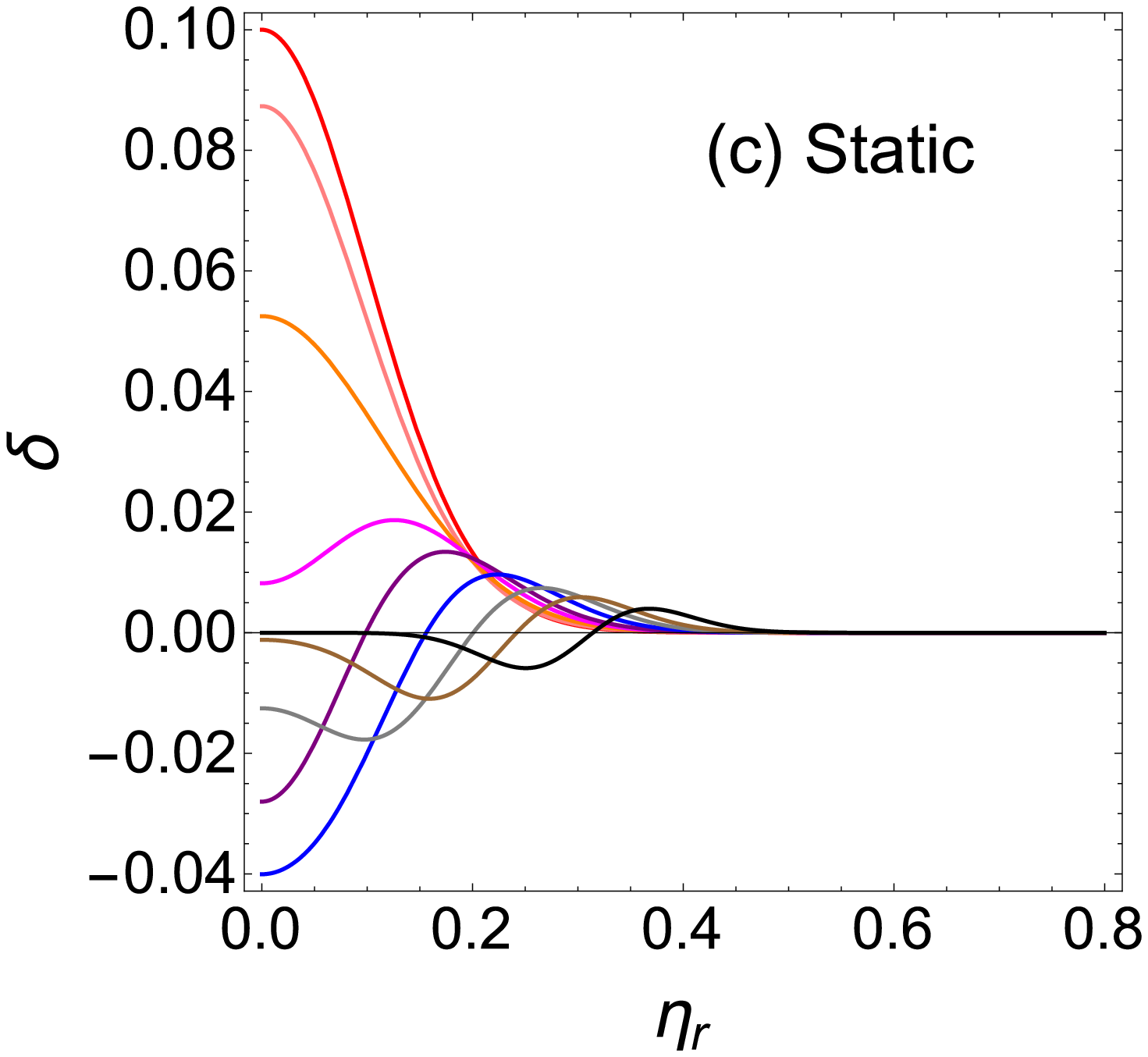}
\vspace{-0.2in}
\caption{(Color online) The early time evolution of 3D sound wave amplitude $\delta=p_1/p_0$ from initial Guassian perturbation at the center  of Bjorken  flow (a), Hubble  flow (b) and static background (c) at evolution time  $\tau/\tau'=$ 0 (red), 0.05 (pink), 0.1 (orange), 0.15 (magenta), 0.2 (purple), 0.3 (blue), 0.4 (gray), 0.5 (brown) and 0.7 (black).} \label{fig_early_tau}
\end{figure}

Finally, as discussed above, we find that the 3D wave amplitude   on top of both Bjorken and Hubble flow and Hubble flow has a structure of a trough following the crest, which is quite different from the 1D longitudinal wave in the Bjorken case. Here we demonstrate that the origin of such difference is due to different dimensions of wave propagation, by showing that the same difference arises  also for 1D and 3D waves on top of a static background without any flow. In that case, a 1D Gaussian perturbation evolves as
\begin{equation}
\delta(t,x)=\frac{\xi}{2(2\pi)^{1/2}\sigma}[e^{-\frac{(x-c_st)^2}{2\sigma^2}}+e^{-\frac{(x+c_st)^2}{2\sigma^2}}],
\end{equation}
while a 3D Gaussian one evolves according to
\begin{equation}
\delta(t,r)=\frac{\xi}{2(2\pi)^{3/2}\sigma^3}[\frac{r-c_st}{r}e^{-\frac{(r-c_st)^2}{2\sigma^2}}+\frac{r+c_st}{r}e^{-\frac{(r+c_st)^2}{2\sigma^2}}].
\end{equation}
The former is always positive, while the latter develops crest/trough structure.  In fact when close to the center of both the Bjorken and the Hubble expansion, the background flow is relatively weak, so the early evolution of perturbation in these cases still keeps a similar patten as in the static background. Hence, we can examine and compare the very early time evolution of perturbations in these cases. Fig.\ref{fig_early_tau} shows the perturbation evolution in these very early moments. Clearly in all three cases, the sound waves  show highly similar crest/trough patterns. One may thus conclude that for the 3D waves in all three cases, the crest/trough patterns are generated early in the evolution near the original perturbation center and subsequently propagating toward wide regions away from the center.

\section{Summary}
\label{s6}
We have studied the evolution of fluctuations in relativistically expanding fluid in the framework of linearized hydrodynamics. The complete and analytic solutions of sound waves on top of the Bjorken flow as well as the Hubble flow backgrounds have been obtained. Regarding fluctuations as perturbations with respect to the background flow, we showed the way to derive the propagation of a  fluctuation in thermodynamic functions such as the pressure. For the often considered Gaussian fluctuation, we obtained analytically the expression of its space-time evolution and saw clearly its propagation on top of the background. We also numerically calculated the rapidity, transverse distance and time dependence of the Gaussian fluctuation and the pressure-pressure correlation which is closely related to the observable correlations in heavy ion collisions. A dedicated study based on the present work and focusing on the  phenomenological applications in heavy ion collisions is in progress and will be reported in a future publication.

\section*{Acknowledgements}
 SS and PZ  acknowledge support from the NSFC (Grant No. 11335005) and the MOST (Grant Nos. 2013CB922000 and 2014CB845400), and JL is supported by the NSF (Grant No. PHY-1352368). JL is also grateful to the RIKEN BNL Research Center for partial support.

\begin{appendix}

\section{Coordinate Transformation for Hubble Flow }
\label{app_1}

In this Appendix we give the details about  the coordinate transformation for the 3D Hubble flow and the corresponding hydrodynamic equations. From the inverse coordinate transformation from $(\tau, \eta, \theta, \phi)$ to $(t, x, y, z)$
\vspace{-0.1in}
\begin{eqnarray}
t &=& \tau~\cosh\eta,~~~~~~~~~~~~~~~~
z ~=~ \tau~\sinh\eta~ \cos\theta,\nonumber\\
x &=& \tau~\sinh\eta~ \sin\theta~ \cos\phi,~~~~
y ~=~ \tau~\sinh\eta~ \sin\theta~ \sin\phi,
\end{eqnarray} 
the metric tensor can be written as \vspace{-0.1in}
\begin{eqnarray}
g_{\mu\nu}&=&\mathrm{Diag}(1,-\tau^2,-\tau^2 \sinh^2\eta, -\tau^2 \sinh^2\eta~\sin^2\theta),\nonumber\\
g^{\mu\nu}&=&\mathrm{Diag}\big(1,-\frac{1}{\tau^2},-\frac{1}{\tau^2 \sinh^2\eta}, -\frac{1}{\tau^2 \sinh^2\eta~\sin^2\theta}\big),
\end{eqnarray}\vspace{-0.05in}
and the four-velocity $(u^\tau, u^\eta, u^\theta, u^\phi)$ is related to the three-velocity $(v^x, v^y, v^z)$ through
\begin{eqnarray}
u^\tau&=&{\gamma}(-v^x \sinh\eta~ \sin\theta~\cos\phi-v^y  \sinh\eta~\sin\theta~\sin\phi-v^z \sinh\eta~\cos\theta
+\cosh\eta),\nonumber\\
u^\eta&=&\frac{\gamma}{\tau}(v^x \cosh\eta~ \sin\theta~\cos\phi+v^y  \cosh\eta~\sin\theta~\sin\phi+v^z \cosh\eta~\cos\theta-\sinh\eta),\nonumber\\
u^\theta&=&\frac{\gamma}{\tau~\sinh\eta}(v^x  \cos\theta~\cos\phi+v^y  \cos\theta~\sin\phi-v^z \sin\theta),\nonumber\\
u^\phi&=&\frac{\gamma}{\tau~\sinh\eta~ \sin\theta}(v^y \cos\phi-v^x \sin\phi).
\end{eqnarray}

From the Affine connections with non-vanishing terms
\begin{eqnarray}
&& \Gamma^\tau_{~\eta\eta}=\tau,\ \ \ \ \ \Gamma^\tau_{~\theta\theta}=\tau\sinh^2\eta,\ \ \ \ \ \Gamma^\tau_{~\phi\phi}=\tau\sinh^2\eta~\sin^2\theta,\nonumber\\
&& \Gamma^\eta_{~\theta\theta}=-\sinh\eta~\cosh\eta,\ \ \ \ \ \Gamma^\eta_{~\phi\phi}=-\sinh\eta~\cosh\eta~\sin^2\theta,\nonumber\\
&& \Gamma^\theta_{~\phi\phi}=-\sin\theta~\cos\theta,\ \ \ \ \ \Gamma^\phi_{~\theta\phi}=\Gamma^\phi_{~\phi\theta}=\frac{\cos\theta}{\sin\theta},\nonumber\\
&& \Gamma^\eta_{~\tau\eta}=\Gamma^\eta_{~\eta\tau}=\Gamma^\theta_{~\tau\theta}=\Gamma^\theta_{~\theta\tau}=
\Gamma^\phi_{~\tau\phi}=\Gamma^\phi_{~\phi\tau}=\frac{1}{\tau},\nonumber\\
&& \Gamma^\theta_{~\eta\theta}=\Gamma^\theta_{~\theta\eta}=
\Gamma^\phi_{~\eta\phi}=\Gamma^\phi_{~\phi\eta}=
\frac{\cosh\eta}{\sinh\eta}
\end{eqnarray}
and the covariant derivatives
\begin{eqnarray}
T^{\alpha\tau}_{~~;\tau}&=&T^{\alpha\tau}_{~~,\tau}
+\delta^\alpha_\eta \Gamma^\eta_{~\eta\tau}T^{\tau\eta}
+\delta^\alpha_\theta \Gamma^\theta_{~\theta\tau}T^{\tau\theta}
+\delta^\alpha_\phi \Gamma^\phi_{~\phi\tau}T^{\tau\phi},\nonumber\\
T^{\alpha\eta}_{~~;\eta}&=&T^{\alpha\eta}_{~~,\eta}
+\Gamma^\eta_{~\tau\eta}T^{\alpha\tau}
+\delta^\alpha_\tau \Gamma^\tau_{~\eta\eta}T^{\eta\eta}
+\delta^\alpha_\eta \Gamma^\eta_{~\eta\tau}T^{\tau\eta}
+\delta^\alpha_\theta \Gamma^\theta_{~\theta\eta}T^{\eta\theta}
+\delta^\alpha_\phi \Gamma^\phi_{~\phi\eta}T^{\eta\phi},\nonumber\\
T^{\alpha\theta}_{~~;\theta}&=&T^{\alpha\theta}_{~~,\theta}
+\Gamma^\theta_{~\tau\theta}T^{\alpha\tau}
+\Gamma^\theta_{~\eta\theta}T^{\alpha\eta}
+\delta^\alpha_\tau\Gamma^\tau_{~\theta\theta}T^{\theta\theta}
+\delta^\alpha_\eta\Gamma^\eta_{~\theta\theta}T^{\theta\theta}
+\delta^\alpha_\phi\Gamma^\phi_{~\phi\theta}T^{\theta\phi}\nonumber\\
&&+\delta^\alpha_\theta \Gamma^\theta_{~\tau\theta}T^{\theta\tau}
+\delta^\alpha_\theta \Gamma^\theta_{~\eta\theta}T^{\theta\eta},\nonumber\\
T^{\alpha\phi}_{~~;\phi}&=&T^{\alpha\phi}_{~~,\phi}
+\Gamma^\phi_{~\tau\phi}T^{\alpha\tau}
+\Gamma^\phi_{~\eta\phi}T^{\alpha\eta}
+\Gamma^\phi_{~\theta\phi}T^{\alpha\theta}
+\delta^\alpha_\tau \Gamma^\tau_{~\phi\phi}T^{\phi\phi}
+\delta^\alpha_\eta \Gamma^\eta_{~\phi\phi}T^{\phi\phi}\nonumber\\&&
+\delta^\alpha_\theta \Gamma^\theta_{~\phi\phi}T^{\phi\phi}
+\delta^\alpha_\phi\Gamma^\phi_{~\tau\phi}T^{\phi\tau}
+\delta^\alpha_\phi\Gamma^\phi_{~\eta\phi}T^{\phi\eta}
+\delta^\alpha_\phi\Gamma^\phi_{~\theta\phi}T^{\phi\theta},
\end{eqnarray}
the full hydrodynamic equations in the frame $(\tau, \eta, \theta, \phi)$ are expressed as
\begin{eqnarray}
T^{\tau\lambda}_{~~;\lambda}&=&T^{\tau\tau}_{~~,\tau}+T^{\tau\eta}_{~~,\eta}+T^{\tau\theta}_{~~,\theta}+T^{\tau\phi}_{~~,\phi}
+\Gamma^\eta_{~\tau\eta}T^{\tau\tau}
+\Gamma^\tau_{~\eta\eta}T^{\eta\eta}
+\Gamma^\theta_{~\tau\theta}T^{\tau\tau}
+\Gamma^\theta_{~\eta\theta}T^{\tau\eta}\nonumber\\&&
+\Gamma^\tau_{~\theta\theta}T^{\theta\theta}
+\Gamma^\phi_{~\tau\phi}T^{\tau\tau}
+\Gamma^\phi_{~\eta\phi}T^{\tau\eta}
+\Gamma^\phi_{~\theta\phi}T^{\tau\theta}
+\Gamma^\tau_{~\phi\phi}T^{\phi\phi},\nonumber\\
T^{\eta\lambda}_{~~;\lambda}&=&T^{\eta\tau}_{~~,\tau}+T^{\eta\eta}_{~~,\eta}+T^{\eta\theta}_{~~,\theta}+T^{\eta\phi}_{~~,\phi}
+3\Gamma^\eta_{~\eta\tau}T^{\tau\eta}
+\Gamma^\theta_{~\tau\theta}T^{\eta\tau}
+\Gamma^\theta_{~\eta\theta}T^{\eta\eta}
+\Gamma^\eta_{~\theta\theta}T^{\theta\theta}\nonumber\\&&
+\Gamma^\phi_{~\tau\phi}T^{\eta\tau}
+\Gamma^\phi_{~\eta\phi}T^{\eta\eta}
+\Gamma^\phi_{~\theta\phi}T^{\eta\theta}
+\Gamma^\eta_{~\phi\phi}T^{\phi\phi},\nonumber\\
T^{\theta\lambda}_{~~;\lambda}&=&T^{\theta\tau}_{~~,\tau}+T^{\theta\eta}_{~~,\eta}+T^{\theta\theta}_{~~,\theta}+T^{\theta\phi}_{~~,\phi}
+3\Gamma^\theta_{~\theta\tau}T^{\tau\theta}
+\Gamma^\eta_{~\tau\eta}T^{\theta\tau}
+3\Gamma^\theta_{~\theta\eta}T^{\eta\theta}
+\Gamma^\phi_{~\tau\phi}T^{\theta\tau}\nonumber\\&&
+\Gamma^\phi_{~\eta\phi}T^{\theta\eta}
+\Gamma^\phi_{~\theta\phi}T^{\theta\theta}
+\Gamma^\theta_{~\phi\phi}T^{\phi\phi},\nonumber\\
T^{\phi\lambda}_{~~;\lambda}&=&T^{\phi\tau}_{~~,\tau}+T^{\phi\eta}_{~~,\eta}+T^{\phi\theta}_{~~,\theta}+T^{\phi\phi}_{~~,\phi}
+3\Gamma^\phi_{~\phi\tau}T^{\tau\phi}
+\Gamma^\eta_{~\tau\eta}T^{\phi\tau}
+3\Gamma^\phi_{~\phi\eta}T^{\eta\phi}
+\Gamma^\theta_{~\tau\theta}T^{\phi\tau}\nonumber\\&&
+\Gamma^\theta_{~\eta\theta}T^{\phi\eta}
+3\Gamma^\phi_{~\phi\theta}T^{\theta\phi}.
\end{eqnarray}

\section{Sound Wave on Top of Hubble Flow}
\label{app_2}

In this Appendix we give the detailed steps for obtaining the sound wave solutions on top of Hubble flow. The sound wave upon the Hubble-expanding system satisfies the evolution equation
\begin{eqnarray}
3\tau^2p_{1,\tau\tau}+{27\tau}p_{1,\tau}+{48}p_1=p_{1,\eta\eta}+2 \frac{\cosh\eta}{\sinh\eta} p_{1,\eta}+\frac{1}{\sinh^2\eta}\big( p_{1,\theta\theta}+\frac{\cos\theta}{\sin\theta}p_{1,\theta}+\frac{1}{\sin^2\theta}p_{1,\phi\phi} \big).
\end{eqnarray}
Expanding the pressure $p_1$ in terms of the spherical harmonic functions,
\begin{eqnarray}
p_1(\tau,\eta,\theta,\phi)&=&\sum_{l,m} p_{l,m}(\tau,\eta) Y_l^m(\theta,\phi),\nonumber\\
3\tau^2\frac{\partial^2 p_{l,m}}{\partial \tau^2}+27\tau\frac{\partial p_{l,m}}{\partial \tau}+48p_{l,m}
&=&\frac{\partial^2 p_{l,m}}{\partial\eta^2}+2 \frac{\cosh\eta}{\sinh\eta}\frac{\partial p_{l,m}}{\partial \eta}-\frac{l(l+1)}{\sinh^2\eta}p_{l,m}
\end{eqnarray}
and then separating the $\tau$ and $\eta$ dependence $p_{l,m}(\tau,\eta)=T_{l,m}(\tau)H_{l,m}(\eta)$ which leads to two independent equations
\begin{eqnarray}
3\tau^2T''_{l,m}+27\tau T'_{l,m}+48T_{l,m}&=&(k^2-1) T_{l,m},\nonumber\\
H''_{l,m}+2 \frac{\cosh\eta}{\sinh\eta} H'_{l,m}-\frac{l(l+1)}{\sinh^2\eta} H_{l,m}&=&(k^2-1) H_{l,m}
\end{eqnarray}
with the solution
\begin{eqnarray}
T_{l,m}&=&\tau^{\pm \mathrm{i}\sqrt{\frac{1-k^2}{3}}-4},\nonumber\\
H_{l,m}&=&c_{l,m} \sinh^{l}\eta~
_2F_1(\frac{l+1+k}{2},\frac{l+1-k}{2},\frac{3}{2}+l,-\sinh^2\eta),
\end{eqnarray}
we have the pressure fluctuation
 \begin{eqnarray}
p_1&=&p_0\sum_{l,m}\int_{-1}^1 \Big[p_{l,m}(k) e^{\pm\mathrm{i}\sqrt{\frac{1-k^2}{3}}\ln(\tau/\tau')} Y_l^m(\theta,\phi)\nonumber\\
&&\times \sinh^{l}\eta~
_2F_1(\frac{l+1+k}{2},\frac{l+1-k}{2},\frac{3}{2}+l,-\sinh^2\eta)\Big]~\mathrm{d}k,
\end{eqnarray}
where $k$ can be either a real number or a pure imaginary number. For the latter, we can do the transformation $k\to ik$ and obtain
\begin{eqnarray}
p_1&=&p_0\sum_{l,m}\int_{-\infty}^\infty \Big[p_{l,m}(k) e^{\pm\mathrm{i}\sqrt{\frac{1+k^2}{3}}\ln(\tau/\tau')} Y_l^m(\theta,\phi)\nonumber\\
&&\times \sinh^{l}\eta~
_2F_1(\frac{l+1+ik}{2},\frac{l+1-ik}{2},\frac{3}{2}+l,-\sinh^2\eta)\Big]~\mathrm{d}k.
\end{eqnarray}•

From the boundary condition $\sinh\eta\to\infty$ in the limit of $r\to\infty$, $p_1\sinh\eta$ should not be larger than $\mathcal{O}(\eta^0)$ at $\eta\to\infty$. Considering the asymptotic behavior of the hypergeometric functions,
\begin{eqnarray}
&&\sinh^{l}\eta~
_2F_1(\frac{l+1+k}{2},\frac{l+1-k}{2},\frac{3}{2}+l,-\sinh^2\eta) \sim e^{(|k|-1)\eta},\nonumber \\
&&\sinh^{l}\eta~
_2F_1(\frac{l+1+ik}{2},\frac{l+1-ik}{2},\frac{3}{2}+l,-\sinh^2\eta) \sim e^{-\eta} e^{ik\eta},
\end{eqnarray}•
the proper pressure fluctuation which satisfies the boundary condition should take the form
\begin{eqnarray}
p_1&=&p_0\sum_{l,m}\int_{-\infty}^\infty \Big[p_{l,m}(k) e^{\pm\mathrm{i}\sqrt{\frac{1+k^2}{3}}\ln(\tau/\tau')} Y_l^m(\theta,\phi)\nonumber\\
&&\times \sinh^{l}\eta~
_2F_1(\frac{l+1+ik}{2},\frac{l+1-ik}{2},\frac{3}{2}+l,-\sinh^2\eta)\Big]~\mathrm{d}k.
\end{eqnarray}

Taking into account the symmetric behavior of the hypergeometric functions with respect to the parameters $a$ and $b$, $_2F_1(a,b;c;x) = ~_2F_1(b,a;c;x)$, we find
\begin{eqnarray}
&&\Big[\sinh^{l}\eta~
_2F_1(\frac{l+1+ik}{2},\frac{l+1-ik}{2},\frac{3}{2}+l,-\sinh^2\eta)\Big]^* \nonumber\\
&=&\Big[\sinh^{l}\eta~
_2F_1(\frac{l+1-ik}{2},\frac{l+1+ik}{2},\frac{3}{2}+l,-\sinh^2\eta)\Big] \nonumber\\
&=&\Big[\sinh^{l}\eta~
_2F_1(\frac{l+1+ik}{2},\frac{l+1-ik}{2},\frac{3}{2}+l,-\sinh^2\eta)\Big]
\end{eqnarray}•
which means that $p_1$ is a real function of $\eta$.

\section{Normalization of the sound wave on top of Hubble flow}
\label{app_3}

In this Appendix we normalize the sound wave showed in Section \ref{4b}. From the asymptotic expression of the radial part of the sound wave at $\eta\to\infty$, \begin{eqnarray}
\widetilde R_l(k,\eta) &\equiv&\sinh^l\eta~_2F_1(\frac{l+1+ik}{2},\frac{l+1-ik}{2},\frac{3}{2}+l,-\sinh^2\eta)
\no  &\to&~~\frac{\sqrt{\pi}2^l\Gamma(l+3/2)}{\sinh(\pi k)}
\Big[ \frac{1}{\Gamma(l+1+ik) \Gamma(1-ik)} + \frac{1}{\Gamma(l+1-ik) \Gamma(1+ik)  }\Big]
\frac{\sin(k\eta)}{\sinh\eta}
\no&& -\frac{\sqrt{\pi}2^l\Gamma(l+3/2)}{\sinh(\pi k)}
\Big[\frac{i}{ \Gamma(l+1+ik)\Gamma(1-ik) } - \frac{i}{\Gamma(l+1-ik)\Gamma(1+ik)  }\Big]
\frac{\cos(k\eta)}{\sinh\eta},
\end{eqnarray}•
its normalization integration derived from the differential equations can be asymptotically expressed as
\begin{eqnarray}
\int_0^\infty \sinh^2 \eta \widetilde R_{l}(k',\eta) \widetilde R_{l}(k,\eta) ~d\eta
&=& \frac{ \sinh^2\eta (\widetilde R'(k,\eta)\widetilde R(k',\eta)-\widetilde R(k,\eta)\widetilde R'(k',\eta)) |_\infty}{k'^2-k^2}\nonumber \\
&  {\to}& \frac{2^{2l+1}\Gamma(l+3/2)^2}{\Gamma(l+1+ik)\Gamma(l+1-ik)}\frac{1}{k\sinh(\pi k)} \left[\frac{\sin[(k'-k)\eta]}{(k'-k)}+\frac{\sin[(k'+k)\eta]}{(k'+k)}\right]_\infty\nonumber \\
&=& \frac{2^{2l+1}\Gamma(l+3/2)^2}{\Gamma(l+1+ik)\Gamma(l+1-ik)}\frac{\pi}{k\sinh(\pi k)} [\delta(k'-k)+\delta(k'+k)].
\end{eqnarray}•

From the relations between the Gegenbauer functions and the Gamma functions and associated Legendre functions,
\begin{eqnarray}
_2F_1(\frac{l+1+ik}{2},\frac{l+1-ik}{2},l+3/2,-\sinh^2\eta) &=&_2F_1(l+1+ik,l+1-ik,l+3/2,-\sinh^2(\eta/2))\nonumber \\
&=& \cosh^{-2l-1}(\eta/2)~_2F_1(1/2+ik,1/2-ik,l+3/2,-\sinh^2(\eta/2))\nonumber\\
&=& \cosh^{-2l-1}(\eta/2)\frac{\Gamma(l+3/2)}{\tanh^{l+1/2}(\eta/2)}
 P_{ik-1/2}^{-l-1/2}(\cosh\eta) \nonumber\\
&=&\frac{\Gamma(l+3/2)2^{l+1/2}}{\sinh^{l+1/2}\eta}
 P_{ik-1/2}^{-l-1/2}(\cosh\eta),
\end{eqnarray}•
the radial part can be written as
\begin{eqnarray}
\widetilde R_l(k,\eta) &\equiv& \sinh^l\eta~ _2F_1(\frac{l+1+ik}{2},\frac{l+1-ik}{2},l+3/2,-\sinh^2\eta) \no
&=&\frac{\Gamma(l+3/2)2^{l+1/2}}{\sinh^{1/2}\eta}
 P_{ik-1/2}^{-l-1/2}(\cosh\eta)  \nonumber\\
&=& {  \frac{\Gamma(2l+2)\Gamma(ik-l)\sinh^{l}\eta}{\Gamma(ik+l+1)} C^{(l+1)}_{ik-l-1}(\cosh\eta)}.
\end{eqnarray}•
For convenient, we modify the radial part with a factor,
\begin{eqnarray}
R_l(k,\eta) &\equiv& \sqrt{\frac{\Gamma(l+1+ik)\Gamma(l+1-ik)}{\pi2^{2l+2}\Gamma(l+3/2)^2}} \widetilde R_l(k,\sinh\eta) \no
&=& \sqrt{\frac{\Gamma(l+1+ik)\Gamma(l+1-ik)}{\pi2^{2l+2}\Gamma(l+3/2)^2}}\sinh^l\eta~ _2F_1(\frac{l+1+ik}{2},\frac{l+1-ik}{2},l+3/2,-\sinh^2\eta) \no
&=& \frac{\Gamma(l+1)}{\sqrt{\Gamma(l+1-ik)\Gamma(l+1+ik)}}
\frac{2^l}{\sinh(\pi k)}\sinh^{l}\eta C^{(l+1)}_{ik-l-1}(\cosh\eta)
\end{eqnarray}•
which satisfies the normalization condition
\begin{eqnarray}
\int_0^\infty \sinh^2 \eta R_{l}(k',\eta) R_{l}(k,\eta) ~d\eta
&=&  \frac{1}{k\sinh(\pi k)} \frac{\delta(k'-k)+\delta(k'+k)}{2}.
\end{eqnarray}•

\section{Sum Rule}
\label{app_4}

In this Appendix we use the relation between the associated Legendre functions and Gegenbauer functions to   prove the summation identity
\begin{eqnarray}
R_0(k,\bar\eta)R_0(k,0)|Y_0^0|^2 = \sum_{l,m}
R_l(k,\eta')R_l(k,\eta)Y_l^m(\theta,\phi)Y_l^m(\theta',\phi')^*.
\end{eqnarray}

From the addition theorem of the spherical harmonic function
\begin{eqnarray}
\sum_{m=-l}^{l} Y_l^m(\theta,\phi)Y_l^m(\theta',\phi')^* =\sqrt{\frac{2l+1}{4\pi}} Y_l^0(\xi,0)
=\frac{2l+1}{4\pi}P_l(\cos\xi)
\end{eqnarray}
with
\begin{eqnarray}
\cos\xi=\cos\theta'\cos\theta+\sin\theta'\sin\theta\cos(\phi-\phi'),
\end{eqnarray}
the identity becomes
\begin{eqnarray}
R_0(k,\bar\eta) = \sum_{l=0}^{\infty}
(2l+1)R_l(k,\eta')R_l(k,\eta)P_l(\cos\xi)
\end{eqnarray}
with
\begin{eqnarray}
\cosh\bar\eta&=&\cosh\eta'\cosh\eta-\sinh\eta'\sinh\eta\cos\xi.
\end{eqnarray}

Representing the radial and angel parts in terms of the Gegenbauer functions,
\begin{eqnarray}
P_l(\cos\xi) &=& C^{(1/2)}_{l}(\cos\xi),\nonumber  \\
R_l(k,\eta)
&=&   \frac{\Gamma(l+1)}{\sqrt{\Gamma(l+1-ik)\Gamma(l+1+ik)}}
\frac{2^l}{\sinh(\pi k)}\sinh^{l}\eta C^{(l+1)}_{ik-l-1}(\cosh\eta),
\end{eqnarray}
the identity further becomes
\begin{eqnarray}
C^{(1)}_{ik-1}(\cosh\bar\eta) &=& \sum_{l=0}^{\infty}
(2l+1)(-4)^l\frac{\Gamma(l+1)^2\Gamma(ik-l)}{\Gamma(l+1+ik)}
 \sinh^{l}\eta C^{(l+1)}_{ik-l-1}(\cosh\eta) \sinh^{l}\eta'C^{(l+1)}_{ik-l-1}(\cosh\eta')C^{(1/2)}_l(\cos\xi) \no
&=& \sum_{l=0}^{\infty}(-1)^{l}
\frac{\Gamma(l+1)\Gamma(ik-l)\Gamma(2l+2)\Gamma(1/2)}{\Gamma(l+1/2)\Gamma(l+1+ik)}
 \sinh^{l}\eta C^{(l+1)}_{ik-l-1}(\cosh\eta) \sinh^{l}\eta'C^{(l+1)}_{ik-l-1}(\cosh\eta')C^{(1/2)}_l(\cos\xi).
\end{eqnarray}
This is exactly the addition theorem of Gegenbauer functions~\cite{gegenbauer}.

\section{Pressure-Pressure Correlation} \label{app_cor}
\subsection{Longitudinal Bjorken Flow}\label{app_cor_BL}
For Guassian perturbation on top of longitudinal Bjorken flow, the correlation can be simplified as
\begin{eqnarray} \label{eq_CC_bj_L}
C_{\delta\delta}(\tau_f,\Delta\eta) &\equiv& \int_{-\infty}^\infty d\xi ~
\delta(\tau_f,\eta-\xi) \delta(\tau_f,\eta+\Delta\eta-\xi)\nonumber \\
&=& \Big(  \frac{\xi}{2\pi} \Big)^2 \Big( \frac{\tau_f}{\tau'} \Big)^{-2/3}
 \int_{-\infty}^{\infty} dk  \int_{-\infty}^{\infty} dk'
e^{-\frac{\sigma^2 k^2}{2}} e^{-\frac{\sigma^2 k'^2}{2}} \no&&\times
\cos[\sqrt{\frac{k^2-1/3}{3}}\ln(\tau_f/\tau')]\cos[\sqrt{\frac{k'^2-1/3}{3}}\ln(\tau_f/\tau')]
 \int_{-\infty}^{\infty} d\xi e^{-i(k+k')\xi}e^{ik'\Delta\eta}\nonumber \\
 &=&  2\pi\Big(  \frac{\xi}{2\pi} \Big)^2 \Big( \frac{\tau_f}{\tau'} \Big)^{-2/3}
 \int_{-\infty}^{\infty} dk e^{-\sigma^2 k^2}\cos^2[\sqrt{\frac{k^2-1/3}{3}}\ln(\tau_f/\tau')]\cos(k\Delta\eta)\nonumber \\
 &=& \frac{\xi^2}{2\pi} \Big( \frac{\tau_f}{\tau'} \Big)^{-2/3}
 \int_{-\infty}^{\infty} dk e^{-\sigma^2 k^2}\cos^2[\sqrt{\frac{k^2-1/3}{3}}\ln(\tau_f/\tau')]\cos(k\Delta\eta).
\end{eqnarray}

\subsection{General Bjorken Flow}\label{app_cor_BG}
For perturbation on top of general Bjorken flow, the integration over transverse distance $\rho$ can be done analytically. From the definition
\begin{eqnarray}
\Omega(\omega)\equiv \int_0^\infty J_0(\omega\rho) \rho d\rho = \lim_{a\to0}\int_0^\infty e^{-a^2\rho^2/2} J_0(\omega\rho) \rho d\rho  = \lim_{a\to0} \frac{1}{a^2}e^{-\frac{\omega^2}{2a^2}}
\end{eqnarray}•
and its property for any function $f(\omega)$ with convergency at $\omega=0$,
\begin{eqnarray}
\int_0^\infty \Omega(\omega) \omega f(\omega) d\omega &=&\lim_{a\to0}  \frac{1}{a^2}\int_0^\infty \omega f(\omega) e^{-\frac{\omega^2}{2a^2}} d\omega \no
&=& \sum_{n=0}^\infty \frac{f^{(n)}(0)}{n!}\lim_{a\to0}  \frac{1}{a^2}\int_0^\infty \omega^{n+1} e^{-\frac{\omega^2}{2a^2}} d\omega \no
&=&  \sum_{n=0}^\infty \frac{f^{(n)}(0)}{n!} \lim_{a\to0}(2a)^{n/2}\Gamma(1+n/2) \no
&=& f(0),
\end{eqnarray}•
we have
\begin{eqnarray}
\bar\delta(\tau_f,\eta)  &=& \iiiint e^{ik\eta} J_0(\omega\rho) W(\omega,k,\tau_f) \rho d\rho d\omega dk d\phi\nonumber\\
&=&2\pi\int_{-\infty}^{\infty} e^{ik\eta} \frac{W}{\omega}(\omega\to0,k,\tau_f) dk
\end{eqnarray}
and
\begin{eqnarray}
W_1(\tau) &=&
\frac{1-c_s^2-2\alpha_k}{2\Gamma(1-\alpha_k)}\left(\frac{c}{2}\right)^{-\alpha_k}
\left(\frac{\tau}{\tau'}\right)^{-\frac{1-c_s^2}{2}-\alpha_k},\nonumber\\
W_2(\tau) &=&
\frac{1-c_s^2+2\alpha_k}{2\Gamma(1+\alpha_k)}\left(\frac{c}{2}\right)^{\alpha_k}
\left(\frac{\tau}{\tau'}\right)^{-\frac{1-c_s^2}{2}+\alpha_k},\nonumber\\
W_3(\tau) &=& (\omega\tau)^{1+c_s^2}
\end{eqnarray}
at $\omega\to 0$ and in turn
\begin{eqnarray}
\frac{W}{\omega}(\omega\to0,k,\tau_f) &=&  \frac{\xi \tau'^2}{8\pi^2}e^{-\frac{\sigma^2k^2}{2}}
\left(\frac{\tau_f}{\tau'}\right)^{-1/3} \no &&\times
\left[\left(1-\frac{1}{\sqrt{1-3k^2}}\right) \left(\frac{\tau_f}{\tau'}\right)^{-\sqrt{1-3k^2}/3}
+\left(1+\frac{1}{\sqrt{1-3k^2}}\right) \left(\frac{\tau_f}{\tau'}\right)^{\sqrt{1-3k^2}/3} \right] \nonumber\\
&\equiv&\frac{\xi \tau'^2}{8\pi^2}e^{-\frac{\sigma^2k^2}{2}}
\left(\frac{\tau_f}{\tau'}\right)^{-1/3}\times F(k,\frac{\tau_f}{\tau'}).
\end{eqnarray}
Consequently, we find the correlation with only one-dimensional integration,
\begin{eqnarray}
C_{\delta\delta}(\tau_f,\Delta\eta)
&\equiv& \int_{-\infty}^\infty d\xi ~
\bar\delta(\tau_f,\eta-\xi) \bar\delta(\tau_f,\eta+\Delta\eta-\xi) \nonumber\\
&=& \Big(  \frac{\xi\tau'^2}{4\pi} \Big)^2 \Big( \frac{\tau_f}{\tau'} \Big)^{-2/3}
 \int_{-\infty}^{\infty} dk  \int_{-\infty}^{\infty} dk'
e^{-\frac{\sigma^2 k^2}{2}} e^{-\frac{\sigma^2 k'^2}{2}}
K(k,\frac{\tau_f}{\tau'})K(k',\frac{\tau_f}{\tau'})
 \int_{-\infty}^{\infty} d\xi e^{-i(k+k')\xi}e^{ik'\Delta\eta} \nonumber\\
 &=&  2\pi\Big(  \frac{\xi\tau'^2}{4\pi} \Big)^2 \Big( \frac{\tau_f}{\tau'} \Big)^{-2/3}
 \int_{-\infty}^{\infty} dk e^{-\sigma^2 k^2}F^2(k,\frac{\tau_f}{\tau'})\cos(k\Delta\eta).
\end{eqnarray}

\subsection{Hubble Flow}\label{app_cor_Hu}
Making the transformation from the frame $(\tau_r, \eta_r, \theta, \phi)$ to the frame $(\tau_r, \eta_z, \rho, \phi)$ leads to the projection
\begin{eqnarray}
\sinh^2\eta_r &\to& \sinh^2\eta_z + \frac{\rho^2}{\tau_r^2}\cosh^2\eta_z, \nonumber\\
\sin^2\theta &\to& \frac{\rho^2}{\tau_r^2\sinh^2\eta_z+\rho^2\cosh^2\eta_z},
\end{eqnarray}
and the evolution of the central Gaussian perturbation is expressed as
\begin{eqnarray}
\delta(\tau_r,\eta_z,\rho,\phi)
&=&\frac{\xi}{2\pi^2}
\int_{-\infty}^\infty \frac{\sin(k\sigma^2)}{k\sigma^2e^{-\sigma^2/2}}e^{-\frac{\sigma^2 k^2}{2}}
\frac{1}{\sqrt{\tau_r^2\tanh^2\eta_z+\rho^2}} \cos[\beta_k \ln(\tau_r/\tau')]
\frac{\cosh^{ik-1}\eta_z}{\tau_r^{ik-1}}\no
&&\times\frac{(\sqrt{\tau_r^2+\rho^2}+\sqrt{\tau_r^2\tanh^2\eta_z+\rho^2})^{ik}
-(\sqrt{\tau_r^2+\rho^2}-\sqrt{\tau_r^2\tanh^2\eta_z+\rho^2})^{ik}}{2i } k~dk.
\end{eqnarray}
The position of the wave-front can be found from the dispersion relation,
\begin{eqnarray}
\frac{\delta \eta}{\delta \tau} &\approx& \frac{c_s}{\tau},\nonumber \\
\eta_f &=& c_s \ln(\tau_r/\tau_r'),\nonumber \\
\rho_f &=& \tau_r \sinh\eta_f = \tau_r\frac{(\tau_r/\tau_r')^{c_s}-(\tau_r/\tau_r')^{-c_s}}{2}.
\end{eqnarray}•

\subsection{Pressure-Pressure Correlation in General}

For a given background flow there will be multiple sound wave solutions in general, and a given initial perturbation will trigger a certain superposition of these solutions which subsequently propagate independently. A particularly important case is the solution from an initial condition of delta-function perturbation  $\sim \delta(t-t')\delta^{(3)}(\vec r -\vec r')$, which can be denoted as the ``Green's function'' $G(t,\vec r ; t', \vec r')$. Suppose space-time dependent perturbations $f(t',\vec r')$ are present for all time $t'<t$, then at time $t$, the sound solution (in terms of pressure) is then
\begin{eqnarray}
p_1 (t, \vec r) = \int_{t'<t,\vec r'} G(t, \vec r; t' , \vec r') \,  f (t', \vec r'),
\end{eqnarray}
the equal time correlations (on average) at given time $t$ is then given by
\begin{eqnarray}
\langle p_1(t,\vec r_1) \, p_1 (t, \vec r_2) \rangle =  \int_{t'<t,\vec r'} \int_{t''<t,\vec r''}  G(t, \vec r_1; t' , \vec r') \,    G(t, \vec r_2; t'' , \vec r'') \,  \langle  f (t', \vec r') \, f (t'', \vec r'')  \rangle .
\end{eqnarray}
If one is considering sound perturbation from only initial fluctuations $f_0(\vec r)$ at time $t_0$, then the above correlations are reduced to
\begin{eqnarray}
\langle p_1(t,\vec r_1) \, p_1 (t, \vec r_2) \rangle =  \int_{\vec r'} \int_{\vec r''}  G(t, \vec r_1; t_0 , \vec r') \,    G(t, \vec r_2; t_0 , \vec r'') \,  \langle f_0 (\vec r') \, f_0 (\vec r'') \rangle .
\end{eqnarray}
If the initial fluctuations at different space points are uncorrelated, $\langle  f_0 (\vec r') \, f_0 (\vec r'') \rangle \to F(\vec r') \delta^{(3)}(\vec r' - \vec r'')$, we have
\begin{eqnarray}
\langle p_1(t,\vec r_1) \, p_1 (t, \vec r_2) \rangle =  \int_{\vec r'}    G(t, \vec r_1; t_0 , \vec r') \,    G(t, \vec r_2; t_0 , \vec r') \,  F(\vec r').
\end{eqnarray}
Alternatively for hydrodynamic fluctuations that occur stochastically and locally in space-time throughout the course of background hydrodynamic evolution, $\langle  f (t', \vec r') \, f (t'', \vec r'')  \rangle \to F(t',\vec r') \delta(t' - t'') \delta^{(3)}(\vec r' - \vec r'')$, one obtains
\begin{eqnarray}
\langle p_1(t,\vec r_1) \, p_1 (t, \vec r_2) \rangle =  \int_{t'<t,\vec r'}    G(t, \vec r_1; t' , \vec r') \,    G(t, \vec r_2; t' , \vec r') \,  F(t', \vec r')  \,\, .
\end{eqnarray}
The above formulation will be very useful for the phenomenological application of the found sound wave solutions, and for completeness we include this discussion here. 

\end{appendix}

\end{document}